\documentclass[10pt,aps,prx,twocolumn,showpacs,superscriptaddress,nobalancelastpage,longbibliography]{revtex4-2}
\pdfoutput=1
\usepackage[utf8]{inputenc}
\usepackage[english]{babel}
\usepackage[T1]{fontenc}
\usepackage{amsmath}
\usepackage{dsfont}
\usepackage{enumitem}
\usepackage{hyperref}
\usepackage{graphicx}
\usepackage{wrapfig}
\usepackage{dcolumn}
\usepackage{bm}
\usepackage{hyperref}
\usepackage[mathlines]{lineno}
\usepackage{amssymb}
\usepackage[f]{esvect}
\usepackage[braket, qm]{qcircuit}  
\usepackage{listings}

\usepackage{xcolor}
\usepackage{tabulary}
\usepackage{booktabs}
\usepackage{amssymb,amsthm}
\usepackage{nicematrix}
\usepackage{relsize}
\usepackage{tikz}
\usepackage{lipsum}
\usepackage{import}
\usepackage{ulem}

\newcommand{\red}[1]{\textcolor{red}{#1}}

\newcommand{\delete}[1]{}

\newcommand{\brm}[1]{\bm{\mathrm{#1}}}
\newcommand{\blm}[1]{\bm{\mathcal{#1}}}
\newcommand{\hl}{\brm{L}_{k}}
\newcommand{\hll}{\brm{L}_{k}^{\ell}}
\newcommand{\hlu}{\brm{L}_{k}^{u}}

\newcommand{\hlln}{\blm{L}_{k}^{\ell}}
\newcommand{\hlun}{\blm{L}_{k}^{u}}
\newcommand{\bk}{\brm{B}_{k}}
\newcommand{\bkdag}{\brm{B}_{k}^{\dagger}}
\newcommand{\bkk}{\brm{B}_{k+1}}
\newcommand{\bkkdag}{\brm{B}_{k+1}^{\dagger}}
\newcommand{\bkn}{\blm{B}_{k}}

\newcommand{\bkkn}{\blm{B}_{k+1}}
\newcommand{\bkkdagn}{\blm{B}_{k+1}^{\dagger}}
\newcommand{\ketbra}[2]{\lvert#1\rangle\langle#2\rvert}

\newcommand{\p}[2]{\mathrm{Poly}_{#1}#2}
\newcommand{\norm}[2]{\left\lVert #1 - #2\right\rVert}
\newcommand{\commentout}[1]{}

\makeatletter
\newsavebox{\@brx}
\newcommand{\llangle}[1][]{\savebox{\@brx}{\(\m@th{#1\langle}\)}%
  \mathopen{\copy\@brx\kern-0.5\wd\@brx\usebox{\@brx}}}
\newcommand{\rrangle}[1][]{\savebox{\@brx}{\(\m@th{#1\rangle}\)}%
  \mathclose{\copy\@brx\kern-0.5\wd\@brx\usebox{\@brx}}}
\makeatother

\newtheorem{theorem}{Theorem}
\newtheorem{lemma}{Lemma}
\newtheorem{corollary}{Corollary}
\newtheorem{definition}{Definition}

\newtheorem*{mainresult*}{Main Result}

\theoremstyle{remark}
\newtheorem{remark}{Remark}

\begin{document}

\title{Topological Signal Processing on Quantum Computers for Higher-Order Network Analysis}

\author{Caesnan M. G. Leditto}
\email{caesnan.leditto@monash.edu}
\affiliation{School of Physics and Astronomy, Monash University, Clayton, VIC 3168, Australia}
\affiliation{Quantum Systems, Data61, CSIRO, Clayton, VIC 3168, Australia}

\author{Angus Southwell}
\affiliation{Quantum for New South Wales, Haymarket, NSW 2000, Australia}
\affiliation{School of Physics and Astronomy, Monash University, Clayton, VIC 3168, Australia}

\author{Behnam Tonekaboni}
\affiliation{Quantum Systems, Data61, CSIRO, Clayton, VIC 3168, Australia}

\author{Gregory A. L. White}
\affiliation{School of Physics and Astronomy, Monash University, Clayton, VIC 3168, Australia}
\affiliation{Dahlem Center for Complex Quantum Systems, Freie Universit\"at Berlin, 14195 Berlin, Germany}

\author{Muhammad Usman}
\affiliation{Quantum Systems, Data61, CSIRO, Clayton, VIC 3168, Australia}
\affiliation{School of Physics, The University of Melbourne, Parkville, VIC 3052, Australia}
\affiliation{School of Physics and Astronomy, Monash University, Clayton, VIC 3168, Australia}

\author{Kavan Modi}
\email{kavan.modi@transport.nsw.gov.au}
\affiliation{Quantum for New South Wales, Haymarket, NSW 2000, Australia}
\affiliation{School of Physics and Astronomy, Monash University, Clayton, VIC 3168, Australia}

\begin{abstract}
Predicting and analyzing global behaviour of complex systems is challenging due to the intricate nature of their component interactions. Recent work has started modelling complex systems using networks endowed with multiway interactions among nodes, known as higher-order networks. Simplicial complexes are a class of higher-order networks that have received significant attention due to their topological structure and connections to Hodge theory. Topological signal processing (TSP) utilizes these connections to analyze and manipulate signals defined on non-Euclidean domains such as simplicial complexes. In this work, we present a general quantum algorithm for implementing filtering processes in TSP and describe its application to extracting network data based on the Hodge decomposition. We leverage pre-existing tools introduced in recent quantum algorithms for topological data analysis and combine them with spectral filtering techniques using the quantum singular value transformation framework. While this paper serves as a proof-of-concept, we obtain a super-polynomial improvement over the best known classical algorithms for TSP filtering processes, modulo some important caveats about encoding and retrieving the data from a quantum state. The proposed algorithm generalizes the applicability of tools from quantum topological data analysis to novel applications in analyzing high-dimensional complex systems.
\end{abstract}
\maketitle 

\section{Introduction}

\begin{figure}
    \centering
    \includegraphics[scale=0.4]{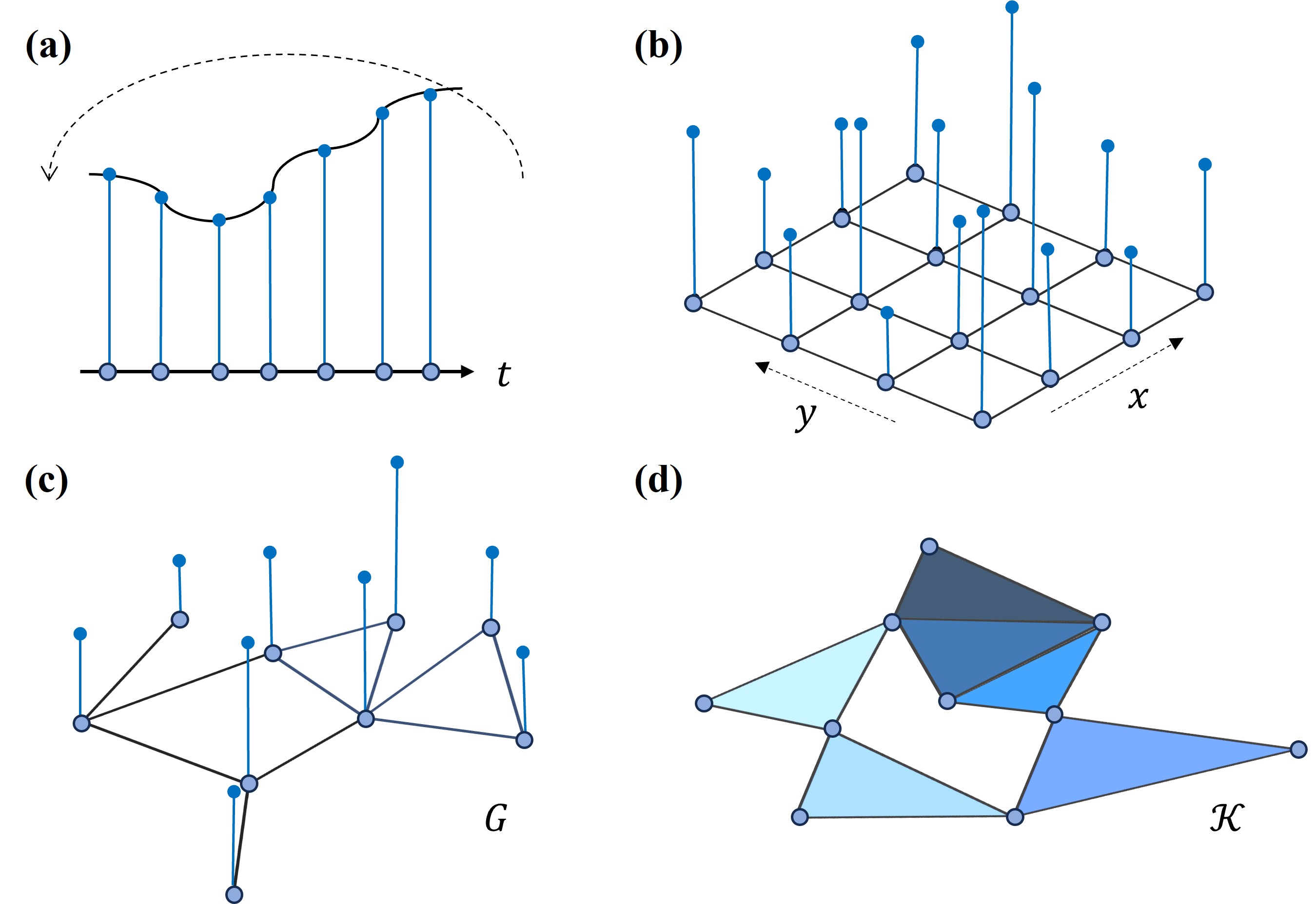}
    \caption{(a) discrete-time signals (1-D signals) with periodic boundary condition, (b) grid-like signals (2-D signals on $x-y$ plane), (c) node signals (signals on a graph $G$), and (d) simplicial signals (signals on a simplicial complex $\mathcal{K}$).}
    \label{various signals fig}
\end{figure}

Recent work in the field of signal processing has focused on understanding signals that are defined on domains other than time or space.
In particular, there has been considerable research on signals which are defined on graphs~\cite{Sandryhaila2013DiscreteGraphs,Sandryhaila2014DiscreteAnalysis,Ortega2018GraphApplications} or, more generally, higher-order networks~\cite{Barbarossa2020TopologicalComplexes,Schaub2021SignalBeyond}. In the graph context, a `signal’ means a value attributed to a node on the graph{, and the edges denote interactions between the elements (vertices) of the system}. This value often comes from real-world observations of some (potentially dynamical) system. 
This includes a discrete 1D signal — such as time series data, 2D signals on a grid — such as colour values in an image, or, more generally, 2D signals on an arbitrary graph. Figure~\ref{various signals fig} (a)-(c) shows illustrations of such signals.
Moving beyond graphs and pairwise interactions, higher-order networks — or \textit{hypergraphs} — can be far more expressive in capturing the interactions of a given system~\cite{Battiston2020NetworksDynamics,Battiston2021TheSystems,Majhi2022DynamicsReview,Bick2023WhatNetworks}. In a hypergraph, each edge may contain an arbitrary number of vertices. Consequently, signals attributed to higher-order networks (such as the edges and hyperedges) can represent inherently multi-party data. In this work, we are particularly interested in categories {similar to} the last case, where a signal is more generally defined {on} the $k$-body interactions ({depicted in the $3$-body case by triangles in Figure~\ref{various signals fig}(d)}) in a higher-order network. {The topology of these higher-order interactions has a non-trivial impact on the evolution of many systems~\cite{Calmon2023DiracSignals}.} 

Equipped with this way of encoding data on vertices and (hyper)edges, graphs and networks can be utilized to study interactions captured in the signals between different elements of a system. The field of graph signal processing (GSP) leverages tools both from graph theory and signal processing to perform standard tasks with signals (such as denoising or filtering) on data defined over non-Euclidean domains. GSP has been successfully applied to understand sensor networks~\cite{Jablonski2017GraphCities}, brain connectivity~\cite{Huang2016GraphSignals,Goldsberry2017BrainPerspective,Huang2018AImaging}, protein interactions~\cite{Segarra2017NetworkTemplates}, and {to} perform image processing~\cite{Sharma2018EfficientProcessing,Cheung2018GraphProcessing}. However, just as graphs are insufficient to describe multi-way interactions, GSP is insufficient as a tool to study more complex systems. As a result, \textit{topological} signal processing (TSP) has recently emerged as a method by which signals may be studied in the context of data defined on higher-order networks. Already, these networks and the corresponding signals have been used to model systems such as the brain~\cite{Lucas2020MultiorderNetworks}, research collaboration networks~\cite{Patania2017TheCollaborations}, and protein-protein interaction networks~\cite{Cang2015AClassification}, where many-body interactions play vital roles in the dynamics of the underlying system.
In this paper, we present a quantum algorithm for filtering and analyzing topological signals defined on a class of higher-order networks known as simplicial complexes. The theoretical complexity bounds of this algorithm show a promising advantage over the best currently known classical methods.

\begin{figure*}[t]
    \centering
    \includegraphics[width=0.8\linewidth]{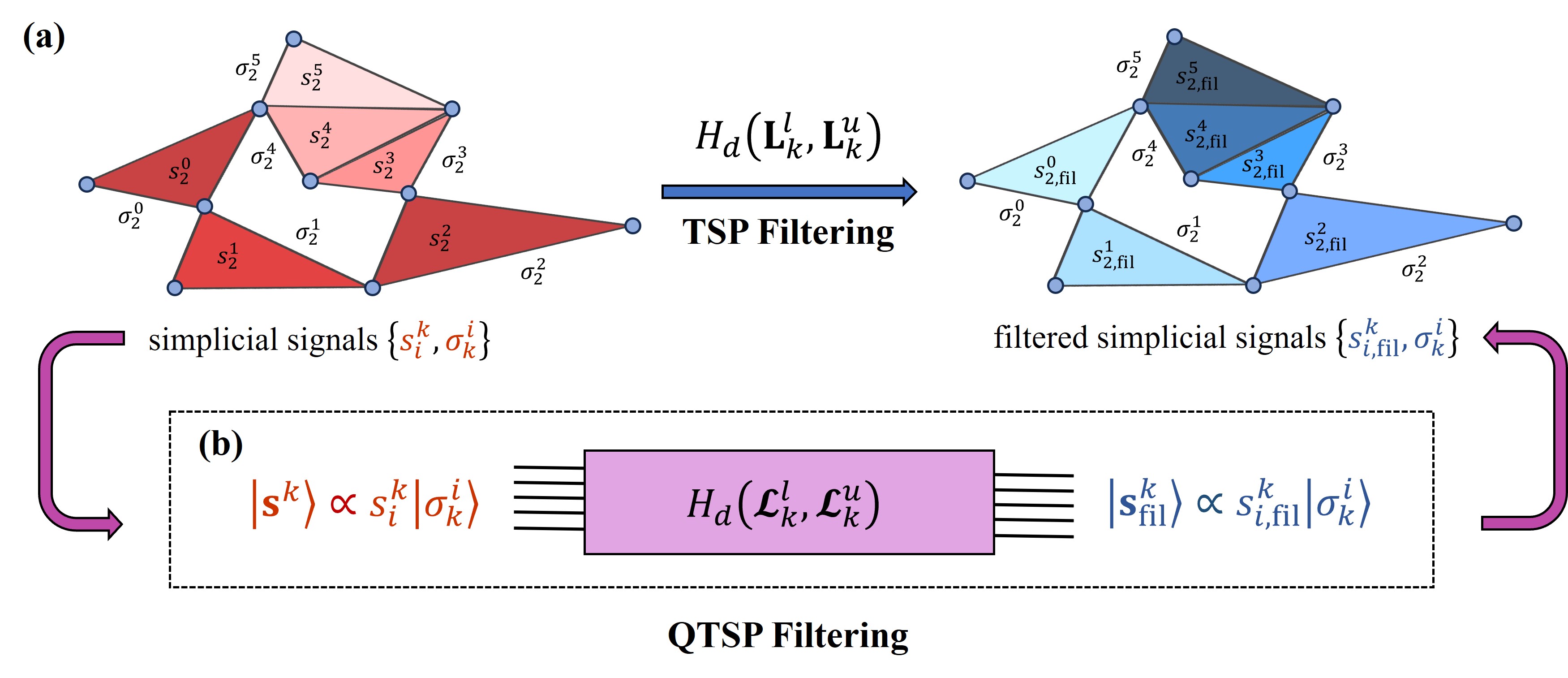}
    \caption{\textbf{(a)} A flowchart diagram for topological signal processing (TSP) filtering using a simplicial filter $H(\hll,\hlu)$ with input simplicial signals $\{s_{i}^{k}\}$ defined on simplices $\{\sigma_{k}^{i}\}$ (with magnitudes represented by colour opacity) and the filtered simplicial signals $\{{s}_{\mathrm{fil}}^{k}\}$. \textbf{(b)} A QTSP filtering process on a quantum computer using a quantum simplicial filter $H(\hlln,\hlun)$ which filters a simplicial signal state $\lvert \brm{s}^k\rangle$, which encodes the signals $\{s_i^k\}$. The full pipeline of the algorithm is comprised of three subsequent subroutines: \textsc{Encode}, \textsc{Filter}, and \textsc{Retrieve}, depicted by the flow of purple arrows.
    }
    \label{QTSP flowchart figure}
\end{figure*}

Higher-order networks are defined mathematically as a pair $(\mathcal{V}, S)$, where $\mathcal{V}$ is the set of vertices (the individual elements being studied), and $S$ is a set of subsets of $\mathcal{V}$ describing the interactions between elements of $\mathcal{V}$. In a standard graph or network, $S$ would be the edge set, a set of $2$-element subsets of $\mathcal{V}${, whereas in higher-order networks, $S$ may contain larger subsets}. Higher{-}order networks contain many rich sub-classes. In particular, we consider here the extensively studied category of \textit{simplicial complexes}. The defining property of a simplicial complex $\mathcal{K} = (\mathcal{V},S)$ is that if some $A \subset S$, then every subset of $A$ is also in $S$. 
Although this is not the most general hypergraph model, {their extra structure allows for the use of tools from algebraic topology.}  
{This} can be incredibly {useful} for analysing signals found on such complexes.
Furthermore, many interesting physically and mathematically relevant systems can be represented by simplicial complexes. Hence, 
{considering this class expands the available toolkit} while maintaining versatility and applicability.

Simplicial complexes and algebraic topology are used extensively in data analysis. Indeed, topological data analysis (TDA) is an umbrella term for a wide range of statistical methods that employ techniques from algebraic topology to study real-world data~\cite{Carlsson2009TOPOLOGYDATA,Patania2017TopologicalData,Wasserman2018TopologicalAnalysis,Chazal2021AnScientists}. By analysing {the shape of data} from a topological viewpoint, TDA can both reduce data dimensionality and benefit from robustness to noise. These methods have seen wide-ranging applications, such as in sensor networks~\cite{deSilva2007CoverageHomology}, financial market prediction~\cite{Gidea2018TopologicalCrashes}, tumour classification~\cite{Qaiser2019FastFeatures}, and analyzing collaboration networks characteristics~\cite{Carstens2013PersistentNetworks}. A popular aspect of TDA is the study of \textit{Betti numbers} of simplicial complexes. The Betti numbers $(\beta_k)_{k\geq0}$ describe, at a high level, the number of $k$-dimensional ``holes'' in the simplicial complex, which in TDA is generated by the data set. These can be used for distinguishing data sets~\cite{Adams2017PersistenceHomology}, for hypothesis testing~\cite{Blumberg2014RobustSpaces}, or as features for machine learning algorithms~\cite{Hensel2021AMethods}.

In tandem with the development of sophisticated classical data analysis algorithms, there has been an urgent need to showcase the utility of quantum computers in real-world scenarios. Consequently, {there has been} a proliferation of ideas in implementing quantum computers for data analysis and machine learning. In recent years, quantum algorithms for TDA (QTDA) have been proposed to efficiently compute the higher-order Betti numbers of simplicial complexes arising in applications of TDA~\cite{Lloyd2016QuantumData,Cade2021ComplexityProblem,Gyurik2022TowardsAnalysis,Akhalwaya2022TowardsComputers,Hayakawa2022QuantumAnalysis,McArdle2022AQubits}. The perceived advantage in contrast to classical algorithms is that the classical complexity of computing the $k$-th Betti number $\beta_{k}$ of a simplicial complex has worst-case exponential scaling in $k$. This severely limits the capability to study the topology of higher-order interactions in networks using classical computers. On the other hand, since the seminal paper by Lloyd, Garnerone, and Zanardi~\cite{Lloyd2016QuantumData}, there has been a flurry of research about quantum algorithms for computing Betti numbers in fault-tolerant quantum computers. Although it was shown that Betti number estimation is $\mathsf{QMA}_1$-hard~\cite{Cade2021ComplexityProblem}, one can relax the problem by approximating a normalized version of $\beta_{k}$. While in early years, several promising results indicated the potential for an exponential (in $k$) speedup over classical algorithms for estimating \textit{approximate} normalized Betti numbers, recent results have tempered these expectations. In particular, Refs.~\cite{Gyurik2022TowardsAnalysis,Schmidhuber2022Complexity-TheoreticAnalysis,berry2023analyzing,McArdle2022AQubits} have argued that the existence of a superpolynomial speedup over classical algorithms for this task is, at best, highly situational. In most cases, the quantum algorithms might provide a polynomial speedup compared to classical algorithms. Importantly, to overcome current quantum error correction overheads, it is highly desirable for this speed-up to be greater than quadratic~\cite{sanders2020compilation,Babbush2021FocusAdvantage}.

In this paper, we introduce a quantum algorithm that adapts the tools used in QTDA to the new task of quantum topological signal processing (QTSP). The problem setting{s} for {QTSP and QTDA are} quite different: QTDA works on point cloud data and defines sequences of simplicial complexes from that data (see an example given in Figure~\ref{fig:QTDA}(a)), whereas QTSP is generally focused on studying signals on a fixed network or complex (such as the example given in Figure~\ref{QTSP flowchart figure}(a)). Beyond this distinction, we can think of the QTSP as a generalization of QTDA that applies to a different range of tasks: QTDA involves projecting states to the kernel of the Hodge Laplacian $\hl$ (given formally in Definition~\ref{hodgelaplacian}) to determine its dimension (see Figure~\ref{fig:QTDA} as an illustration of the algorithm); in QTSP, we analyze the image of $\hl$ as well as the kernel (and thus the whole codomain) through the lens of what is known as the Hodge decomposition. {Previous works in QTDA~\cite{berry2023analyzing,Hayakawa2022QuantumAnalysis,McArdle2022AQubits} often implement} spectral filtering to construct a projection to the kernel $\mathrm{ker}(\hl)$. The goal of QTSP is to, given a TSP filtering task (formally defined in Section~\ref{sec:background}) and some data, construct a quantum circuit that implements this filter on a quantum representation of the data (Figure~\ref{QTSP flowchart figure}(b)). Here, the quantum filter operator is designed to manipulate not just the kernel of the Hodge Laplacian, but also its image, in a way that respects the topology of the space on which the data is defined. On the other hand, if the filter is just a projection to $\mathrm{ker}(\hl)$, then we can recover the QTDA task of estimating Betti numbers as an application of QTSP.

In QTSP, we assume access to a quantum input model given by an oracle that encodes the signal as a quantum state. Given such access, we construct a quantum algorithm to efficiently perform TSP filtering on a fault-tolerant quantum computer. Here, we informally state our main result, which is given formally in Section~\ref{sec:main} as Theorem~\ref{QTSP filtering}. 
\begin{mainresult*}
    We present a quantum algorithm for filtering topological signals defined on the $k$-dimensional simplices of a simplicial complex, given access to a state preparation oracle and upon successful postselection of the ancilla qubits. If the simplicial complex is a clique complex, the copmlexity of this algorithm scales linearly in $k$. In comparison, the complexity of performing the filtering on a classical computer scales exponentially in $k$.
\end{mainresult*}
\noindent This theoretical speed-up comes with some important caveats to be addressed before making any claims of a practical speed-up, which would likely be problem-dependent. In Section~\ref{sec:main}, we state the cost of implementing the filtering process (specifically in Theorem~\ref{QTSP filtering}). We also discuss the hurdles between our complexity results and a practical speed-up in performing an end-to-end implementation of our algorithm on (classical or quantum) data. Additionally, we showcase an application of this result to extracting signal subcomponents based on the Hodge decomposition{, which we discuss in more detail in Sections~\ref{sec:background} and~\ref{sec:main}.}

{A brief} background to topological signal processing and the specific description of the class of filtering tasks is given in Section~\ref{sec:homology}. This class of tasks is very general, and {given that higher-order networks of interest may contain thousands or millions of vertices}, {a superpolynomial} advantage has huge potential. On the more cautioned side, signal processing on higher-order networks is a relatively new field, which means that concrete applications to test this quantum advantage in higher dimensions are {not well} understood. Our algorithm also only applies to clique complexes, a specific class of simplicial complexes defined by an underlying graph. Furthermore, this algorithm suffers the classic pitfalls of many quantum linear algebra (QLA) algorithms in data encoding and retrieval costs, and the costs of these steps may limit the speed-up in practice. Nonetheless, the generality of our framework lends itself to plenty of opportunity for a concrete quantum advantage to be achieved in the future.

The remainder of the paper is laid out as follows. In Section~\ref{sec:background}, we briefly overview the background mathematics and notations required. Then, in Section~\ref{sec:main}, we {state} the main results of our paper and discuss the advantages and limitations of our algorithm. Finally, in Section~\ref{sec:details}, we discuss the precise details of our implementation and prove the necessary results about its complexity and {correctness}. 
\section{Background of the Simplicial Filter Design Problem}\label{sec:background}

\subsection{Simplicial complexes and Hodge Laplacian}\label{sec:homology}

Here we provide a concise overview of simplicial homology and delve into essential tools commonly employed in the realms of TDA and TSP. For a more extensive discussion of these subjects, see Ref.~\cite{Lim2020HodgeGraphs}. Topology has found practical applications across {a huge range of} domains, such as biology{, computer science, quantum field theory, and statistics}. This utility arises from the fact that both physical and abstract systems can be mapped onto topological spaces, enabling the adoption of topological approaches as an alternate analysis method. Simplicial complexes in particular have gained substantial attention as models that capture complex data representations or interactions, especially in the context of higher-order networks. 

We start from the notion of a simplicial complex, simplices, and their topological properties. Let $\mathcal{V} = \{v_1, \dots, v_n\}$ be a non-empty set called the vertex set. A simplicial complex $\mathcal{K}$ is a set of subsets of $\mathcal{V}$ with two properties: (a) $\mathcal{K}$ contains all singleton subsets, that is, $\{v\} \in \mathcal{K}$ for all $v \in \mathcal{V}$; (b) if $\sigma \in \mathcal{K}$, then every subset of $\sigma$ is also in $\mathcal{K}$. We say that $\sigma$ is a $k$-simplex {(short for $k$-dimensional simplex)} when $|\sigma| = k+1$. If $\sigma' \subset \sigma$ with $q+1 < k+1$ elements, then $\sigma'$ is called a ($q$-)face of $\sigma$. 

We focus on a particular class of simplicial complexes that piqued the interest of the quantum community working on TDA, which is the class of \textit{clique complexes}. A clique complex is a simplicial complex whose structure is entirely determined by an underlying graph $G$. Formally, a clique complex ${\mathcal{K} =} \mathcal{K}(G)$ is a simplicial complex where a set {$\{v_{i_0}, \dots, v_{i_k}\}$} is a $k$-simplex in $\mathcal{K}$ if and only if {$\{v_{i_0}, \dots, v_{i_k}\}$} is a $(k+1)$-clique (a set of $k+1$ mutually adjacent vertices) in $G$. It is straightforward to see that this is a simplicial complex: each vertex forms a $1$-clique on its own, and an induced subgraph of a clique is a clique. We focus on this class of complexes as their structure allows for efficient quantum circuit implementations of some otherwise difficult functions. 

Modelling systems and their signals using simplicial complexes, instead of general hypergraphs, allows the use of tools from algebraic topology. In order to build the algebraic objects of interest later, one associates an orientation to each simplex in the complex.
An orientation $\epsilon(\sigma)$ of a simplex $\sigma$ is the parity of an ordering of the vertices {$\epsilon$} (as an element of the permutation group $S_n$).  Let $j\in[k+1]:=\{0,\cdots,k\}$ and $i_{j}\in[n]:=\{0,\cdots,n-1\}$. An orientation of a simplex $\sigma = \{v_{i_0}, \dots, v_{i_k}\}$ is an ordering of the vertices, which we denote $\left[ v_{i_0}, \dots, v_{i_k}\right]$, where two orderings are equivalent if they differ by an even number of transpositions. Thus, there are only two distinct orientations for each simplex, often referred to as positive and negative. For our purposes, we define the lexicographic ordering of each simplex to be positive. 

One particular class of algebraic objects that mathematically underpins our work is the space of $k$-chains. These are defined as formal sums of simplices in the complex.
For our purposes, the space of $k$-chains $\mathfrak{C}_k := \mathfrak{C}_k(\mathcal{K})$ of a simplicial complex $\mathcal{K}$ is (isomorphic to) a real vector space where the basis is a collection of $k$-simplices in $\mathcal{K}$, i.e., $\mathfrak{C}_k:=\big\{\sum_{i=0}^{n_k-1}c_i\sigma_k^i\mid c_i\in \mathbb{R}\big\}$, where $n_k$ is the number of $k$-simplices in $\mathcal{K}$.  The different spaces $\mathfrak{C}_k(\mathcal{K})$ associated with different dimensions $k$ are related to each other by maps known as boundary operators. The \textit{$k$-th boundary operator} $\partial_k:\mathfrak{C}_k \to \mathfrak{C}_{k-1}$ is a linear map such that for $\sigma_k=[v_{i_0},\cdots,v_{i_k}]\in \mathfrak{C}_k$,
\begin{eqnarray}\label{boundary operator}
    \partial_k\sigma_k&:=&\sum_{j=0}^{k}\,(-1)^j\,[v_{i_0},\cdots,v_{i_{j-1}},v_{i_{j+1}}\cdots,v_{i_k}].
\end{eqnarray}
In each summand, a different vertex $v_{i_j}$ is removed from the simplex. That is, $\partial_k$ maps a $k$-simplex to a formal sum of its $(k-1)$-faces, which are the boundary of the $k$-simplex in a topological sense. Since this map acts linearly on the space of $k$-chains, we can represent it by a matrix. Let $\bm{\mathrm{B}}_k$ be the matrix representation of $\partial_k$, using the canonical bases of $k$-simplices and $(k-1)$-simplices for $\mathfrak{C}_{k}(\mathcal{K})$ and $\mathfrak{C}_{k-1}(\mathcal{K})$ respectively. 
The entries of $\bm{\mathrm{B}}_k$ are
\begin{eqnarray}\label{boundary matrix}
    \left(\bm{\mathrm{B}}_k\right)_{ij}=\begin{cases}
        1,\quad&\mbox{if $i\neq j$, $\sigma_{k-1}^i\subset\sigma_k^j$, $\epsilon(\sigma_{k-1}^i)\sim \epsilon(\sigma_k^j)$}\\
        -1,\quad&\mbox{if $i\neq j$, $\sigma_{k-1}^i\subset\sigma_k^j$, $\epsilon(\sigma_{k-1}^i)\nsim \epsilon(\sigma_k^j)$}\\
        0,\quad&\mbox{otherwise},
    \end{cases}\nonumber\\
\end{eqnarray}
where $\epsilon(\sigma_{k-1}^i)\sim \epsilon(\sigma_k^j)$ means $\sigma_{k-1}^i$ and $\sigma_k^j$ have the same orientation {(both positive or both negative)} and $\epsilon(\sigma_{k-1}^i)\nsim \epsilon(\sigma_k^j)$ means $\sigma_{k-1}^i$ and $\sigma_k^j$ have different orientations~\cite{Goldberg2002Complexes}. {Thus, the} $\pm 1$ entries account for the orientation of each $k$-simplex relative to its boundary $(k-1)$-faces. In this paper, we generally refer to the boundary operator by its matrix representation $\brm{B}_k$, and the sign of $(\brm{B}_k)_{ij}$ is determined by the placement of the vertex $\sigma_{k}^j \backslash \sigma_{k-1}^i$ in the simplex $\sigma_{k}^j$, as given by Eq.~\ref{boundary matrix}.

The adjoint (equivalently, transpose) of the boundary operator $\bk^\mathrm{T}$ is called the coboundary operator and naturally maps from $\mathfrak{C}_{k-1}$ to $\mathfrak{C}_{k}$. The boundary and coboundary operators are used to construct the central object of our consideration: the Hodge Laplacian. The Hodge Laplacian matrix $\bm{\mathrm{L}}_k$ is defined as
\begin{eqnarray}\label{hodgelaplacian}
\bm{\mathrm{L}}_k:=\begin{cases}
\bm{\mathrm{B}}_{\mathrm{k+1}}\bm{\mathrm{B}}_{k+1}^\mathrm{T},\quad&\mbox{for $k=0$}\\
\bk^\mathrm{T}\bm{\mathrm{B}}_k+\bm{\mathrm{B}}_{k+1}\bm{\mathrm{B}}_{k+1}^\mathrm{T},\quad&\mbox{for $1<k<n-2$}\\
\bk^\mathrm{T}\bm{\mathrm{B}}_k,\quad&\mbox{for $k=n-1$}
\end{cases}\nonumber\\
\end{eqnarray}
where superscript $\mathrm{T}$ indicates the matrix transpose operation. Furthermore, {the lower and upper Hodge Laplacian operators are defined} as $\bm{\mathrm{L}}_k^l:=\bk^\mathrm{T}\bm{\mathrm{B}}_k$ and $\bm{\mathrm{L}}_k^u:=\bm{\mathrm{B}}_{k+1}\bm{\mathrm{B}}_{k+1}^\mathrm{T}$ respectively. Hence, the lower and upper Hodge Laplacians $\bm{\mathrm{L}}_k^l$ and $\bm{\mathrm{L}}_k^u$ are given by
\begin{eqnarray}
    \bm{\mathrm{L}}_k^{l}&=&\begin{cases}
        \mathrm{deg}_{l}(\sigma_k^i),\quad&\mbox{if $i=j$}\\
        1,\quad&\mbox{if $i\neq j$, $\sigma_k^i\sim_{l}\sigma_k^j$, $\epsilon(\sigma_k^i)\sim \epsilon(\sigma_k^j)$}\\
        -1,\quad&\mbox{if $i\neq j$, $\sigma_k^i\sim_{l}\sigma_k^j$, $\epsilon(\sigma_k^i)\nsim \epsilon(\sigma_k^j)$}\\
        0,\quad&\mbox{otherwise},
    \end{cases}\nonumber\\
\\
    \bm{\mathrm{L}}_k^{u}&=&\begin{cases}
        \mathrm{deg}_{u}(\sigma_k^i),\quad&\mbox{if $i=j$}\\
        1,\quad&\mbox{if $i\neq j$, $\sigma_k^i\sim_{u}\sigma_k^j$, $\epsilon(\sigma_k^i)\sim \epsilon(\sigma_k^j)$}\\
        -1,\quad&\mbox{if $i\neq j$, $\sigma_k^i\sim_{u}\sigma_k^j$, $\epsilon(\sigma_k^i)\nsim \epsilon(\sigma_k^j)$}\\
        0,\quad&\mbox{otherwise},
    \end{cases}\nonumber\\
\end{eqnarray}
where $\mathrm{deg}_{l}(\sigma_k^i)$ ($\mathrm{deg}_{u}(\sigma_k^i)$) is the number of $(k-1)$-simplices ($(k+1)$-simplices) which are faces of $\sigma_k^i$ (which have $\sigma_k^i$ as a face). The sign $\sim_{l}$ is a relation between $k$-simplices sharing a $(k-1)$-face, and $\sim_u$ is a relation between faces of a $(k+1)$-simplex (described in detail in~\cite{Goldberg2002Complexes}). {Remarkably}, the definitions of $\hl^\ell$ and $\hl^u$ do not depend on the choice of the orientation of the simplices in the complex. Note that $\hl$, $\hl^\ell$, and $\hl^u$ are all real, symmetric matrices. This operator $\hl$ and its constituent elements form the building blocks for the topological signal processing operations we examine in this paper. 

\subsection{Topological signal processing (TSP)}

Here we formally introduce the mathematical background to TSP. We use the language and terminology used in Refs.~\cite{Barbarossa2020TopologicalComplexes,Yang2021FiniteComplexes}. The aim of signal processing is to {modulate the various components} --- determined by some structured set of frequency ranges --- of the signal. This requires the design of a \textit{filter}: an operator with appropriate input-output relations to transform the signal into something more desirable.
In general, the nature of such a filter will be specific to the task at hand. 
For example, in discrete-time signal processing, a typical filtering task is suppressing high frequency noise. Denoising is achieved by utilising a shift operator and attenuating higher frequency components of the signal by weighting the signal in the frequency domain using the {eigenvalues of the shift operator}. In the framework of topological signal processing, the Hodge Laplacian $\hl$ plays the role of the shift operator. 

The formalism and topological properties of a simplicial signal defined on a simplicial complex are as follows. Let $\mathcal{K}$ be a simplicial complex with $n_k$ $k$-simplices. A \textit{simplicial signal} is a $k$-cochain{, which is a linear functional} $s^k: \mathfrak{C}_k(\mathcal{K}) \to \mathbb{R}$.
This definition of a simplicial signal is equivalent to assigning a value $s^k_i\in\mathbb{R}$ to each $k$-simplex $\sigma_{k}^{i}\in\mathcal{K}$ for all $i\in[n_{k}]$. Since the space of $k$-cochains is an $n_k$-dimensional vector space over $\mathbb{R}$ (where $n_k$ is the number of $k$-simplices in $\mathcal{K}$), this gives rise to an isomorphism between $\mathfrak{C}_k$ and $\mathbb{R}^{n_k}$. Therefore, the simplicial signal can be represented by a vector $\bm{\mathrm{s}}^k\in\mathbb{R}^{n_k}$. In this language, the $i$-th component of the signal vector is simply the \textit{signal value} $s^{k}_i$ associated with the $i$-th simplex $\sigma_k^i$, for some ordering of the simplices. For example, a simplicial complex could describe the collaboration network of researchers, where a set of researchers form a simplex if they all have a joint publication. A natural $k$-simplicial signal on this network would be the number of joint publications of each group of $k+1$ authors, which can be considered as a vector.

The vector space $\mathbb{R}^{n_{k}}$ can be represented as the direct sum of three orthogonal subspaces defined in terms of the boundary maps and the Laplacian. This is known as the \textit{Hodge decomposition}, which states that 
\begin{align*}
    \mathbb{R}^{n_k} = \ker(\hl) \oplus \rm{im}(\bk^T) \oplus \rm{im}(\bkk).
\end{align*}
The three subspaces $\mathrm{ker}(\bm{\mathrm{L}}_k)$, $\mathrm{im}(\bk^\mathrm{T})$, and $\mathrm{im}(\bm{\mathrm{B}}_{k+1})$ are known in TSP literature as the \textit{harmonic}, \textit{gradient}, and \textit{curl} subspaces, respectively. These names are drawn from the analogously named Helmholtz-Hodge decomposition in the context of discrete vector fields, which is a special case of the Hodge decomposition for edge signals (i.e. the case $k=1$).
Since $\hl$ is a Hermitian matrix, the eigenvectors of $\hl$ can be used to construct an orthonormal basis of $\mathbb{R}^{n_k}$. Each eigenvector belongs to precisely one of these orthogonal spaces.
Thus, each $\brm{s}^{k}\in\mathbb{R}^{n_{k}}$ can be decomposed into three orthogonal components
\begin{eqnarray}\label{Hodge decomposition of simplicial signals}
    \bm{\mathrm{s}}^k&:=&\bm{\mathrm{s}}^k_{\mathrm{H}}+\bm{\mathrm{s}}^k_{\mathrm{G}}+\bm{\mathrm{s}}^k_{\mathrm{C}},
\end{eqnarray}
where $\bm{\mathrm{s}}^k_{\mathrm{H}}\in\mathrm{ker}(\bm{\mathrm{L}}_k)$, $\bm{\mathrm{s}}^k_{\mathrm{G}}\in\mathrm{im}(\bk^\mathrm{T})$, and $\bm{\mathrm{s}}^k_{\mathrm{C}}\in\mathrm{im}(\bm{\mathrm{B}}_{k+1})$. The orthogonality of these components allows them to be modulated somewhat independently by {applying functions to} the eigenvalues of the eigenvectors of the respective spaces.

Incorporating signal processing terminology from discrete-time signals, a notion of ``frequency'' of a simplicial signal has been defined that relates to the eigenvalues of $\brm{L}_{k}$. Let $\bm{\mathrm{L}}_k=\brm{U}\brm{\Lambda}_k\brm{U}^\mathrm{T}$ be an eigendecomposition of the Hodge Laplacian, where $\bm{\mathrm{\Lambda}}_k$ is a diagonal matrix where $\left(\bm{\mathrm{\Lambda}}_k\right)_{ii}=\lambda_{k,i}$ is an eigenvalue of $\bm{\mathrm{L}}_k$ for each $i\in[n_{k}]$, and $\brm{U}$ is a unitary matrix composed of the eigenvectors of $\bm{\mathrm{L}}_k$ in the chosen basis. Then the \textit{simplicial Fourier transform} is defined as $\Tilde{\bm{\mathrm{s}}}_k:=\brm{U}^\mathrm{T}\bm{\mathrm{s}}_k$. This paves the way for a filtering process analogous to discrete-time signal processing by modulating the frequencies.

{The standard definition of a \textit{simplicial filter} follows from exploring this connection between discrete signals and the Hodge Laplacian.} As the graph Laplacian $\brm{L}_{0}:=\brm{L}_{G}$ plays the role of filter operator primitive in GSP, the Hodge Laplacian $\hl$ has similar functionality in TSP. {This filter is a map from the space of $k$-cochains to itself. Keeping in mind the aforementioned isomorphism to $\mathbb{R}^{n_k}$, this can be considered as a map from $\mathbb{R}^{n_k}$ to itself. Consequently, a} filter is defined by its action on the basis of eigenvectors of the Hodge Laplacian, which is equivalent to defining a function $g:\mathbb{R} \to \mathbb{R}$ by its action on the eigenvalues $\{\lambda_{k,i}\}$ of $\hl$. Let us first define 
\begin{gather*}
    \mathrm{Eig}^{\mathrm H}(\brm{L}_{k}):=\left\{\lambda_{k,i}^{\mathrm H}\right\}_{i\in[n_{k}^{\mathrm H}]},\;\mathrm{Eig}^{\mathrm G}(\brm{L}_{k}):=\left\{\lambda_{k,i}^{\mathrm{G}}\right\}_{i\in[n_{k}^{\mathrm G}]},\\\mathrm{Eig}^{\mathrm C}(\brm{L}_{k}):=\left\{\lambda_{k,i}^{\mathrm C}\right\}_{i\in[n_{k}^{\mathrm C}]}
\end{gather*}
to be the sets of harmonic, gradient, and curl eigenvalues of $\brm{L}_{k}$ respectively, where $n_{k}^{\mathrm H}$, $n_{k}^{\mathrm G}$, and $n_{k}^{\mathrm C}$ are the numbers of eigenvalues in each set. We denote by $\bm{0}_{k}^{\mathrm H}$ a square matrix with all zero entries with size equal to the dimension of $\mathrm{ker}(\hl)$. It follows from the Hodge decomposition that $\bm{\mathrm{\Lambda}}_{k}=\bm{0}_{k}^{\mathrm H}\oplus\bm{\mathrm{\Lambda}}_{k}^{\mathrm G}\oplus\bm{\mathrm{\Lambda}}_{k}^{\mathrm C}$, where $\bm{\mathrm{\Lambda}}_{k}^{\mathrm G}$ and $\bm{\mathrm{\Lambda}}_{k}^{\mathrm C}$ denote diagonal matrices with entries $\left(\brm{\Lambda}_{k}^{\mathrm{G}}\right)_{ii}=\lambda_{k,i}^{\mathrm G}$ for all $i\in[n_{k}^{\mathrm G}]$ and $\left(\brm{\Lambda}_{k}^{\mathrm{C}}\right)_{ii}=\lambda_{k,i}^{\mathrm C}$ for all $i\in[n_{k}^{\mathrm C}]$. 

Due to the decomposition $\hl=\hll+\hlu$, the action of Hodge Laplacian to $\brm{s}^{k}$ performs ``filtering'' operations, which shifts-and-sums signal values across lower and upper adjacent simplicies, respectively. Thus, the \textit{TSP filtering process} can be seen as transforming signal values of $\brm{s}^{k}$ by the action of \textit{simplicial filter}. Formally, a simplicial filter $H_d(\brm{L}_{k}^{\ell},\brm{L}_{k}^{u})$ is defined as a multivariate polynomial of total degree $d:=\mathrm{max}\{d_{\ell},d_{u}\}$, $H_d(\brm{L}_{k}^{\ell},\brm{L}_{k}^{u}):\mathbb{R}^{n_k}\longrightarrow\mathbb{R}^{n_k}$ given by~\cite{Yang2022SimplicialFilters}
\begin{eqnarray}\label{TSPfilter}
       H_d(\brm{L}_{k}^{\ell},\brm{L}_{k}^{u}):=h_0\bm{\mathrm{I}}+\sum_{i_\ell=1}^{d_\ell}h_{i_\ell}^\ell\Bigr(\bm{\mathrm{L}_k^\ell}\Bigr)^{i_\ell}+\sum_{i_u=1}^{d_u}h_{i_u}^u\Bigr(\bm{\mathrm{L}_k^u}\Bigl)^{i_u}\nonumber\\
\end{eqnarray}
where $h_0,h_{i_\ell}^\ell,h_{i_u}^u\in\mathbb{R}$ are called the \textit{simplicial filter coefficients}. Alternatively, a simplified version of this filter has also been defined~\cite{Ebli2020SimplicialNetworks,Yang2021FiniteComplexes}, which is given by a degree-$d$ real-valued polynomial of the Hodge Laplacian
\begin{eqnarray}
    H_d(\bm{\mathrm{L}_k})&=&\sum_{i=0}^{d}h_i\big(\bm{\mathrm{L}_k}\big)^i,
\end{eqnarray}
where $h_i\in\mathbb{R}$. This is accommodated in Eq.~(\ref{TSPfilter}) by setting $h_{i}^\ell = h_{i}^u$ for all $i \leq d$. However, the simplified version of the simplicial filter is less expressive and less able to distinguish signals from the different subspaces described by the Hodge decomposition~\cite{Yang2021FiniteComplexes}. Hence, for certain tasks (such as extracting the signal subcomponents $\brm{s}^k_{\mathrm{H}}$, $\brm{s}^k_{\mathrm{G}}$, and $\brm{s}^k_{\mathrm{C}}$ {given in Eq.~(\ref{Hodge decomposition of simplicial signals})}), $H_d\left(\hl\right)$ does not give accurate filtered signals. Therefore, we focus on the type of simplicial filter defined in Eq.~(\ref{TSPfilter}).

Using the eigendecomposition of $\hl$, the filter can be expressed as $H_d(\bm{\mathrm{L}_k^\ell, \hl^u})=\brm{U}\tilde{H}_d\big(\bm{\Lambda}_k\big)\brm{U}^\mathrm{T}$,
where $\tilde{H}$ is called the \textit{filter frequency response} of $H_d(\bm{\mathrm{L}_k}^\ell, \hl^u)$. If we decompose $\widetilde{H}_d\big(\brm{\Lambda}_k\big)$ into its harmonic, gradient, and curl parts, this can be written as $\widetilde{H}_d=\widetilde{H}_d^{\mathrm H}\oplus\widetilde{H}_d^{\mathrm G}\oplus\widetilde{H}_d^{\mathrm C}$, where
\begin{eqnarray}
\widetilde{H}_d^{\mathrm H}\big(\bm{\Lambda}_k^{\mathrm H}\big)&=&h_{0}\brm{I},\nonumber\\
\widetilde{H}_d^{\mathrm G}\big(\bm{\Lambda}_k^{\mathrm G}\big)&=&h_0\bm{\mathrm{I}}+{\sum_{i_\ell=1}^{d_\ell}h_{i_\ell}^\ell\Bigl(\bm{\mathrm{\Lambda}_k}^{\mathrm{G}}\Bigr)^{i_\ell}},\\
\widetilde{H}_d^{\mathrm C}\big(\bm{\Lambda}_k^{\mathrm C}\big)&=&h_0\bm{\mathrm{I}}+\sum_{i_u=1}^{d_u}h_{i_u}^u\Bigr(\bm{\mathrm{\Lambda}_k}^{\mathrm{C}}\Bigl)^{i_u}\nonumber.
\end{eqnarray}
Having defined the filter frequency response of a simplicial filter, one can now describe the filter design construction~\cite{Yang2021FiniteComplexes,Yang2022SimplicialFilters}, which is central to the field of topological signal processing. Given desired frequency responses $g^{\mathrm{H}}(x),g^{\mathrm{G}}(x),g^{\mathrm{C}}(x)$ for all $x$ in $\mathrm{Eig}^{\mathrm H}(\brm{L}_{k})$, $\mathrm{Eig}^{\mathrm G}(\brm{L}_{k})$, $\mathrm{Eig}^{\mathrm C}(\brm{L}_{k})$ respectively{, and $\delta_{g^{\mathrm H}},\delta_{g^{\mathrm G}},\delta_{g^{\mathrm C}}\in\mathbb{R}_{+}$,} the filter design construction is the task of finding $\widetilde{H}_d\big(\brm{\Lambda}_k\big)$ (accordingly $H_d(\hl^\ell, \hl^u)$) such that
\begin{eqnarray}\label{simplicial filter design problem}
    \Bigl\lVert g^{\mathrm H}(x)-\widetilde{H}_d^{\mathrm H}(x)\Bigr\rVert\leq\delta_{g^{\mathrm H}},&\quad\forall x\in\mathrm{Eig}^{\mathrm H}(\brm{L}_{k}),\nonumber\\
    \Bigl\lVert g^{\mathrm G}(x)-\widetilde{H}_d^{\mathrm G}(x)\Bigr\rVert\leq\delta_{g^{\mathrm G}},&\quad\forall x\in\mathrm{Eig}^{\mathrm G}(\brm{L}_{k}),\\
    \Bigl\lVert g^{\mathrm C}(x)-\widetilde{H}_d^{\mathrm C}(x)\Bigr\rVert\leq\delta_{g^{\mathrm C}},&\quad\forall x\in\mathrm{Eig}^{\mathrm C}(\brm{L}_{k}).\nonumber
\end{eqnarray}
As $H_d(\hl^\ell, \hl^u)$ can be characterized by its coefficients, the above problem can be translated to the problem of finding $\{h_{0},\{h_{i_{\ell}}\}_{i_{\ell}\in[d_{\ell}]},\{h_{i_{u}}\}_{i_{u}\in[d_{u}]}\}$ that construct polynomials satisfying the above conditions.

Simplicial filter design construction can be solved by solving least square problems. However, this method requires explicit knowledge of the eigenvalues of $\hl$ and suffers from numerical instability~\cite{Yang2022SimplicialFilters}. In practice, eigendecompositions are computationally expensive, so other methods are used to determine the precise filter coefficients for a desired frequency response. One approach that alleviates such issues is known as the Chebyshev filter design~\cite{Yang2022SimplicialFilters}. In this approach, the functions $g^{\mathrm{G}},g^{\mathrm{C}}$ are assumed to be continuous functions with domains $[0,\lambda_{k,\mathrm{max}}^{\mathrm{G}}]$ and $[0,\lambda_{k,\mathrm{max}}^{\mathrm{C}}]$ respectively. These functions are then approximated using a series of Chebyshev polynomials. Since the gradient and curl frequency responses affect the harmonic subspace, it is required that $g^{\mathrm{G}}(0)=g^{\mathrm{C}}(0)=\mathfrak{c}_{0}$ (for some constant $\mathfrak{c}_0$) to ensure the correct response on the harmonic subspace (i.e., $\tilde{H}^{\mathrm{H}}(x) = \mathfrak{c}_0$). Let $T_{i}$ be the $i$-th Chebyshev polynomial of the first kind and $\lambda_{\mathrm{G,max}}$, $\lambda_{\mathrm{C,max}}$ are the maximum eigenvalues of $\hll$ and $\hlu$ respectively. Then the \textit{Chebyshev filter}~\cite{Yang2022SimplicialFilters} is defined as
\begin{align}\label{Chebyshev filter}
    H_{\mathrm{Cheb}}(\hl^\ell, \hl^u) := &\sum_{i=0}^{d_\ell} \mathfrak{c}_{i}^\ell T_{i}\left( \frac{2\,\hl^\ell}{\lambda_{\mathrm{G,max}}}-{\brm{I}}\right)\nonumber \\&+ \sum_{i=0}^{d_u} \mathfrak{c}_{i}^u T_{i}\left( \frac{2\,\hlu}{\lambda_{\mathrm{C,max}}}-{\brm{I}}\right) - \mathfrak{c}_0{\brm{I}},
\end{align}
such that $H_d^{\mathrm{G}}(x):=\sum_{i_{\ell}=0}^{d_{\ell}}\mathfrak{c}_{i_{\ell}}^{\ell}T_{i_{\ell}}(x)$ and $H_d^{\mathrm{C}}(x):=\sum_{i_{u}=0}^{d_{u}}\mathfrak{c}_{i_{u}}^{u}T_{i_{u}}(x)$ for all $x\in[0,1]$ satisfying Eq.~\eqref{simplicial filter design problem}. Equivalent to Eq.~\eqref{TSPfilter}, the Chebyshev filter is characterized by its coefficients $\{\mathfrak{c}_{0},\{\mathfrak{c}_{i_{\ell}}^{\ell}\}_{i_{\ell}\in[d_{\ell}+1]},\{\mathfrak{c}_{i_{u}}^{u}\}_{i_{u}\in[d_{u}+1]}\}$. The classical advantages of this design are two-fold. Firstly, it avoids directly computing all of the eigenvalues of $\hl$, which is expensive (even the largest eigenvalues of $\hll$ and $\hlu$ can be bounded from above rather than explicitly computed). Secondly, this framework allows for the filtering algorithm to be implemented distributively in the classical setting. The latter can significantly reduce the classical computational burden for local computation, albeit at the expense of having exponentially many networked computers~\cite{ShumanChebyshevProcessing,Yang2022SimplicialFilters}. 

Generally, the filtering operation is computationally challenging in higher-order network analysis. In standard implementations \cite{Yang2021FiniteComplexes,Yang2022SimplicialFilters}, the computational complexity grows linearly in the number of multi-way interactions captured in the higher-order network. In real-world scenarios, these large higher-order interactions, modelled in higher-dimensional simplices, have been detected in collaboration~\cite{Ebli2020SimplicialNetworks,Baccini2022WeightedTopology}, social~\cite{Krishnagopal2021SpectralLaplacians}, and brain networks~\cite{Lucas2020MultiorderNetworks}. Explicitly, the number of possible $k$-wise interactions, given by $\binom{n}{k}$, grows exponentially in $k$ (for the practically relevant ranges of $k$), and thus the number of possible higher-order interactions in these large networks is enormous. To combat this exponential relationship between the dimension of the simplicial complex and the classical complexity, we aim to reduce the TSP filtering cost by utilizing quantum computers. In this paper, we do not aim to solve the filter design construction and consider that the polynomials of the filter are already given, instead focusing on applying this filter to a given signal $\brm{s}^k$ (encoded in a quantum state). In the following subsection, we re-express the simplicial filtering process in the quantum linear algebra framework and briefly describe how we modify tools from QTDA to implement a quantum analogue of the filtering operation on quantum computers. Then we give our main results on the complexity of our algorithm in Section~\ref{sec:main}.

\subsection{Quantum linear algebra framework}

Both QTDA and QTSP can be considered as part of the class of quantum linear algebra (QLA) algorithms. QLA constitutes a broad church of quantum algorithms with perceived applications in machine learning and data analysis. The breakthrough of quantum algorithms for linear algebra problems started from the seminal work by Harrow, Hassidim, and Lloyd~\cite{Harrow2009QuantumEquations} that{, given a Hermitian matrix $\brm{A}$ and a (state) vector $\ket{\brm{v}_\mathrm{{in}}}$,} used quantum phase estimation and Hamiltonian simulation to prepare a state $\lvert \widetilde{\brm{v}}_{\mathrm{sol}}\rangle$, an approximate solution to $\brm{A}\lvert \brm{v}_{\mathrm{sol}}\rangle=\lvert\brm{v}_{\mathrm{in}}\rangle$. Since then, QLA concepts have been applied to a wide range of topics {in} big data analysis and machine learning, including support vector machines~\cite{Rebentrost2014QuantumClassification}, principle component analysis~\cite{Lloyd2014QuantumAnalysis}, recommendation systems~\cite{Kerenidis2017QuantumSystem}, and clustering~\cite{KerenidisQ-means:Learning,Kerenidis2021QuantumClustering}. 

However, as pointed out by Aaronson~\cite{Aaronson2015ReadPrint}, even under the favourable assumptions that there exist efficient algorithms that encode $\lvert \brm{v}_{\mathrm{in}}\rangle$ and $\brm{A}$ in a quantum circuit, many QLA algorithms such as the HHL algorithm only ``solve'' the desired linear algebra problem efficiently in the sense that they can rapidly prepare a quantum state $\lvert \widetilde{\brm{v}}_{\mathrm{sol}}\rangle$, a quantum state representing an approximation of $\brm{v}_{\rm{sol}}$. {At this point, tomography must be performed} to retrieve {the} information encoded in $\lvert \widetilde{\brm{v}}_{\mathrm{sol}}\rangle$ to obtain the solution. This brings additional costs and sources of error. With this in mind, the whole pipeline of QLA, including the step to obtain a (classical) description of a state $\ket{\hat{\brm{v}}_{\mathrm{sol}}}$ which approximates the state $\lvert \widetilde{\brm{v}}_{\mathrm{sol}}\rangle$, can be described as follows.
\begin{enumerate}
    \item A classical data $\textsc{Input}_c$ is given in the form of a list of vector components $\{v^i_{\mathrm{in}}\}$ of a data vector $\brm{\bm{\mathrm{v}}}_{\mathrm{in}}$ and a list of entries $\{A_{ij}\}$ of a matrix $\brm{A}$ that implements a linear transformation to $\brm{v}_{\mathrm{in}}$. Given a $\textsc{Task}$ to solve or accomplish {and a vector $\brm{v}_{\mathrm{in}}$}, one can obtain the solution to {the} a given linear algebra task, ${\brm{\bm{\mathrm{v}}}}_{\mathrm{sol}}$, by employing a classical algorithm $\textsc{Alg}_{c}(\brm{A})$.
    \item In \textsc{Encode} step, the set of vector components $\{v^i_{\mathrm{in}}\}$ is encoded in the quantum state $\lvert\brm{\bm{\mathrm{v}}}_{\mathrm{in}}\rangle$, either in the amplitudes, basis states, or both{, creating the input state} $\textsc{Input}_q:=\{\lvert\brm{\bm{\mathrm{v}}}_{\mathrm{in}}\rangle\}$.
    \item A quantum algorithm $\textsc{Alg}_{q}$ starts by encoding the matrix $\brm{A}$ {(or an approximation of $\brm{A}$)} as a unitary operator $\brm{U}_{\brm{A}}$ that is implementable in a quantum circuit. Then, it takes $\textsc{Input}_q$ to solve a given task by transforming $\lvert\brm{\bm{\mathrm{v}}}_{\mathrm{in}}\rangle\mapsto\lvert\widetilde{\brm{\bm{\mathrm{v}}}}_{\mathrm{sol}}\rangle:=\textsc{Alg}_{q}(\brm{U}_{\brm{A}})\lvert\brm{\bm{\mathrm{v}}}_{\mathrm{in}}\rangle$ such that $\big\lVert\lvert\brm{\bm{\mathrm{v}}}_{\mathrm{sol}}\rangle-\lvert\widetilde{\brm{\bm{\mathrm{v}}}}_{\mathrm{sol}}\rangle\big\rVert\leq\delta$, where $\delta$ is an error from approximating the solution. In other words, $\textsc{output}_q:=\lvert\widetilde{\brm{\bm{\mathrm{v}}}}_{\mathrm{sol}}\rangle$ is the approximate solution to a given task. 
    \item An additional \textsc{Retrieve} step is required in order to obtain classical information about the approximate solution encoded in $\lvert\widetilde{\brm{\bm{\mathrm{v}}}}_{\mathrm{sol}}\rangle$. This step normally consists of some variant of state tomography techniques. From this one obtains a classical description of an approximation $\ket{\hat{\brm{v}}_{\mathrm{sol}}}$ to $\ket{\brm{v}_{\mathrm{sol}}}$, which carries an additional error $\varepsilon$ such that $\norm{\ket{\hat{\brm{v}}_{\mathrm{sol}}}}{\lvert\widetilde{\brm{v}}_{\mathrm{sol}}\rangle}\leq\varepsilon$. As a result, one has $\norm{\ket{\brm{v}_{\mathrm{sol}}}}{\ket{\hat{\brm{v}}_{\mathrm{sol}}}}\leq\delta+\varepsilon$. 
\end{enumerate}

Our algorithm for TSP filtering on quantum computers fits into this QLA paradigm, specifically in step 3. Given access to the state preparation oracle, we encode the simplicial signals in a quantum state $\ket{\brm{v}_{\mathrm{in}}} = \lvert\brm{s}^k\rangle$ and build a circuit that has the action of a polynomial of the lower and upper Laplacians on the subspace containing $\lvert\brm{s}^k\rangle$. Due to nuances in implementing the quantum algorithm, our filters are {inspired by} but distinct from the Chebyshev filter design. In our case, the upper and lower Laplacian are {each} rescaled such that their eigenvalues are in $[0,1]$ so that they can be embedded in unitary matrices. In kind, we assume that $g^{\mathrm{G}},g^{\mathrm{C}}: [0,1] \to [-1,1]$. We discuss this in more detail in the following section.

\section{A Quantum Implementation of Topological Signal Processing}\label{sec:main}
\label{QTSP overview}

This section formally states our main result{, where we} give a general framework for implementing simplicial filters on fault-tolerant quantum computers. In TSP, a solution to filter design construction connects the desired frequency responses $\{g^{\mathrm{H}},g^{\mathrm{G}},g^{\mathrm{C}}\}$ with a set of polynomial coefficients. The filtering algorithm then aims to output a filtered simplicial signal given input simplicial signal data $\brm{s}^k$ and the filter as described in Eq.~(\ref{TSPfilter}). To make such a construction more amenable to implementation on a quantum computer, we focus on a subset of filtering problems inspired by the Chebyshev filter design given in Eq.~(\ref{Chebyshev filter}). Here, we require that the frequency responses $g^{\mathrm{G}}$ and $g^{\mathrm{C}}$ are continuous, and even more strictly we assume these functions are (at least approximated by) polynomials. As the eigenvalues of any operator embedded in a unitary matrix are bounded by $1$ and $\hl^\ell,\hl^u$ are Hermitian, we impose that $g^\mathrm{G},g^{\mathrm{C}}: [0,1] \to [-1,1]$. We also rescale the upper and lower Hodge Laplacians so that their eigenvalues are bound by $1$. This is similar but not identical to the rescaling present in the Chebyshev filter design. We also impose that $g^{\mathrm{G}}(0) = g^{\mathrm{C}}(0) = g^{\mathrm{H}} = \mathfrak{h}_0$ for some real constant $\mathfrak{h}_0\in[0,1]$. Given a set of continuous frequency responses satisfying these properties, we present an algorithm for performing the TSP filtering operation on a fault-tolerant quantum computer. The precise details of the filter are given in Eq.~(\ref{eqn:quantumfilter}). This section gives an overview of this quantum algorithm and of its properties. Then, we give the specifics and detailed analysis of the algorithm in Section~\ref{sec:details}.

Below we sketch a high-level picture of our QTSP algorithm, presented in the QLA structure {outlined in Section~\ref{sec:background}}.

\begin{enumerate}
    \item Classical data representing a topological signal $\bm{\mathrm{s}}^k \in \mathbb{R}^{n_k}$ is given in the form of a list of signal values $s^k_i$ for each $k$-simplex $\sigma_k^i$, for $i\in[n_k]$. The frequency responses are functions $g^{\mathrm{G}}, g^{\mathrm{C}}:[0,1] \to [-1,1]$ such that $g^{\mathrm{G}}(0) = g^{\mathrm{C}}(0) = g^{\mathrm{H}} = \mathfrak{h}_0$ for some real constant $\mathfrak{h}_0\in[0,1]$. We assume $g^{\mathrm{G}}$ and $g^{\mathrm{C}}$ are (at least approximated by) polynomials with coefficients $\{\mathfrak{h}_{i_{\ell}}^{\ell}\}_{i_{\ell}\in[d_{\ell}+1]}$ and $\{\mathfrak{h}_{i_{u}}^{u}\}_{i_{u}\in[d_{u}+1]}$ respectively. Hence, a classical input is defined as
    \begin{eqnarray}
        \textsc{Input}_c&:=&\Big\{\brm{s}^{k},\{\mathfrak{h}_0,\mathfrak{h}_{i_{\ell}}^{\ell},\mathfrak{h}_{i_{u}}^{u}\}\Big\},
    \end{eqnarray}
    for $i_{\ell}\in[d_{\ell}+1],i_{u}\in[d_{u}+1]$.
    \item We assume access to a state preparation unitary $\brm{U}_{\mathrm{prep}}$ which implements the \textsc{Encode} step. This step takes input $\brm{s}^{k} \in \textsc{Input}_c$ and a state $\ket{\bar{0}}:=\ket{0\cdots 0}$ and outputs the state $\lvert \bm{\mathrm{s}}^k\rangle$, which we call the \textit{simplicial signal state}. Given a basis state $\ket{\sigma^i_k}$ to represent each simplex $\sigma_k^i \in \mathcal{K}$, we define $\lvert \brm{s}^k\rangle$ as 
    \begin{eqnarray}\label{simplicialquantumtsignal}
        \brm{U}_{\mathrm{prep}}\ket{\bar{0}}=\lvert\brm{s}^k\rangle&:=&\frac{1}{\lVert \brm{s}^k\rVert_{2}}\sum_{i=0}^{n_k-1} \,s^k_i\:\lvert\sigma_k^i\rangle,
    \end{eqnarray}
    where $\lVert \brm{s}^k\rVert_{2}^2:=\sum_{i={0}}^{n_k-1} (s^k_i)^2$ and $n_k$ is the number of $k$-simplices in $\mathcal{K}$. In this representation, the simplices are encoded in the basis states following two previously used approaches described in Ref.~\cite{McArdle2022AQubits}, which are called the \textit{compact} and \textit{direct} approaches. We review these in Appendix~\ref{state preparation}. 
    
    \item The \textsc{Filter} step takes $\{\mathfrak{h}_0,\mathfrak{h}_{i_{\ell}}^{\ell},\mathfrak{h}_{i_{u}}^{u}\}\in \textsc{Input}_c$ to constructs a quantum circuit implementing a simplicial filter $H_d\left( \hll/\alpha_{k}^2, \hlu/\alpha_{k+1}^2 \right)$ and then applies it to $\ket{\brm{s}^{k}}\ket{0}^{\otimes a}$ to output 
    \begin{eqnarray}\label{pre postselected simplicial signal state}
        \ket{0}^{\otimes a}\,H_d\left( \frac{\hll}{\alpha_{k}^2}, \frac{\hlu}{\alpha_{k+1}^2} \right)\lvert\brm{s}^k\rangle+\ket{\perp},
    \end{eqnarray}
    where
    \begin{eqnarray}\label{eqn:quantumfilter}
         H_d\left( \frac{\hll}{\alpha_{k}^2}, \frac{\hlu}{\alpha_{k+1}^2} \right)&:=&\sum_{i_\ell=0}^{d_\ell}\mathfrak{h}_{i_\ell}^l\left(\frac{\hll}{\alpha_{k}^2}\right)^{i_\ell}\nonumber\\&&+\sum_{i_u=0}^{d_u}\mathfrak{h}_{i_u}^u \left(\frac{\hlu}{\alpha_{k+1}^2}\right)^{i_u}-\mathfrak{h}_{0}\brm{I}.\nonumber\\
    \end{eqnarray}
    Here, $\alpha_{k},\alpha_{k+1}\in\mathbb{R}_{+}$ are the rescaling factors ($\alpha_k = \sqrt{n}$ in the direct approach and $\alpha_k = \sqrt{(n+1)(k+1)}$ in the compact approach), for all $k\in[n]$, and $\ket{\perp} = \sum_{x\neq 0} \ket{x}\ket{\text{Garb}_x}$ is an unnormalized state that corresponds to the ancilla being in a nonzero state, which vanishes after successful postselection in the \textsc{Retrieve} step. Specifically, we use quantum singular value transformation and the linear combination of unitaries (LCU) to obtain a projected unitary encoding of $H_d\left( \hll/\alpha_{k}^2, \hlu/\alpha_{k+1}^2 \right)$, which will be discussed further in Section~\ref{sec:details}.
    
    \item The \textsc{Retrieve} step implements a postselection to the ancilla qubits $\ket{0}^{\otimes a}$ in Eq.~\eqref{pre postselected simplicial signal state} to output the \textit{filtered simplicial signal state} given by
    \begin{eqnarray}
        \lvert\brm{s}_{\mathrm{fil}}^k\rangle:=\frac{1}{\mathcal{N}}\,H_d\left( \frac{\hll}{\alpha_{k}^2}, \frac{\hlu}{\alpha_{k+1}^2} \right)\lvert\brm{s}^k\rangle
    \end{eqnarray}
    with probability of success $O(\mathcal{N}^2)$, with $\mathcal{N}:=\big\|H_d\left( \hll/\alpha_{k}^2, \hlu/\alpha_{k+1}^2 \right)\lvert\brm{s}^k\rangle\big\|_2$. One could also employ a subsequent amplitude amplification technique to boost the success probability. Afterwards, state tomography is used to obtain an approximation $\hat{\brm{s}}^k_{\mathrm{fil}}$ to the filtered signal encoded in amplitudes of $\lvert\brm{s}_{\mathrm{fil}}^k\rangle$.
\end{enumerate}

Put this way, the QTSP filtering process is simply a state preparation protocol for outputting   $\lvert\brm{s}_{\mathrm{fil}}^k\rangle$ given the PUE of a simplicial filter and the signal state $\lvert\brm{s}^k\rangle$.
Now we formally state our main result, which we call QTSP filtering, as follows.
\begin{theorem}[QTSP filtering]\label{QTSP filtering}
    Let $G$ be the underlying $n$-vertex graph of a clique complex $\mathcal{K}(G)$, and let $d = \max\{d_\ell, d_u\}$ be the order of the simplicial filter $H_d\left( \hll/\alpha_{k}^2, \hlu/\alpha_{k+1}^2 \right)$. There exists a quantum circuit implementing \textsc{Filter} with:
    \begin{enumerate}
        \item[(a)] (\textbf{Direct approach}) $O\left(dn\log n\right)$ non-Clifford gate depth and $O(n)$ space; 
        \item[(b)] (\textbf{Compact approach}) $O\left(dkn^2\log(n) {\log(\log (n))}\right)$ non-Clifford gate depth and $O(k\log n)$ space.
    \end{enumerate}
    The probability of success of outputting filtered state $\ket{\brm{s}^{k}_{\mathrm{fil}}}$ is $O(\mathcal{N}^2)$, where $\mathcal{N}:=\big\|H_d\left( \hll/\alpha_{k}^2, \hlu/\alpha_{k+1}^2 \right)\lvert\brm{s}^k\rangle\big\|_2$.
\end{theorem}
\noindent The proof of this theorem is given in Sec.~\ref{sec:details}. The filtered state $\ket{\brm{s}^{k}_{\mathrm{fill}}}$ is given by postselecting the ancilla qubits in the \textsc{Filter} step.

The computational cost of the classical TSP filtering algorithm is $O(dn_kD)$~\cite{Yang2021FiniteComplexes}, where $n_k$ is the number of $k$-simplices in the complex and $D$ is the maximum degree of a simplex in the complex (equivalently, the maximum number of non-zero, off-diagonal entries in a row of $\hl$). An $n$-vertex simplicial complex may have as many as $\binom{n}{k+1}$ $k$-simplices, and thus the classical filtering complexity scales exponentially in $k$ in the worst case. In comparison, our QTSP filtering algorithm scales polynomially in $k$. This suggests that QTSP filtering could be a promising avenue for achieving a super-quadratic speedup over performing TSP filtering on a classical computer.

However, there are some notable caveats to any claims of a {practical} speedup. Theorem~\ref{QTSP filtering} only accounts for the complexity of the PUE of $H_d(\hl^\ell/\alpha_\ell^2, \hl^u/\alpha_u^2)${, that is, the complexity of the \textsc{Filter} step}. In order to recover the filtered signal $\ket{\brm{s}^k_{\mathrm{fil}}}$ one still needs to perform post-selection. In this work, we do not account for the complexity of this or any related amplitude amplification. 
We also do not consider the costs of creating the quantum state $\ket{\brm{s}^k}$ or recovering the desired information from $\ket{\brm{s}^k_{\mathrm{fil}}}$ (i.e., the \textsc{Encode} or \textsc{Retrieve} steps). Some forms of data (such as a signal concentrated entirely on a single simplex) are computationally straightforward to encode in a quantum state. However, even in general it is worth noting that the space complexity of efficient state preparation methods (such as the one given by Ref.~\cite{Zhang2022QuantumApplications}) have an ancilla cost comparable to the number of bits needed to store the vector on a classical computer. In terms of retrieving the filtered signal, the tomography methods of van Apeldoorn, Cornelissen, Gily\'en, and Nannicini~\cite{vanApeldoorn2022QuantumUnitaries} allow for efficient $\ell_\infty$-norm approximations of $\ket{\brm{s}^k_{\mathrm{fil}}}$ to be obtained. This is particularly useful if a small number of amplitudes are expected to be large in the filtered signal. {On the other hand, the output from the QTSP algorithm could potentially be used as the input for another quantum algorithm, such as a clustering algorithm, sidestepping the need for tomography on the signal state itself.}

We also note that the scaling factors of $\alpha_k$ and $\alpha_{k+1}$ in Eq.~(\ref{eqn:quantumfilter}) are not present in Eqs.~(\ref{TSPfilter}) or (\ref{Chebyshev filter}). Thus, in order to make a precise comparison between the complexities of quantum and classical TSP for the same task, the filter coefficients for the quantum filter need to be adjusted to account for this rescaling. In comparison to the {Chebyshev filter design, the rescaling factor given in Eq.~(\ref{Chebyshev filter}) is roughly $\lambda_{\rm{G,\max}}/2, \lambda_{\rm{C,\max}}/2 = O(n)$~\cite{Friedman1998ComputingLaplacians}.} Thus, for simplicial complexes with a large maximum eigenvalue, the rescaling factors inherent in this design are similar to the rescaling factors in the quantum simplicial filters. The specifics of {these rescalings, as well as encoding and retrieval costs,} can vary greatly between applications; thus, we do not compare the complexities between classical and quantum algorithms for general simplicial filters. However, the broad range of possible applications of TSP increases the likelihood that there exist applications for which these hurdles can be overcome, and a practical speed-up can be obtained.

We give a specific application of Theorem~\ref{QTSP filtering} related to the Hodge decomposition. This decomposition plays an important role in many areas ranging from {analysis of} biological networks to computer vision~\cite{Bhatia2013TheSurvey,Lim2020HodgeGraphs}, especially in lower-dimensional cases (i.e., $k=1,2$). Its applications to higher-order network analysis remain mostly unexplored. We present a QTSP-based projection algorithm that {(approximately)} projects a signal $\brm{s}^k$ (represented by its simplicial signal state $\lvert\brm{s}^k\rangle$) onto its gradient, curl, or harmonic components as described in Eq.~(\ref{Hodge decomposition of simplicial signals}). Using our QTSP framework given in Theorem~\ref{QTSP filtering}, we define quantum simplicial filters that approximate the functions $\Pi^{\mathrm{G}}$, $\Pi^{\mathrm{C}}$ which implement projections onto the gradient and the curl spaces (noting that $\Pi^{\mathrm{H}} = \brm{I} - \Pi^{\mathrm{G}} - {\Pi}^{\mathrm{C}}$). Recall the definition of $\alpha_k$ given in the $\textsc{Filter}$ step, that is, $\alpha_k = \sqrt{n}$ for the direct approach and $\alpha_k = \sqrt{(n+1)(k+1)}$ for the compact approach.
\begin{theorem}[QTSP filtering for subcomponent projections]\label{QTSP Hodge filtering}
    Let $\xi_{\mathrm{min}}^{\ell},\xi_{\mathrm{min}}^{u}$ be the smallest non-zero singular values of the boundary operators $\brm{B}_{k}$ and $\brm{B}_{k+1}$ respectively. Let $\kappa_{\ell},\kappa_{u}$ be such that $1/\kappa_{\ell}\in\left(0,\xi_{\mathrm{min}}^{\mathrm{\ell}}/\alpha_{k}\right)$, and $1/\kappa_{u}\in\left(0,\xi_{\mathrm{min}}^{\mathrm{u}}/\alpha_{k+1}\right)$.
    Given a simplicial signal state $\lvert\brm{s}^k\rangle$ and  $\varepsilon_{\ell},\varepsilon_{u}\in(0,1/2)$, there exist quantum circuits implementing quantum simplicial filters $H_{d_{\ell}+1}\left(\hll/\alpha_{k}^2\right)$ and $H_{d_{u}+1}\left(\hlu/\alpha_{k+1}^2\right)$, with orders $d_\ell+1 \in O(\kappa_{\ell}^{2}\log((n\kappa_{\ell})^{2}/(\alpha_{k}^2\varepsilon_{\ell})))$ and $d_u+1 \in O(\kappa_{u}^{2}\log((n\kappa_{u})^{2}/(\alpha_{k+1}^2\varepsilon_{u})))$ respectively, such that
    \begin{align*}
        &\norm{\Pi^{\mathrm{G}}}{2\kappa_{\ell}^2\,H_{d_{\ell}+1}\left(\hll/\alpha_{k}^2\right)}\leq\varepsilon_{\ell}, \:\mbox{and}\\ &\norm{\Pi^{\mathrm{C}}}{2\kappa_{u}^2\,H_{d_{u}+1}\left(\hlu/\alpha_{k+1}^2\right)}\leq\varepsilon_{u}.
    \end{align*}
    These circuits have complexity
    \begin{enumerate}
        \item[(a)] (\textbf{Direct approach}) $O\left(dn\log n\right)$ non-Clifford gate depth and $O(n)$ space,
        \item[(b)] (\textbf{Compact approach}) $O\left(dkn^2\log(n) {\log(\log (n))}\right)$ non-Clifford gate depth and $O(k\log n)$ space,
    \end{enumerate}
    where $d \in \{d_\ell, d_u\}$.
\end{theorem}
\noindent The proof of this theorem is given in Sec.~\ref{sec:QTSP Hodge filtering}, and naturally follows from Theorem~\ref{QTSP filtering} by choosing the appropriate filter $H_d$.

The subcomponent extraction problem is important in applications such as preference ranking. For example, given pairwise preference data between $n$ elements, one well-studied algorithm for obtaining a global ranking of all elements is HodgeRank~\cite{Jiang2011StatisticalTheory}. This global ranking is obtained by considering the pairwise preference data as an edge signal $\brm{s}^1$ and projecting onto the gradient component $\brm{s}^1_G$, which corresponds to a pairwise ranking that is ``globally consistent'' with the desired total ranking.
Theorem~\ref{QTSP Hodge filtering} states that a filter to perform this projection can be implemented efficiently (polynomially in $n$, $k$, $d$) on a quantum computer for arbitrary dimension $k$ under certain conditions. This gives a potential application of QTSP filtering beyond what is commonly explored in TSP. We give the details of the application of our algorithm to Hodge decomposition filtering in Section~\ref{sec:QTSP Hodge filtering}. In Ref.~\cite{leditto2024quantumhodgeranktopologybasedrank}, quantum algorithms for HodgeRank are discussed in significantly more detail. 

In comparison to Theorem~\ref{QTSP filtering}, the complexity of implementing the filters described in Theorem~\ref{QTSP Hodge filtering} has an extra dependence on the smallest singular value of the appropriate boundary operator (which is the square root of the smallest eigenvalue of the corresponding lower or upper Laplacian). This is because the order of the filter required to achieve an accuracy of $\varepsilon$ depends on the ratio between the smallest and largest eigenvalues of the corresponding Laplacian. Thus, we note that the complexity is only polynomial in $n$ and $k$ if the spectral gap of the Laplacian is bounded from below by some polynomial in $n$ and $k$. 

While Theorem~\ref{QTSP filtering} does not itself require any assumptions about the spectral gap of $\hl$, certain applications, such as that given in Theorem~\ref{QTSP Hodge filtering} (or more generally any application requiring a pseudoinverse) will require a non-vanishing spectral gap. This is somewhat more relaxed than the assumptions required for QTDA, which always requires such a spectral gap, but it is a concern depending on the precise application. In general, the spectral gap of the Hodge Laplacian (and consequently its lower and upper constituents) can be exponentially small in $k$~\cite{Steenbergen2012AComplexes}, and so the applicability of Theorem~\ref{QTSP Hodge filtering} must be analysed on a case-by-case basis depending on the underlying simplicial complex.

It is also worth noting that most of our {algorithm} described by Theorem~\ref{QTSP filtering} does not rely on $\mathcal{K}$ being a clique complex. The restriction to clique complexes is only required so that we can explicitly construct polynomial-depth ``membership function{s}'' $\brm{P}_{k}$ that ha{ve} the action 
\begin{align}\label{eq:projector}
    \brm{P}_k \ket{\sigma_k}\ket{0}\ket{0}^{\otimes a_p} = \ket{\sigma_k}\ket{\mathds{1}\{\sigma_{k}\in\mathcal{K}\} }\ket{0}^{\otimes a_p},
\end{align}
where $\mathds{1}\{\sigma_{k}\in\mathcal{K}\} = 1$ if and only if $\sigma_k \in \mathcal{K}$ and is $0$ otherwise, $\ket{\sigma_k}$ is a basis state representing a possible $k$-simplex (some subset of the vertex set $\mathcal{V}$), and $a_p$ is the number of ancilla qubits used. We do not know how to efficiently implement such a membership oracle for a general simplicial complex. Since these membership functions are implemented $O(d)$ times to create a polynomial of degree $d$, an efficient implementation of this subroutine is vital to an efficient filtering algorithm. In the case of a clique complex, efficient constructions for membership functions are known. In the case of direct encoding, we employ the circuit given by Berry et al.~\cite{berry2023analyzing}. We then adapt this to provide an analogous circuit that works for the compact encoding scheme; the exact construction is given in Appendix~\ref{sec:proj}.

In the next section, we give the details of the algorithm and prove the complexities given in Theorems~\ref{QTSP filtering} and \ref{QTSP Hodge filtering}. First, we define the simplicial signal state $|\brm{s}^k\rangle$ and the direct and compact encodings and review the {projected unitary encoding} of the boundary operator given by McArdle, Gily\'en, and Berta~\cite{McArdle2022AQubits}. We focus on the filter construction for the compact encoding scheme; the results necessary to prove the claims about the direct encoding scheme are given in Appendix~\ref{sec:boundaryoperatorconstruction}.

\section{Complexity Analysis of the Filtering Process}\label{sec:details}

In this section, we give a detailed description of how to implement the \textsc{Filter} step, as described in Section~\ref{sec:main}, on a quantum computer for a general polynomial filter, and prove Theorem~\ref{QTSP filtering}. The tools and techniques we use to build our applications exist in the literature and are adapted from well-known quantum algorithms. The main tools are the {projected unitary encoding} of the boundary operator~\cite{Akhalwaya2022TowardsComputers,berry2023analyzing,McArdle2022AQubits}, {quantum singular value transformation}~\cite{Gilyen2019QuantumArithmetics}, and {linear combinations of unitaries}~\cite{Childs2017QuantumPrecision}.

\begin{figure*}[t]
    \centering
    \includegraphics[scale=0.48]{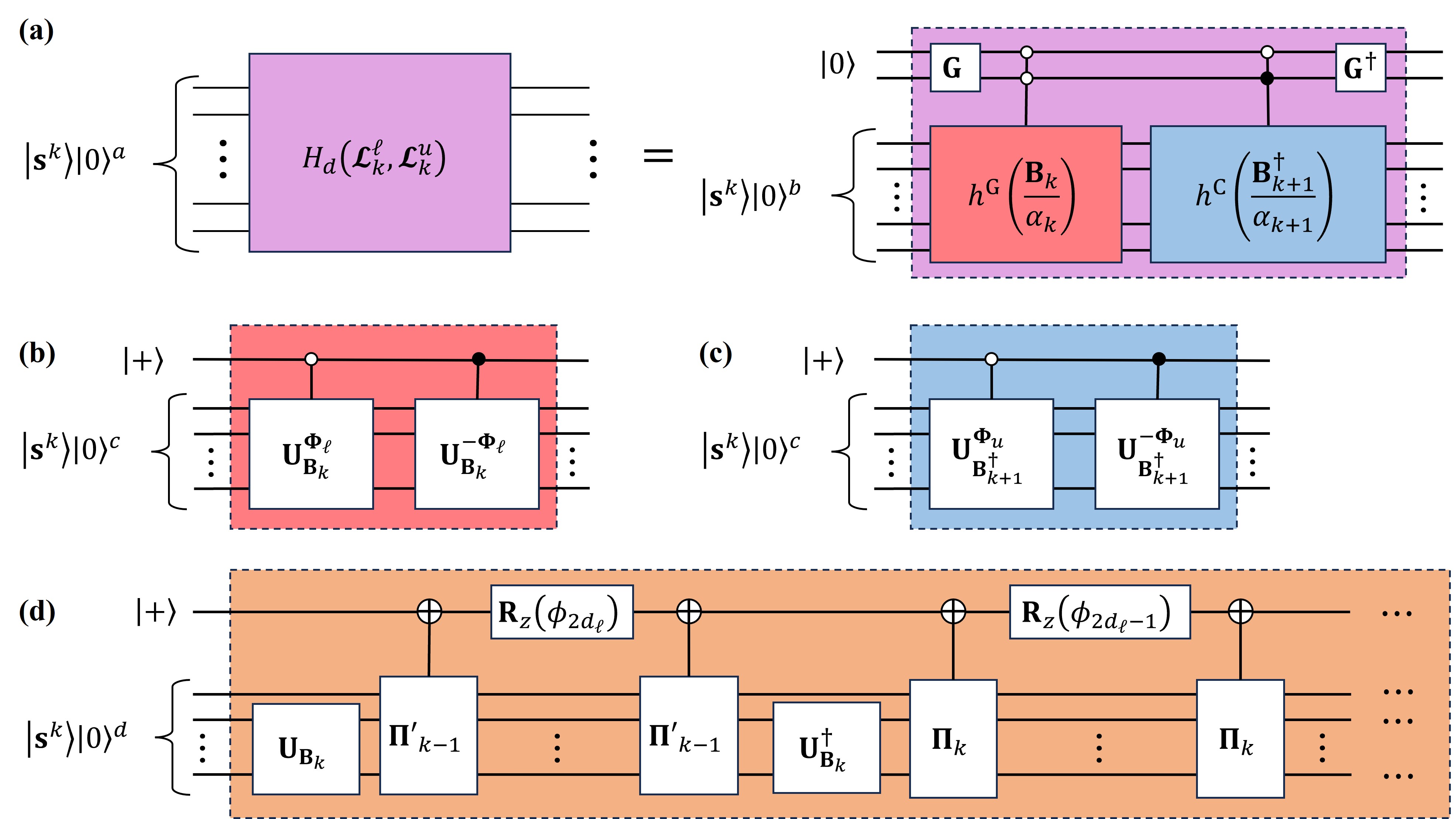}
    \caption{The general QTSP filtering circuit. \textbf{(a)} A quantum simplicial filter $H_d\Bigl(\hlln,\hlun\Bigr)$ is a linear combination of two projected unitary encodings (PUES) of $h^{\mathrm{G}}\Bigl(\bk/\alpha_{k}\Bigr)$ and $h^{\mathrm{C}}\Bigl(\bkk/\alpha_{k+1}\Bigr)$. \textbf{(b)} A PUE of a real-valued polynomial transformation $h^{\mathrm{G}}\Bigl(\bk/\alpha_{k}\Bigr)$. \textbf{(c)} A PUE of a real-valued polynomial transformation $h^{\mathrm{C}}\Bigl(\bkkdag/\alpha_{k+1}\Bigr)$. \textbf{(d)} The QSVT circuit $\brm{U}^{\bm{\Phi}_{\ell}}_{\bk}$ generating degree-$2d_{\ell}$ polynomial transformation of $\bk/\alpha_{k}$.}
    \label{QTSP filter design figure}
\end{figure*}


A simplicial filter $H_{d}(\hl^\ell, \hl^u)$ is comprised of polynomials of $\hll$ and $\hlu$, as given in Eq.~(\ref{TSPfilter}). As described in the \textsc{Filter} step in Section~\ref{sec:main}, we {implement simplicial filters of the form}
\begin{eqnarray}\label{our QTSP filter}
    H_d(\hlln, \hlun)&:=&\sum_{i_\ell=0}^{d_\ell}\mathfrak{h}_{i_\ell}^l\Bigr(\bkn\Bigl)^{2i_\ell}\nonumber\\&&+\sum_{i_u=0}^{d_u}\mathfrak{h}_{i_u}^u\Bigr(\bkkdagn\Bigl)^{2i_u}-\mathfrak{h}_{0}\brm{I},
\end{eqnarray}
where $\bm{\mathcal{B}}_k=\bk/\alpha_k$ is defined in Eq.~(\ref{compact boundary operator}) for the compact encoding, $\hlln:=\bm{\mathcal{B}}_{k}^\dagger\bkn$, and $\hlun:=\bkkn\bkkdagn$. 
This multivariable polynomial of $\hlln$ and $\hlun$ can be constructed by implementing the LCU method~\cite{Childs2017QuantumPrecision} and the QSVT~\cite{Gilyen2019QuantumArithmetics}, which is a versatile quantum algorithm used to construct polynomial transformations of any matrix in a quantum circuit. For completeness, we briefly review the LCU method and QSVT algorithm in Appendi{ces}~\ref{PUE of arbitrary matrices} and~\ref{quantum algorithm for constructing polynomials}.

The combination of the LCU method and QSVT enables us to construct different degree and parity polynomials of different matrices in a single call to a \textsc{Select} operation. To this end, we give a slight generalization of Lemma 52 in Ref.~\cite{Gilyen2019QuantumArithmetics} combined with the construction of general polynomials given in Ref.~\cite{Dong2021EfficientProcessing}{, which produces} a projected unitary encoding of a linear combination of complex-valued polynomials. Here, we define a set of matrices $\left\{\brm{A}_{i}\right\}_{i\in[m]}$, all with the same dimensions, acting on the same Hilbert space.

\begin{lemma}[PUE of a Linear Combination of Complex-Valued Polynomials]\label{LCU of many polynomials}
Let $\left\{f_i(x)\right\}_{i \in [m]}$ be a set of complex-valued polynomials such that $\Re(f_{i}(x))=g_{\Re,i}(x)$ and $\Im(f_{i}(x))=h_{\Im,i}(x)$ are real-valued polynomials of degree $d_{\Re,i}$ and $d_{\Im,i}$ with the same parity. Suppose that $\lvert g_{\Re,i}(x)\rvert\leq 1$ and $\lvert h_{\Im,i}(x)\rvert\leq 1$ for all $x\in[-1,1]$ and $i \in [m]$. Let $\brm{U}_{\brm{A}_{i}}$ be a $(1,a_{i},0)$-PUE of $\brm{A}_{i}$ {for each $i \in [m]$}. Then there exist sequences of phase factors $\Bigl\{\bm{\Phi}_{g_{i}},\bm{\Phi}_{h_{i}}\Bigr\}_{i\in[m]}$ implementing a $\left(\beta,a+\lceil\log(2m)\rceil+3,0\right)$-PUE of $\sum_{i=0}^{m-1} c_i\, f_{i}(\brm{A}_i)$, where $c_{i}\in\mathbb{R}_{+}$, $\beta:=\lVert c\rVert_1$, and $a:=\left(\sum_{i=0}^{m-1}a_{i}\right)$. 

Moreover, let $\delta\in\mathbb{R}_{+}$ be arbitrary and define $d=\mathrm{max}\left(d_{\Re,0},\cdots,d_{\Re,m-1},d_{\Im,0},\cdots,d_{\Im,m-1}\right)$. Then using $O(m\cdot\mathrm{poly}\left(d,\log(1/\delta\right)))$ classical computation time, sequences of phase factors $\Bigl\{\bm{\Phi}'_{g_{i}},\bm{\Phi}'_{h_{i}}\Bigr\}_{i\in[m]}$ can be found which implement a $\left(\beta,a+\lceil\log(2m)\rceil+3,\beta\delta\right)$-PUE of $\sum_{i=0}^{m-1} c_i\, f_{i}(\brm{A}_i)$.
\end{lemma}

The proof is given in Appendix~\ref{app:lemma1proof}. For our purposes, we employ this lemma to construct a PUE of $H_d(\hlln, \hlun)$ from PUEs of $\bkn$ and $\bkkdagn$ (defined in Eq.~(\ref{compact boundary operator}) for the compact encoding). The basic outline of this design is as follows: assume that the frequency responses $g^{\mathrm{G}}, g^{\mathrm{C}}$ are polynomials of degree at most $d$ on $[0,1]$ such that $g^{\mathrm{G}}(0) = g^{\mathrm{C}}(0) = \mathfrak{h}_0$ and $|g^{\mathrm{G}}(x)|\leq 1,|g^{\mathrm{C}}(x)|\leq 1$. 
{We suppose that $g^{\mathrm{G}}$ and $g^{\mathrm{C}}$ are given by $\sum_{i_\ell=0}^{d_\ell}\mathfrak{h}_{i_\ell}^lx^{i_\ell}$ and $\sum_{i_u=0}^{d_u}\mathfrak{h}_{i_u}^ux^{i_u}$ respectively.} Then, given the coefficients of {these} polynomials, {a PUE of the filter given in Eq.~(\ref{our QTSP filter})} is constructed by combining {Lemma~\ref{LCU of many polynomials} with the PUEs of $\bkn$ and $\bkkdagn$.} The construction of a quantum simplicial filter is given in the following lemma. Recall that $\textsc{Input}_c$, as defined in Section~\ref{sec:main}, is the set of classical inputs used in the quantum computation: the signal $\brm{s}^k$ as a vector to be encoded in a quantum state, as well as the filter coefficients $\{\mathfrak{h}_0, \mathfrak{h}_{i_\ell}^\ell, \mathfrak{h}_{i_u}^u \}$ that define $H$, which are used to (classically) calculate the phase angles used in QSVT. Recall that $a_p$ is the number of ancilla qubits used in the simplex identifying oracle $\brm{P}_k$.

\begin{lemma}[PUE of quantum simplicial filter]\label{QTSP Chebyshev filter design}
Given $\textsc{Input}_{c}$, there exists a quantum circuit implementing a $(\beta,a_{k}+a_{k+1}+a_{p}+6,0)$-PUE of {the} quantum simplicial filter $H_d\left(\hlln,\hlun\right)${, where $\beta:=2+\mathfrak{h}_{0}$ and $a_{p}$ is the ancilla cost in the membership function. This quantum circuit uses} $4d_{\ell}$ calls {each} to $\brm{U}_{\bk}$ {and} $\brm{U}_{\bk}^{\dagger}$, $4d_{u}$ calls {each} to $\brm{U}_{\bkk}$ {and} $\brm{U}_{\bkk}^{\dagger}$, $8d_{\ell}$ calls {each} to $\mathrm{C}_{\Pi'_{k-1}}\mathrm{NOT}$ {and} $\mathrm{C}_{\Pi_{k}}\mathrm{NOT}$, $8d_{u}$ calls {each} to $\mathrm{C}_{\Pi'_{k}}\mathrm{NOT}$ {and} $\mathrm{C}_{\Pi_{k+1}}\mathrm{NOT}${,
and $O(\max\{d_{\ell},d_{u}\})$ calls to single qubit rotation gates.}
\end{lemma}

The query complexity of the PUE of the quantum simplicial filter, and thus the proof of Theorem~\ref{QTSP filtering}, follows immediately from Lemma~\ref{QTSP Chebyshev filter design} and is given below for completeness {(with the direct encoding details given in Appendix~\ref{sec:boundaryoperatorconstruction}).}
\begin{proof}[Proof of Theorem~\ref{QTSP filtering}]
    We assume we are given the simplicial state $\lvert\brm{s}^k\rangle$ from the \textsc{Encode} step.
    Given filter coefficients $\{\mathfrak{h}_0,\mathfrak{h}_{i_{\ell}}^{\ell},\mathfrak{h}_{i_{u}}^{u}\}$ for $i_{\ell}\in[d_{\ell}+1]$ and $i_{u}\in[d_{u}+1]$ as part of $\textsc{Input}_{c}$, Lemma~\ref{QTSP Chebyshev filter design} states that there exists a quantum circuit that implements a PUE of $H_d(\hlln, \hlun)$. This lemma also gives the query complexity and ancilla costs in terms of $d_\ell, d_u$ calls to the PUEs of $\bkn$ and $\bkkdagn$, and the costs of implementing $\brm{U}_{\bk}$, $\brm{U}_{\bkk}$, and  $\mathrm{C}_{\Pi_k}\mathrm{NOT}$. The \textsc{Filter} step as described in Section~\ref{sec:main} simply consists of applying the PUE of $H_d\left(\hlln,\hlun\right)$, to the state $\lvert\brm{s}_{k}\rangle$.     
    Thus, the complexity cost of  implementing the \textsc{Filter} step is given as follows: 
    \paragraph{Direct approach.} The unitary operators $\brm{U}_{\bk}$ and $\brm{U}_{\bkk}$ each have non-Clifford gate depth $O(\log n)$. {Each} $\mathrm{C}_{\Pi_k}\mathrm{NOT}$ {or $\mathrm{C}_{\Pi'_k}\mathrm{NOT}$} requires one call to a $Z$-rotation and two calls to the $k$-simplex-identifying function $\brm{P}_{k}$, which gives a non-Clifford gate depth $O(n\log n)$~\cite{Kerenidis2022QuantumStates,McArdle2022AQubits}. Therefore, the total non-Clifford gate depth of the quantum simplicial filter is $O(dn\log n)$, where $d=\mathrm{max}(d_{\ell},d_{u})$. The number of ancilla qubits used in this circuit is $O(n)$, dominated by the application of $\brm{P}_{k}$.
    \paragraph{Compact approach.} In this approach, both the operators $\brm{U}_{\bk}$ and $\brm{U}_{\bkk}$ have non-Clifford gate depth $O(k\log\log(n))$. {Each} $\mathrm{C}_{\Pi_k}\mathrm{NOT}$ {or $\mathrm{C}_{\Pi'_k}\mathrm{NOT}$} requires one call to a $Z$-rotation and two calls to the $k$-simplex-identifying function $\brm{P}_{k}$. The implementation of $\brm{P}_k$ given in Appendix~\ref{sec:proj} has non-Clifford gate depth $O(n^{2}k\log n)$. Therefore, the total non-Clifford depth of the quantum simplicial filter is $O(dkn^2\log n\log\log n)$. Lastly, the number of ancilla qubits used in this circuit is $O(k)$, again dominated by implementing the simplex-identifying circuit.
    \paragraph{Postselection cost.} Additionally, because, in the QTSP filtering, the filtered state is encoded in the subspace of the larger Hilbert space, we require a postselection step to ensure the output state is in the right subspace. The postselection cost depends on the success probability of outputting the filtered state, which is upper-bounded by the squared of the norm of the filtered signal $H_{d}\big(\bkn,\bkkdagn\big)\ket{\brm{s}^{k}}$. Thus, the total query complexity increases by a factor of $O(\mathcal{N}^2)$ to output the correct state.
\end{proof}


\begin{figure*}[t]
    \centering
    \includegraphics[scale=0.5]{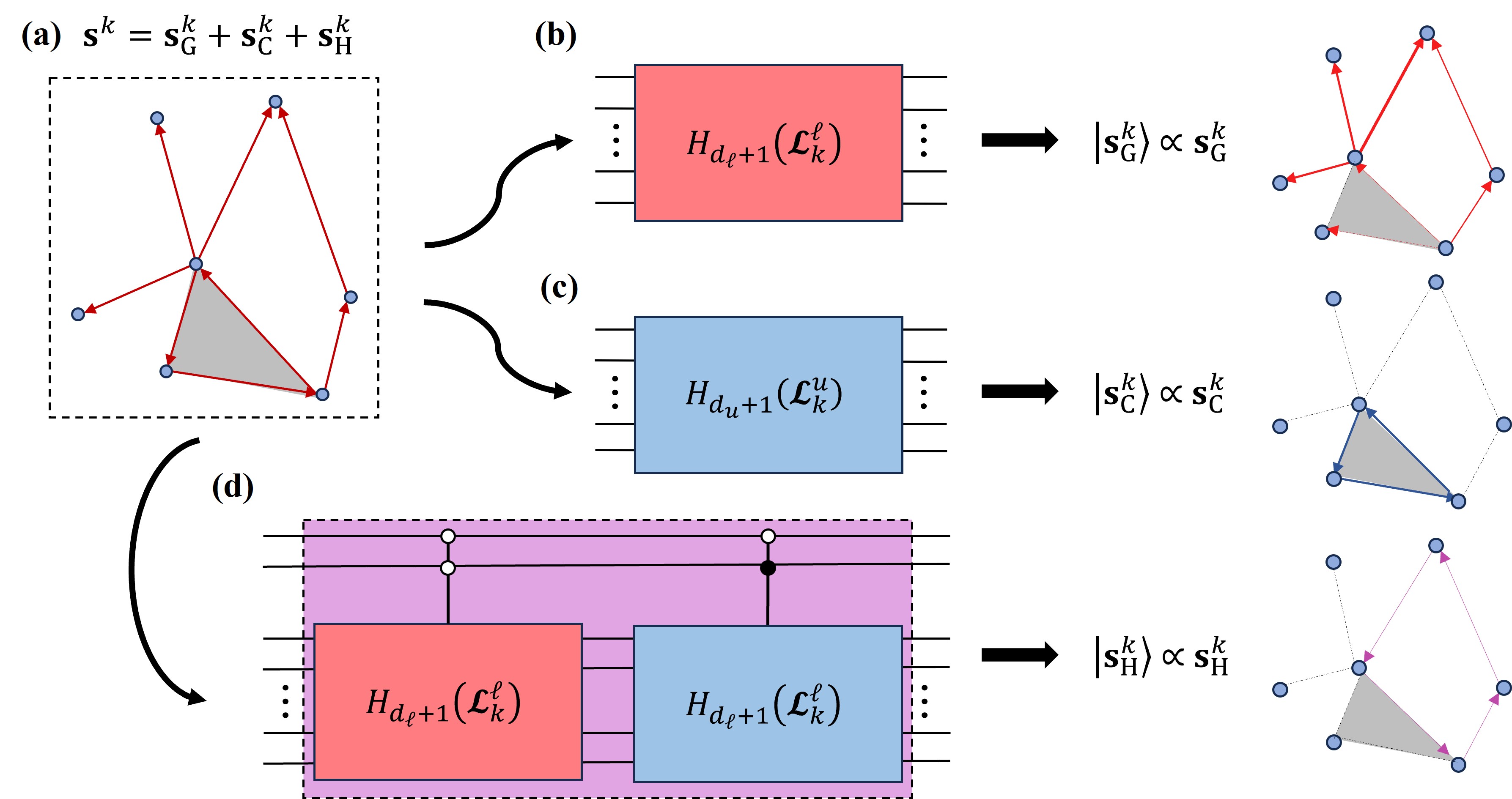}
    \caption{QTSP filtering for subcomponents extraction. \textbf{(a)} The input simplicial signal state $\lvert\brm{s}^{k}\rangle$ decomposed into three orthogonal signal states based on Hodge decomposition. \textbf{(b)} A quantum simplicial filter projecting $\lvert\brm{s}^{k}\rangle$ into the gradient part $\lvert\brm{s}^{k}_{\mathrm{G}}\rangle$. \textbf{(c)} A quantum simplicial filter projecting $\lvert\brm{s}^{k}\rangle$ into the curl part $\lvert\brm{s}^{k}_{\mathrm{C}}\rangle$. \textbf{(d)} A quantum simplicial filter combining gradient and curl projectors employed to project $\lvert\brm{s}^{k}\rangle$ into the harmonic part $\lvert\brm{s}^{k}_{\mathrm{H}}\rangle$.}
    \label{fig:QTSP Hodge decomposition filtering}
\end{figure*}

\section{Application: Filtering for Hodge decomposition}\label{sec:QTSP Hodge filtering}

Here we prove Theorem~\ref{QTSP Hodge filtering} about signal subcomponent extraction based on the Hodge decomposition. Given a signal $\brm{s}^k \in \mathbb{R}^{n_k}$, the Hodge decomposition states that $\brm{s}^k = \brm{s}^k_\mathrm{H} + \brm{s}^k_\mathrm{G} + \brm{s}^k_\mathrm{C}$, where $\brm{s}^k_\mathrm{H} \in \ker(\hl)$, $\brm{s}^k_\mathrm{G} \in \mathrm{im}(\bk^\dagger)$, and $\mathrm{s}^k_\mathrm{C} \in \mathrm{im}(\bkk)$. We design quantum simplicial filters that {approximately} extract each of these {(normalized)} components. We do this by employing Theorem~\ref{QTSP filtering} and applying a filter that approximates a projection onto {each} of the subspaces defined by the Hodge decomposition. 
We do this by modifying a construction for the Moore-Penrose pseudoinverse of a projected unitary encoding, in our case of the upper or lower Laplacians respectively. Constructions for implementing the Moore-Penrose pseudoinverse of a matrix $\brm{A}$, which is denoted by $\brm{A}^+$, via QSVT are given by Refs.~\cite{Gilyen2019QuantumArithmetics,Hayakawa2022QuantumAnalysis}.

The main idea of extracting subcomponents based on {the} Hodge decomposition is to project $\brm{s}^{k}$ to the gradient and curl subspaces, $\mathrm{im}(\bk^\dagger)$ and $\mathrm{im}(\bkk)$. {Mathematically speaking, this} is done by implementing projection operators $\Pi^{\mathrm{G}}$ and $\Pi^{\mathrm{C}}$ with the action $\Pi^{\mathrm{G}}\brm{s}^{k}=\brm{s}_{\mathrm{G}}^{k}$ and $\Pi^{\mathrm{C}}\brm{s}^{k}=\brm{s}_{\mathrm{C}}^{k}$. It has been shown \cite{Jiang2011StatisticalTheory,Barbarossa2020TopologicalComplexes,Lim2020HodgeGraphs} that these projectors are equivalent to
\begin{align}\label{Hodge decomposition projectors}
    &\Pi^{\mathrm{G}}=\bkdag\left(\brm{L}_{k-1}^{u}\right)^{+}\bk,\;\mbox{and}\nonumber\\ &\Pi^{\mathrm{C}}=\bkk\left(\brm{L}_{k+1}^{\ell}\right)^{+}\bkkdag,
\end{align}
where $\brm{L}_{k-1}^{u}=\bk\bkdag$ and $\brm{L}_{k+1}^{\ell}=\bkkdag\bkk$. This can be seen by recalling that $\bk\bkk = \brm{0}$ for all $k \geq 0$ and {the fact} that the pseudoinverse finds the (Euclidean) norm-minimising solution. {We note here again that we can also construct the projector onto the harmonic space by $\Pi^{\mathrm{H}} = \brm{I} - \Pi^{\mathrm{G}} - \Pi^{\mathrm{C}}$.}

One way to implement the Moore-Penrose pseudoinverse of a matrix via QSVT is to implement a polynomial $f:[-1,1]\to[-1,1]$ which closely approximates $1/x$ on a domain that contains the singular values of the matrix in question. Such a polynomial was given by Gily\'en, Su, Low, and Wiebe~\cite{Gilyen2019QuantumArithmetics} (Theorem 41), which is a modification of a polynomial given by Childs, Kothari, and Somma~\cite{Childs2017QuantumPrecision}.
To construct the projection operators given in Eq.~(\ref{Hodge decomposition projectors}), we modify this polynomial slightly and then apply it to the PUE of the boundary operator using QSVT. Notably, if {$g_d$} is the {degree-$d$} polynomial approximation to $1/x$, then we implement the polynomial ${H_{d+1}}:[-1, 1] \to [-1,1]$ defined by ${H_{d+1}(x^2) = x^2 g_d(x^2)}$. It follows (shown in detail in Appendix~\ref{Proof of Lemma Hodge decomposition filter}) that {for sufficiently large $d$,}
\begin{align}
    {H}_{d+1}(\hll) = \bk^\dagger\: {g_d}\left(\brm{L}_{k-1}^{u} \right) \bk \approx \Pi^{\mathrm{G}}.
\end{align}
{Similarly, for sufficiently large $d$ it follows that ${H_{d+1}}(\hlu) \approx \Pi^{\mathrm{C}}$.} Consequently, the projection operators are also simply polynomials of $\bk$ and $\bkkdag$: if ${g_d}$ is a degree-$d$ polynomial, then ${H_{d+1}}$ is a degree-$(2d+2)$ polynomial of $\bk$ or $\bkkdag$ (equivalently, a degree-$(d+1)$ polynomial of $\hll$ or $\hlu$).
In our QTSP filtering framework, we can construct such a polynomial using Lemma~\ref{QTSP Chebyshev filter design} and calculate the complexity {using} Theorem~\ref{QTSP filtering}. As in Theorem~\ref{QTSP Hodge filtering}, choose $\kappa_{\ell}$ such that $\kappa_\ell^{-1} \in (0, \xi^\ell_{\min}/\alpha_k)$ and $\kappa_{u}$ such that $\kappa_u^{-1} \in (0, \xi^u_{\min}/\alpha_{k+1})$, where $\xi^\ell_{\min}$ and $\xi^u_{\min}$ are the minimum singular values of $\brm{B}_k$ and $\brm{B}_{k+1}$ respectively. To account for the approximation error and the rescaling inherent in the PUE of the projector,
we state the existence of polynomials approximating the gradient and curl projections in the QTSP filtering framework as follows.

\begin{lemma}[Polynomial Approximation of Subcomponent Projection]\label{Hodge decomposition filter}
    Let $\Pi^{\mathrm{G}},\Pi^{\mathrm{C}}$ be projection operators to the $\mathrm{im}\Bigl(\bkdag\Bigr)$ and $\mathrm{im}\Bigl(\bkk\Bigr)$ subspaces respectively. Then {given $\varepsilon_{\ell},\varepsilon_{u}\in \left( 0, \frac{1}{2}\right)$, there exist} $d_{\ell}\in O(\kappa_{\ell}^{2}\log((n\kappa_{\ell})^{2}/(\alpha_{k}^2\varepsilon_{\ell})))$ and $d_{u}\in O(\kappa_{u}^{2}\log((n\kappa_{u})^{2}/(\alpha_{k+1}^2\varepsilon_{u})))$ {such that there} exist degree-$2d_{\ell}+2$ and $2d_{u}+2$ polynomials of $\bkn$ and $\bkkdagn$, denoted by $H_{{d_\ell}+1}\left(\hlln\right)$ and $H_{{d_u}+1}\left(\hlun\right)$, satisfying
    \begin{align*}
        &\norm{\Pi^{\mathrm{G}}}{2\kappa_{\ell}^2\,H_{{d_\ell}+1}\left(\hlln\right)}\leq\varepsilon_{\ell}, \;\mbox{and}\\ &\norm{\Pi^{\mathrm{C}}}{2\kappa_{u}^2\,H_{{d_u}+1}\left(\hlun\right)}\leq\varepsilon_{u}{.}
    \end{align*}
\end{lemma}
\noindent The proof of the lemma is given in Appendix~\ref{Proof of Lemma Hodge decomposition filter}{. The proof is an application and straightforward modification of the results by Hayakawa~\cite{Hayakawa2022QuantumAnalysis} and Gily\'en, Su, Low, and Wiebe~\cite{Gilyen2019QuantumArithmetics} about using QSVT to implement the Moore-Penrose pseudoinverse}.
Theorem~\ref{QTSP Hodge filtering} {then} follows by combining
{Lemma~\ref{Hodge decomposition filter} {and} Theorem~\ref{QTSP filtering} to calculate the cost of implementing the simplicial filter described in the lemma using our framework. The proof of Theorem~\ref{QTSP Hodge filtering} is given below}.
\begin{proof}[Proof of Theorem~\ref{QTSP Hodge filtering}]
   Given the polynomial coefficients of $H_{{d_\ell}+1}\left(\hlln\right)$ and $H_{{d_u}+1}\left(\hlun\right)$ from Lemma~\ref{Hodge decomposition filter} ({given in} Appendix~\ref{Proof of Lemma Hodge decomposition filter}), Lemma~\ref{QTSP Chebyshev filter design} {guarantees} that there exist sequences of phase factors $\Phi_{\mathrm{G}}\in\mathbb{R}^{2d_{\ell}+2}$ and $\Phi_{\mathrm{C}}\in\mathbb{R}^{2d_{u}+2}$, and the corresponding phased alternating sequences $\brm{U}_{\bk}^{\Phi_{\mathrm{G}}}$ and $\brm{U}_{\bkkdag}^{\Phi_{\mathrm{C}}}$ {that construct} a $\left(2\kappa_{\ell}^{2},a_{\ell}+1,\varepsilon_{\ell}\right)$-PUE of $\Pi^{\mathrm{G}}$ and a $\left(2\kappa_{u}^{2},a_{u}+1,\varepsilon_{u}\right)$-PUE of $\Pi^{\mathrm{C}}$. {Then the complexity of each quantum simplicial filter follows from Theorem~\ref{QTSP filtering} and the bounds on $d_\ell$ and $d_u$ given in Lemma~\ref{Hodge decomposition filter}.}
\end{proof}

It is worth noting that in certain applications of the Hodge decomposition (for example, statistical ranking~\cite{Jiang2011StatisticalTheory}) recovery of the projected vector is not the last step in the process. In the case of statistical ranking, the desired vector is actually a vector $\brm{s}^{k-1}$ such that $\bk^{T}\brm{s}^{k-1} = \brm{s}^k_{\mathrm{G}}$, whose existence is guaranteed by the Hodge decomposition. By implementing slightly different polynomials (specifically {$H'_{d+1}(x) = x g_d(x^2)$} instead of {$H_{d+1}(x^2) = x^2 g_d(x^2)$}, see {Remark~\ref{rem:alt-projector} in} Appendix~\ref{Proof of Lemma Hodge decomposition filter} for details), we can implement a PUE of an operator that outputs $\brm{s}^{k-1}$ instead of $\brm{s}^k_{\mathrm{G}}$. This allows one to skip the step of solving for $\brm{s}^{k-1}$ using $\bk$ and $\brm{s}^k_{\mathrm{G}}$, which could be expensive when $n_k$, the number of $k$-simplices, is very large.  This application is discussed in the context of statistical ranking in~\cite{leditto2024quantumhodgeranktopologybasedrank}.

\section{Discussion}

This work presents a quantum algorithm implementing TSP filtering processes in higher-order networks modelled by simplicial complexes {using QSVT and tools from QTDA}. The {type of filter we implement, described in Eq.~(\ref{eqn:quantumfilter})} is inspired by the Chebyshev filter design given in~\cite{Yang2022SimplicialFilters}. {W}hen the number of simplicial signals scales exponentially in the number of vertices, our algorithm runs linearly in the dimension $k$ of {the} simplices considered. In comparison, the {worst-case} complexity of the {best-known algorithm for general} classical TSP filtering tasks is exponential in the dimension. 

Nevertheless, {we do not make any rigorous claims of a quantum speedup, as} this potential advantage comes with some caveats that must be {first} accounted for. Firstly, a filtering task must be chosen so that the encoding and retrieving of the simplicial signals encoded in quantum states can be done efficiently. This includes post-selecting on the ancilla register being in the correct state, which is a complexity that is not accounted for in this work. Secondly, we also require that the task is, in some rough sense, insensitive to the scaling factor of $\|\brm{s}^k\|^{-1}$ in the signals and the {factors of $\alpha_{k}$ and $\alpha_{k+1}$} introduced in the {filter}. Applications to statistical ranking which satisfy some of these conditions, and some which avoid explicit post-selection, are explored more in ~\cite{leditto2024quantumhodgeranktopologybasedrank}.

A natural future direction for research would be a rigorous comparison between the classical and quantum algorithms, accounting for the aforementioned caveats in full to determine the nature of any advantage. Approximate filtering algorithms do not seem to be well studied in the classical literature, but would be a more apt comparison point than the exact algorithms given in Refs.~\cite{Yang2021FiniteComplexes,Yang2022SimplicialFilters} that we have used as a reference point. For example, Black and Nayyeri~\cite{Black2022HodgeComplexes} gave a method for solving the equation $\brm{L}_1 \brm{x} = \brm{b}$ which runs nearly linearly in $n$ and polynomially in $\beta_1$, the first Betti number. A deeper exploration of classical approximation algorithms could be combined with looking at other methods for building quantum simplicial filters, such as multivariable QSP as described by Rossi and Chuang~\cite{Rossi2022MultivariableOracle}, in order to determine the optimal way to build quantum simplicial filters.

As TSP has potential applications in analyzing sensor data, another future direction would be to find quantum sensor data~\cite{Proctor2018QuantumSensorNetworks,Zhang2021DistributedQuantumSensing} with the form of a simplicial signal state (e.g., signal aggregating some signals from networks~\cite{Rubio2020QuantumSensingLinearFunction,Qian2021FieldQuantumSensorNetworks,Bringewatt2021QuantumSensingMultipleFunctions}) and to use the QTSP framework for analyzing this data. Such an application would sidestep the difficulties inherent in the \textsc{Encode} step. In terms of more general applications, TSP on classical computers has been mostly studied for low dimensional cases, i.e. $k \leq 2$. As such, {a broader} avenue for future research would be to explore the possible applications of TSP to higher-order networks when $k > 2$. This is a large, open-ended question of independent interest, but it would also help inform the aforementioned analysis of the possibility of a practical speedup using the QSVT framework for classical data.

{Finally, one application of classical TSP that has {recently} received some attention is in the area of neural networks, where simplicial filters are used to construct simplicial neural networks, which are generalisations of graph neural networks~\cite{Ebli2020SimplicialNetworks}. The idea is to add a nonlinear function $w$ to alter the filtering process such that {the} input-output relation{ship is}
\begin{eqnarray}
    \brm{s}_{\mathrm{out}}^k=w\left(H(\hll,\hlu)\brm{s}_{\mathrm{in}}^k\right),
\end{eqnarray}
where $\brm{s}_{\mathrm{in}}^k,\brm{s}_{\mathrm{out}}^k$ are the input and output simplicial signals. Then, one optimizes the filter coefficients to approximate the desired output simplicial signal. It is suggested that this nonlinearity introduced in the neural network approach allows for more expressive modelling of higher-order interactions~\cite{Ebli2020SimplicialNetworks,roddenberry21a,Keros2022Dist2Cycle:Localization}. It would be interesting to see if simplicial neural networks could be implemented more efficiently on fault-tolerant quantum computers by integrating QSVT and nonlinear transformation of amplitudes from Refs.~\cite{guo2021nonlinear,rattew2023nonlinear}.}

\section*{Acknowledgements}
The authors thank Yuval R. Sanders for several technical discussions. C.M.G.L. is supported by the Monash Graduate Scholarship and CSIRO Top-Up Scholarship. 

\bibliographystyle{unsrt}
\bibliography{Main_QTSP}

\onecolumngrid
\newpage
\appendix
\section{Projected unitary encodings of arbitrary matrices}\label{PUE of arbitrary matrices}

Here we describe general methods {from the literature} used to implement non-unitary operators in quantum circuits. We give a brief overview of the linear combination of unitaries, block encodings, and projected unitary encodings for unfamiliar readers, as we use these to build quantum circuits that implement {polynomials of the upper and lower Laplacians $\hl^\ell$ and $\hl^u$. This is a brief summary of ideas from Refs.~\cite{Berry2015HamiltonianParameters,Berry2015SimulatingSeries,Gilyen2019QuantumArithmetics,Martyn2021GrandAlgorithms}, and we refer interested readers to these papers for more details.}

If the desired operator is Hermitian, a common approach to implementing it in a quantum circuit is to do a Hamiltonian simulation~\cite{Harrow2009QuantumEquations}, {where a Hermitian matrix $\brm{H}$ is encoded} in a unitary matrix $\brm{U}_{\brm{H}}=\mathrm{e}^{\imath \brm{H}t}$. The linear combination of unitaries (LCU) method~\cite{Berry2015HamiltonianParameters,Berry2015SimulatingSeries} is one of the avenues for implementing Hamiltonian simulation which has had wide applicability in implementing many different operators in quantum circuits. In this paper, we use this to combine polynomials of the lower and upper Hodge Laplacians to create our desired quantum simplicial filter.

The description of the LCU technique, as given in Refs.~\cite{Berry2015HamiltonianParameters,Berry2015SimulatingSeries} is as follows. Let $\brm{A}=\sum_{i=0}^{m-1} c_i\brm{A}_i$ be a linear combination of a set of unitaries $\brm{A}_{i}$ with a set of coefficients $c_{i}\in\mathbb{R}_{+}$. Let $(\brm{G}\otimes\brm{I}^{s})$  be a so-called $\textsc{Prepare}$ operation such that 
\begin{eqnarray}\label{prepare operation}
    \brm{G}\lvert 0\rangle^{\otimes\log_2 n}= \frac{1}{\sqrt{\lVert c\rVert_{1}}}\sum_{i=1}^{n}\sqrt{c_i}\lvert i\rangle=:\lvert \brm{G}\rangle,
\end{eqnarray}
where $\lVert c\rVert_{1}:=\sum_{i=1}^n |c_i|$. {The $\textsc{Select}$ operation is then defined as}
\begin{eqnarray}\label{select operation}
    \brm{W}&:=&\sum_{i=1}^{n}\;\lvert i\rangle\langle i\rvert\otimes \brm{A}_i.
\end{eqnarray}
Then the transformation $\lvert\psi\rangle\longmapsto\frac{\brm{A}\lvert\psi\rangle}{\lVert \brm{A}\lvert\psi\rangle\rVert}$ can be done by applying $\textsc{Prepare}^\dagger\cdot\textsc{Select}\cdot\textsc{Prepare}$ to $\lvert 0\rangle^{\otimes\log_2 n}\otimes\lvert\psi\rangle$ {and} post-selecting the ancilla qubit register {on the state $\lvert 00\dots0\rangle$}. The success probability of post-selection is $\frac{\lVert \brm{A}\lvert\psi\rangle\rVert^2}{\lvert c\rvert^2}$.

An improvement of the LCU technique in terms of precision and complexity is given by the block-encoding (BE) technique. This technique {was} first developed within the framework of qubitization~\cite{Low2019HamiltonianQubitization}, which directly encodes a Hermitian matrix $\brm{H}$ into the unitary operator $\brm{U}_{\brm{H}}$ as
\begin{eqnarray*}
    \brm{U}_{\brm{H}}=\begin{bmatrix}
           \brm{H}/\alpha&\ast\\
           \ast&\ast
    \end{bmatrix},
\end{eqnarray*}
{where the $*$ denotes a block of the matrix $\brm{U}_{H}$ with arbitrary entries that ensure that $\brm{U}_{H}$ is unitary.}
We follow the presentation of block-encoding given {in} Ref.~\cite{Gilyen2019QuantumArithmetics}.
\begin{definition}[Block-Encoding (BE)]\label{block encoding}
Suppose $\brm{H}$ is a Hermitian matrix. A $(\alpha,a,\varepsilon)$-block encoding of a Hermitian matrix $\brm{H}$ is a {unitary matrix} $\brm{U}_{\brm{H}}$ such that
\begin{eqnarray}
    \Bigl\lVert \brm{H}-\alpha\biggl(\langle 0\rvert^{\otimes a}\otimes\brm{I}^{s}\biggr)\brm{U}_{\brm{H}}\biggl(\lvert 0\rangle^{\otimes a}\otimes\brm{I}^{s}\biggr)\Bigr\rVert\leq\varepsilon,
\end{eqnarray}
where $\alpha\in\mathbb{R}_{+}$ is the scaling factor, $a\in\mathbb{N}$ is the number of ancilla qubits, and $\varepsilon\in\mathbb{R}_{+}$ is an arbitrary small number representing the error (in operator norm). 
\end{definition}
\noindent As is {done} in the above definition, {the notation} $\brm{I}^{x}$ {is used} to denote the identity operator acting on ancilla qubit register which has $x$ qubits, and $\brm{I}^{s}$ is the identity operator acting on the system qubit register. 

The idea of a block encoding of a Hermitian matrix can be generalised to any arbitrary matrix $\brm{A}$ using what is called a projected unitary encoding of $\brm{A}$. Let $\brm{A}\in\mathbb{C}^{m'\times m}$ be a $(m'\times m)$-matrix with complex entries. Then the \textit{singular value decomposition} (SVD) of $\brm{A}$ is given by $\brm{A}=\brm{V}_{L}\brm{\Sigma} \brm{V}_{R}^\dagger$, where $\brm{\Sigma}\in\mathbb{R}^{m'\times m}$ is a diagonal matrix with {entries} $\brm{\Sigma}_{ii}:=\xi_{i}\geq 0${} for all $i\in[\mathrm{min}\{m',m\}]$, which are called 
the \textit{singular values} of $\brm{A}$, and $\left\{\brm{V}_{L}\in\mathbb{C}^{m'\times m'},\brm{V}_{R}\in\mathbb{C}^{m\times m}\right\}$ are a pair of unitary matrices in which each column of $\brm{V}_{L}$ is called a left singular vector $\lvert v_{l}\rangle$ and each row of $\brm{V}_{R}$ is called a right singular vector $\langle v_{r}\rvert$.
\begin{definition}[Projected Unitary Encoding (PUE)]\label{projected unitary encoding}
Suppose $\brm{A}\in\mathbb{C}^{m_{l}\times m_{r}}$ and the SVD of $\brm{A}$ is written as $\brm{A}:=\sum_{i=1}^{\mathrm{min}\{m_l,m_r\}}\xi_i \lvert v_l^i\rangle\langle v_r^i\rvert$. Let $\Pi':=\sum_{i=1}^{m_l}\lvert v_l^i\rangle\langle v_l^i\rvert$ and  $\Pi:=\sum_{i=1}^{m_r}\lvert v_r^i\rangle\langle v_r^i\rvert$ be orthogonal projectors projecting all vectors into the space spanned by {the} left and right singular vectors respectively. A $(\alpha,a,\varepsilon)$-projected unitary encoding of matrix $\brm{A}$ is {a unitary matrix}
\begin{eqnarray*}
    \brm{U_A}=\begin{bNiceMatrix}[first-row,first-col]
            & \Pi&\\
            \Pi'&\brm{A}/\alpha&\ast\\
            &\ast&\ast
    \end{bNiceMatrix},
\end{eqnarray*}
{where $\Pi, \Pi'$ locate $\brm{A}$ within $\brm{U}_{\brm{A}}$} such that
\begin{eqnarray*}
    \Bigl\lVert \brm{A}-\alpha\Bigl(\langle 0\rvert^{\otimes a}\otimes\brm{I}^{s}\Bigr)\Bigl(\Pi'\brm{U_A}\Pi\Bigr)\Bigl(\lvert 0\rangle^{\otimes a}\otimes\brm{I}^{s}\Bigr)\Bigr\rVert\leq\varepsilon,
\end{eqnarray*}
where $\alpha\in\mathbb{R}_{+}$ is the scaling factor, $a\in\mathbb{N}$ is the number of ancilla qubits being used, and $\varepsilon\in\mathbb{R}_{+} \cup \{0\}$ is an arbitrary small number representing the error. 
\end{definition}
\noindent Within this framework, a $(\alpha,a,\varepsilon)$-BE of $\brm{H}$ can be seen as a $(\alpha,a,\varepsilon)$-PUE of $\brm{H}$ with $\Pi'=\Pi=\lvert 0\rangle\langle 0\rvert^{\otimes a}\otimes\brm{I}^{s}$. 

\section{Quantum algorithm for constructing polynomials {(QSVT)}}\label{quantum algorithm for constructing polynomials}
{In this section we give a brief background to the quantum singular value transformation, as well as a statement of the result itself.}
This paper {uses the} notations defined in {Ref.}~\cite{Gilyen2019QuantumArithmetics}. {Let $\mathbb{C}[x]$ be the space of all complex-valued polynomial functions} such that for $f\in\mathbb{C}[x]$, $f(x)=\sum_{i=0}^{d-1}\,c_{i}x^{i}$ and $f^\ast(x)=\sum_{i=0}^{d-1}\,c_{i}^\ast x^{i}$, where $c_{i}\in\mathbb{C}$ for all $i\in[d]$ and $\cdot^\ast$ is the complex conjugation operation. Additionally, the real and imaginary parts of $f$ are denoted as $\Re(f)(x)=\sum_{i=0}^{d-1}\,\Re(c_{i})^\ast x^{i}$ and $\Im(f)(x)=\sum_{i=0}^{d-1}\,\Im(c_{i})^\ast x^{i}$, respectively. If $\brm{A}=\sum_{i=1}^{\mathrm{min}\{m',m\}}\xi_i \lvert v_l^i\rangle\langle v_r^i\rvert$ is the SVD of a matrix $\brm{A}$, then given $f\in\mathbb{C}[x]$, a degree-$d$ polynomial transformation of a 
matrix $\brm{A}$ is described as $f(\brm{A})=\sum_{i=0}^{\mathrm{min}\{m',m\}-1}f(\xi_{i})\lvert v_r^i\rangle\langle v_r^i\rvert$ when $d$ is even and $f(\brm{A})=\sum_{i=0}^{\mathrm{min}\{m',m\}-1}f(\xi_{i})\lvert v_l^i\rangle\langle v_r^i\rvert$ when $d$ is odd. 

The idea of polynomial transformation originated from the \textit{quantum signal processing} (QSP) theorem~\cite{Low2017OptimalProcessing,Low2019HamiltonianQubitization}, which builds a matrix consisting of special complex-valued polynomials using a sequence of reflection operators.

\begin{theorem}[Quantum Signal Processing (QSP) by Reflection, Corollary 8~\cite{Gilyen2019QuantumArithmetics}]\label{QSP}
Let $\brm{R}(x)$ be a parametrized family of single qubit reflections, where
\begin{eqnarray*}
    \brm{R}(x)&:=&\begin{bmatrix}
            x&\sqrt{1-x^2}\\
            \sqrt{1-x^2}&-x
    \end{bmatrix}
\end{eqnarray*}
for all $x\in[-1,1]$. There exists a sequence of phase factors $\bm{\Phi}:=\left(\phi_1,\cdots,\phi_d\right)\in\mathbb{R}^{d}$ such that
\begin{eqnarray*}
    \brm{U}^{\bm{\Phi}}_{x}&:=&\prod_{i=1}^d \biggl(\mathrm{e}^{\imath\phi_i\brm{Z}}\,\brm{R}(x)\biggr)\\
    &=&\begin{bmatrix}
            P(x)&\ast\\
            \ast &\ast
        \end{bmatrix},
\end{eqnarray*}
where $P\in\mathbb{C}^{d}[x]$ satisfies the following properties:
\begin{enumerate}
    \item[C1.] The parity of $P(x)$ is $d\bmod 2$;
    \item[C2.] $\lvert P(x)\rvert\leq 1$, for all $x\in[-1,1]$, whereas $\lvert P(x)\rvert\geq 1$ otherwise;
    \item[C3.] If $d$ is even, $P(\imath x)P^\ast(\imath x)\geq 1$, for all $x\in\mathbb{R}$.
\end{enumerate}
\end{theorem}
\noindent The relation between the QSP theorem and Jordan's lemma, stating that the product of two reflection operators is a rotation, can be found in {Ref.}~\cite{Gilyen2019QuantumArithmetics}. Many standard quantum computing subroutines such as mplitude amplification can be seen as a polynomial transformation of $x$~\cite{Martyn2021GrandAlgorithms}{.}

{The quantum singular value transformation} is a generalization of \textit{quantum eigenvalue transformation} (QET)~\cite{Low2019HamiltonianQubitization,Martyn2021GrandAlgorithms}, which {combines QSP with} \textit{qubitization} {to implement polynomials of} Hermitian matrices. 
\begin{theorem}[Quantum Eigenvalue Transformation, Theorem 3~\cite{Martyn2021GrandAlgorithms}]\label{QETU}
Let $\brm{U}_{\brm{H}}$ be a $(\alpha,a,\varepsilon)$-block encoding of a Hermitian matrix ${\brm{H}}$ as defined in Definition~\ref{block encoding}. Let $\mathrm{Poly}(x)$ be a degree-$d$ polynomial satisfying {C1--C3} in Theorem~\ref{QSP}. Then there exist{s} a {sequence} of phase factors $\bm{\Phi}:=\left(\phi_1,\cdots,\phi_d\right)\in\mathbb{R}^{d}$ such that 
\begin{eqnarray}
    \mathrm{Poly}({\brm{H}}/\alpha)=\Bigl(\langle 0\rvert^{\otimes a}\otimes \brm{I}\Bigr)\Bigl(\Pi \brm{U}_{\brm{H}}^{\bm{\Phi}}\Pi\Bigr) \Bigl(\lvert 0\rangle^{\otimes a}\otimes \brm{I}\Bigr)
\end{eqnarray}
where $\Pi:=\lvert 0\rangle\langle 0\rvert^{\otimes a} \otimes \brm{I}$, and
\begin{eqnarray*}
    \brm{U}_{\brm{H}}^{\bm{\Phi}}:=\begin{cases}
        \prod_{i=1}^{d/2} \biggl(\mathrm{e}^{\bm{\mathrm{i}}\phi_{2i-1}(2\Pi-\brm{I})}\brm{U}_{\brm{H}}^\dagger\,\mathrm{e}^{\bm{\mathrm{i}}\phi_{2i}(2\Pi-\brm{I})}\brm{U}_{\brm{H}}\biggr),& \mbox{for $d$  even},\\
        \mathrm{e}^{\bm{\mathrm{i}}\phi_{1}(2\Pi-\brm{I})}\brm{U}_{\brm{H}}\prod_{i=1}^{(d-1)/2} \biggl(\mathrm{e}^{\bm{\mathrm{i}}\phi_{2i}(2\Pi-\brm{I})}\brm{U}_{\brm{H}}^\dagger\,\mathrm{e}^{\bm{\mathrm{i}}\phi_{2i+1}(2\Pi-\brm{I})}\brm{U}_{\brm{H}}\biggr),& \mbox{for $d$ odd}.
    \end{cases}
\end{eqnarray*} 
\end{theorem}
\noindent The matrix representation of $\brm{U}_{\brm{H}}^{\bm{\Phi}}$ is described by
\begin{eqnarray}
    \brm{U}_{\brm{H}}^{\bm{\Phi}}=
    \begin{bmatrix}
       \mathrm{Poly}(\brm{H}/\alpha)&\ast\\
            \ast&\ast
    \end{bmatrix},
\end{eqnarray}
where {again} the $*$ denotes a block of the matrix $\brm{U}_{\brm{H}}^{\bm{\Phi}}$ with arbitrary entries that ensure that $\brm{U}_{\brm{H}}^{\bm{\Phi}}$ is unitary and $\mathrm{Poly}(\brm{H}/\alpha)$ that the function $\mathrm{Poly}$ is applied to the eigenvalues of $\brm{H}/\alpha$.

While the QET algorithm only implements a polynomial transformation of {Hermitian matrices}, the {quantum singular value transformation}~\cite{Gilyen2019QuantumArithmetics} constructs a polynomial transformation for any matrix (that is, a polynomial transformation of the singular values of the matrix). We state the QSVT theorem based on {the} presentation in {Ref.}~\cite{Martyn2021GrandAlgorithms}.

\begin{theorem}[Quantum Singular Value Transformation (QSVT), Theorem 4~\cite{Martyn2021GrandAlgorithms}]\label{QSVT}
Let $\brm{U_A}$ be a $(\alpha,a,\varepsilon)$-PUE (as in Defintion~\ref{projected unitary encoding}) of a matrix $\brm{A}$. Let $\mathrm{Poly}(x)$ be a degree-$d$ polynomial satisfying C1--C3 in Theorem~\ref{QSP}. Then there exists a sequence of phase factors $\bm{\Phi}:=\left(\phi_1,\cdots,\phi_d\right)\in\mathbb{R}^{d}$ such that 
\begin{eqnarray}
    \mathrm{Poly}^{(\mathrm{SV})}(\brm{A}/\alpha)=\begin{cases}
        \Bigl(\langle 0\rvert^{\otimes a}\otimes \brm{I}^{s}\Bigr)\Bigl(\Pi \brm{U}^{\bm{\Phi}}_{\brm{A}}\Pi\Bigr) \Bigl(\lvert 0\rangle^{\otimes a}\otimes \brm{I}^{s}\Bigr),& \mbox{for $d$ even},\\
        \Bigl(\langle 0\rvert^{\otimes a}\otimes \brm{I}^{s}\Bigr)\Bigl(\Pi' \brm{U}^{\bm{\Phi}}_{\brm{A}}\Pi\Bigr) \Bigl(\lvert 0\rangle^{\otimes a}\otimes \brm{I}^s\Bigr),& \mbox{for $d$ odd},\\
    \end{cases}
\end{eqnarray}
where $\Pi'$, $\Pi$ are defined in Definition~\ref{projected unitary encoding}, and the phased alternating sequence {$\brm{U}^{\bm{\Phi}}_{\brm{A}}$ is given by}
\begin{eqnarray*}
    \brm{U}^{\bm{\Phi}}_{\brm{A}}:=\begin{cases}
        \prod_{i=1}^{d/2} \biggl(\mathrm{e}^{\imath\phi_{2i-1}(2\Pi-\brm{I}^{s})}\brm{U}_{\brm{A}}^\dagger\,\mathrm{e}^{\imath\phi_{2i}(2\Pi'-\brm{I}^{s})}\brm{U_A}\biggr),& \mbox{for $d$ even},\\
        \mathrm{e}^{\imath\phi_{1}(2\Pi'-\brm{I}^{s})}\brm{U_A}\prod_{i=1}^{(d-1)/2} \biggl(\mathrm{e}^{\imath\phi_{2i}(2\Pi-\brm{I}^{s})}\brm{U}_{\brm{A}}^\dagger\,\mathrm{e}^{\imath\phi_{2i+1}(2\Pi'-\brm{I}^{s})}\brm{U_A}\biggr),& \mbox{for $d$ odd}.
    \end{cases}
\end{eqnarray*} 
Moreover, $\brm{U}_{\brm{A}}^{\Phi}$ can be implemented by calling to $\brm{U}_{\brm{A}}$ and $\brm{U}_{\brm{A}}^{\dagger}$ $d$ times, and $\mathrm{C}_{\Pi'}\mathrm{NOT}$ and $\mathrm{C}_{\Pi}\mathrm{NOT}$ $2d$ times, where
\begin{eqnarray*}
    \mathrm{C}_{\Pi}\mathrm{NOT}&:=& \Pi\otimes\brm{X}+\left(\brm{I}-\Pi\right)\otimes\brm{I}.
\end{eqnarray*}
\end{theorem}
\noindent The matrix representation of $\brm{U}_{\brm{A}}^{\bm{\Phi}}$ is described by
\begin{eqnarray}
    \brm{U}_{\brm{A}}^{\bm{\Phi}}=
    \begin{bmatrix}
       \mathrm{Poly}(\brm{A}/\alpha)&\ast\\
            \ast&\ast
    \end{bmatrix}.
\end{eqnarray}

Although the above method only constructs complex-valued polynomials satisfying conditions C1--C3 in Theorem~\ref{QSP}, combining the LCU method and QSVT algorithm {allows one to implement a wider range of polynomials,} especially real-valued polynomials. Based on the fact that the real part of any complex-valued function {is given by} $\Re(f(x))=1/2(f(x)+f^\ast(x))$, one can find a real-valued polynomial $\mathrm{Poly}_{f_{\Re}}(\brm{A}):=\Re(\mathrm{Poly}_{f}(\brm{A}))$ from a linear combination of $\brm{U}_{\brm{A}}^{\bm{\Phi}_{f}}$ and $\brm{U}_{\brm{A}}^{-\bm{\Phi}_{f}}$.

\begin{corollary}[QSVT by Real-Valued Polynomials, Corollary 10 and 18~\cite{Gilyen2019QuantumArithmetics}]\label{QSVT real polynomial}
Let $f_{\Re}\in\mathbb{R}^{d}[x]$ be a (either even or odd parity) real-valued polynomial of degree $d$ satisfying $\lvert f_{\Re}(x)\rvert\leq1$ for all $x\in[-1,1]$. Let $\brm{U}_{\brm{A}}$,$\Pi'$,$\Pi$ be as in Theorem~\ref{QSVT}. Then there exists $\bm{\Phi}_{f}\in\mathbb{R}^{d}$ such that
\begin{eqnarray*}
   f_{\Re}(\brm{A})&=&\biggl(\langle +\rvert\otimes\Bigl(\langle 0\rvert^{\otimes a}\otimes \brm{I}^{s}\Bigr)\biggr)\brm{W}_{f_{\Re}}\biggl(\lvert + \rangle\otimes \Bigl(\lvert 0\rangle^{\otimes a}\otimes \brm{I}^{s}\Bigr)\biggr)
\end{eqnarray*}
where 
\begin{eqnarray*}
    \brm{W}_{f_{\Re}}&:=&\begin{cases}
        \lvert 0\rangle\langle 0\rvert\otimes \Pi \brm{U}_{\brm{A}}^{\bm{ \Phi}_{f}}\Pi+\lvert 1\rangle\langle 1\rvert\otimes \Pi \brm{U}_{\brm{A}}^{-\bm{\Phi}_{f}}\Pi,&\quad\mbox{for $d$ is even}\\
        \lvert 0\rangle\langle 0\rvert\otimes \Pi'\brm{U}_{\brm{A}}^{\bm{ \Phi}_{f}}\Pi+\lvert 1\rangle\langle 1\rvert\otimes \Pi'\brm{U}_{\brm{A}}^{-\bm{\Phi}_{f}}\Pi,&\quad\mbox{for $d$ is odd}.
        \end{cases}
\end{eqnarray*}
Moreover, given $f_{\Re}\in\mathbb{R}^{d}[x]$ and $\delta > 0$, using $O(\mathrm{poly}(d,\log(1/\delta))$ classical computation time one can find a sequence of phase factors $\bm{\Phi'}_{f}\in\mathbb{R}^{d}$ and a corresponding polynomial $\mathrm{Poly}_{f_{\Re}}$ such that
\begin{eqnarray*}
    \Bigl\lVert f_{\Re}(\brm{A})-\mathrm{Poly}_{f_{\Re}}(\brm{A})\Bigr\rVert\leq \delta.
\end{eqnarray*}
\end{corollary}
\noindent Note that one can also find an approximation of a purely imaginary-valued polynomial by observing that $\imath\Im(f(x))=1/2(f(x)-f^\ast(x))$. Therefore, one can use $W_{\imath f_{\Im}}:=(\brm{Z}\otimes \brm{I}^{a+s+1})W_{f_{\Re}}$ to construct $\mathrm{Poly}_{\imath f_{\Im}}(\brm{A}/\alpha)$.

One important application of this algorithm is implementing the Moore-Penrose pseudoinverse, a generalisation of the inverse matrix to arbitrary matrices. For an arbitrary matrix $\brm{A}$ with singular value decomposition $\brm{A} = \brm{W}\bm{\Sigma}\brm{V}^\dagger$, the Moore-Penrose pseudoinverse of $\brm{A}$ is given by $\brm{A}^+ := \brm{V} \bm{\Sigma}^+ \brm{W}^\dagger$, where $\bm{\Sigma}^+$ is the diagonal matrix obtained by inverting the nonzero entries of $\bm{\Sigma}$, which are the singular values of $\brm{A}$, and leaving the zero entries unchanged. QSVT can be used to implement unitaries approximating the pseudoinverse of a matrix $\brm{A}$ encoded in some unitary $\brm{U}_{\brm{A}}$ by approximating $1/x$ with polynomials. Such an approximation is given in Theorem 41 of Ref.~\cite{Gilyen2019QuantumArithmetics}, as well as Appendix C of Ref.~\cite{Martyn2021GrandAlgorithms}.
For $\kappa>1$, $\varepsilon\in(0,1/2)$, and $x\in[-1,1]\backslash[-1/\kappa,1/\kappa]$, they show that there exists a polynomial $g^+\in\mathbb{R}^d[x]$ with degree $d\in O\left(\kappa\log(\kappa/\varepsilon)\right)$ that $\frac{\varepsilon}{2\kappa}$-approximates $\frac{1}{2\kappa x}$ on the domain $[-1,1] \backslash \left[ -1/\kappa, 1/\kappa\right]$. Furthermore, $|g^+(x)|\leq 1$ for all $x \in [-1,1]$ and $g^+(0) = 0$. Thus, given $g^+$ with $\kappa:=\kappa(\brm{A})$, Corollary~\ref{QSVT real polynomial} guarantees that there exists a sequence of phase factors $\Phi_{+}\in\mathbb{R}^{d}$ and $\brm{U}_{\brm{A}}^{\Phi_{+}}$ such that $\left(\bra{0}^{\otimes a}\otimes\brm{I}^{s}\right)\brm{U}_{\brm{A}}^{\Phi_{+}}\left(\ket{0}^{\otimes a}\otimes\brm{I}^{s}\right)=g^+(\brm{A})$ satisfies $\left\|\brm{A}^{+}-2\kappa\, g^+(\brm{A})\right\|\leq\varepsilon$.

\section{A brief overview of QTDA}
\begin{figure*}[t]
    \centering
    \includegraphics[scale=0.55]{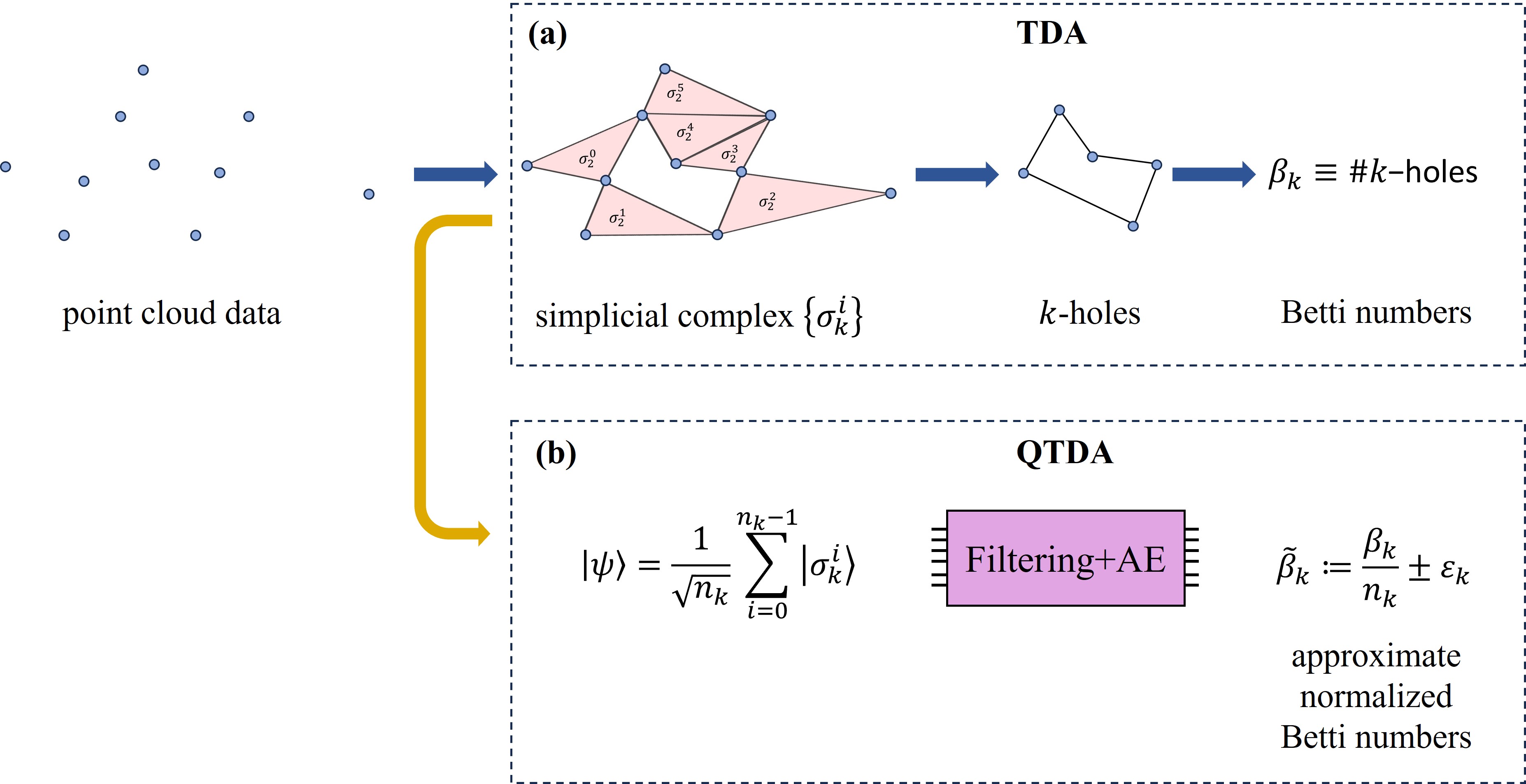}
    \caption{(a) A high-level picture of the TDA pipeline. The left-to-right flow of blue arrows describes taking some point cloud data, constructing a simplicial complex from this data, extracting the kernel of the corresponding Hodge Laplacian, and counting the dimension of the kernel to determine the Betti numbers; (b) a high-level picture of the QTDA pipeline utilizing eigenstate filtering and amplitude estimation (AE) algorithms to determine approximate, normalized Betti numbers.}
    \label{fig:QTDA}
\end{figure*}

TDA utilizes the topological invariants, extracted from the simplicial structure of point cloud data, as signatures of the dataset. The key subroutine of this method is an algorithm to find the dimension of the kernel of the Hodge Laplacian, known as the Betti number. One begins with mapping the data in a high-dimensional space and constructing a simplicial complex in such a space. The next step is then constructing the Hodge Laplacian of the simplicial complex and obtaining its kernel. The last step is computing the rank of the kernel. Figure~\ref{fig:QTDA}(a) shows the high-level picture of this subroutine. In recent QTDA algorithm given by McArdle, Gily{\'e}n, and Berta~\cite{McArdle2022AQubits}, given classical information of the underlying graph, they tackle such a problem by encoding the simplices as an initial quantum state, constructing the corresponding boundary operation in a quantum circuit, projecting the initial quantum state to the kernel of the Hodge Laplacian via a QSVT-based eigenstate filtering, and finding the dimension of the Hodge Laplacian by utilizing a QSVT-based amplitude estimation algorithm. As the quantum algorithms rescale the Betti number, one can only estimate the approximation of the normalized Betti numbers. We depict this pipeline in Figure~\ref{fig:QTDA}(b).

\section{Simplicial signal state representation}\label{state preparation}
The first step is to implement an \textsc{Encode} subroutine to encode the simplicial signals $\brm{s}^{k} \in \mathbb{R}^{n_k}$ as a quantum state $\lvert\brm{s}^{k}\rangle$, which we formally define in Eq.~\eqref{simplicialquantumtsignal}. Broadly speaking, there are two approaches for encoding simplex information in the QTDA literature: a ``direct'' encoding and a ``compact'' encoding. In this work, we slightly generalize these encodings {to also} encode the signals $\brm{s}^k$, rather than just the structure of the simplicial complex. This allows us to encode the signal information using no extra qubits, at the cost of making the signal state more difficult to prepare in general.

The encoding schemes we consider here represent simplices as basis elements and signals as their corresponding amplitudes. To this end, we first define the \textit{$k$-simplicial state $\lvert \sigma_{k}^{i}\rangle$}, which is a (non-specific) basis encoding for the $i$-th $k$-simplex $\sigma_{k}^{i}$. 
A simplicial signal can be represented as a collection of pairs $\{(s^k_i,\sigma_k^i)\}_{i\in[n_{k}]}$, where $s_{i}^{k}$ is the signal value on the simplex $\sigma_k^i$. Here $\lvert \brm{s}^k\rangle$ is a superposition of all $k$-simplicial states with amplitudes representing the (normalised) signal values $\{s_{i}^{k}\}$. Note that this definition is quite general, as $\ket{\sigma_k^i}$ is as-of-yet undefined.

\begin{figure}
    \centering
    \includegraphics[scale=0.37]{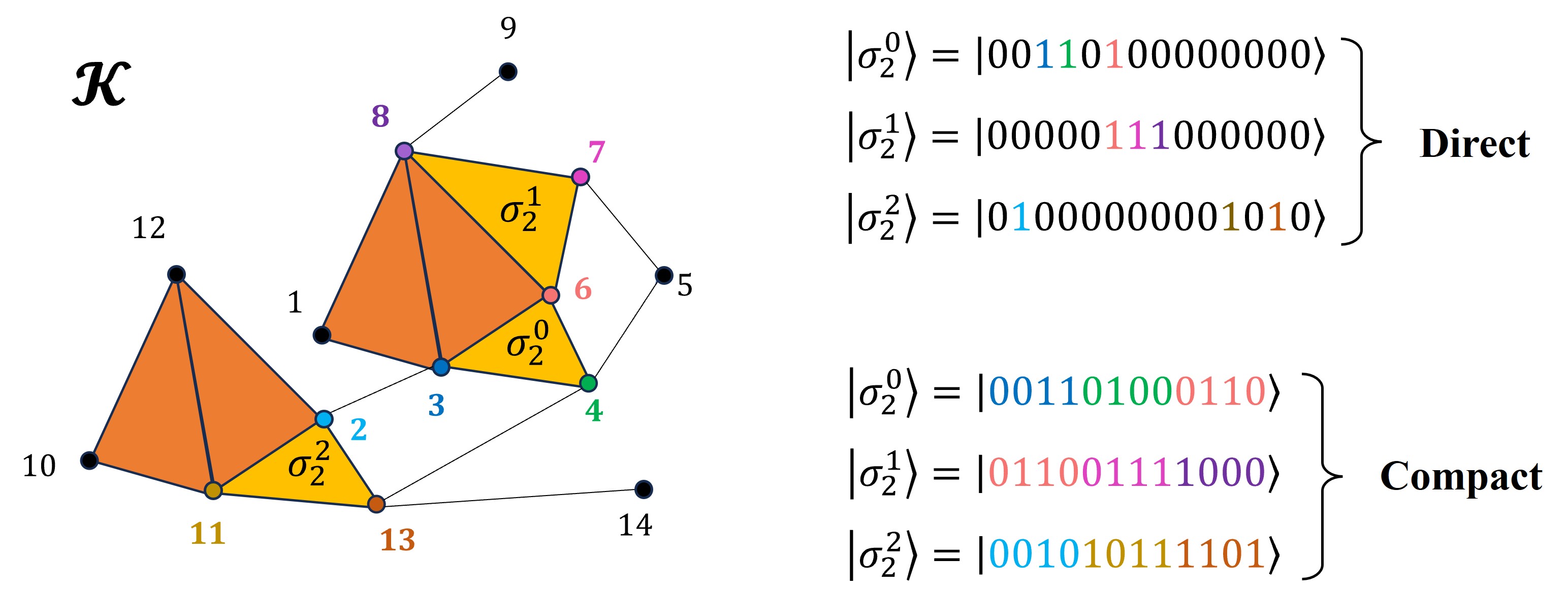}
    \caption{An illustrative example of the direct and compact encodings for three different $2$-simplices ($3$-cliques) in the simplicial complex $\mathcal{K}$, which is given on the left.}
    \label{fig:Direct vs Compact Encoding}
\end{figure}

As previously mentioned, there are two common approaches to representing a $k$-simplex $\sigma_{k}$ as a corresponding ($k$-)simplicial state $\lvert \sigma_{k}\rangle$. These are known as the {\it direct} and {\it compact} encodings and are given respectively in Refs.~\cite{berry2023analyzing} and~\cite{McArdle2022AQubits}. These encodings refer specifically to the qubit representations of simplices $\{\sigma_k^i\}$, which is the pertinent information to QTDA. However, the inclusion of amplitudes $\{s_i^k\}$ is unique to QTSP; in QTDA, these amplitudes are always uniform. The choice of basis encoding affects the construction of the whole algorithm, as it informs the construction of the quantum implementations of the boundary operators. Here we briefly describe both of these basis encodings. {An illustrative example of these two encodings for a small complex is given in Figure~\ref{fig:Direct vs Compact Encoding}.} Recall that the vertex set (equivalently the $0$-simplices) of a complex $\mathcal{K}$ are denoted $\{v_1, \dots, v_n\}$.

\paragraph{Direct approach.} The direct encoding was first proposed in Ref.~\cite{Lloyd2016QuantumData} and is defined as follows on an $n$-qubit register. 
{A $k$-simplicial state $\ket{\sigma_k}$ {corresponding to a simplex $\sigma_k = \{v_{i_0}, \dots, v_{i_k}\}$} is an $n$-qubit basis state such that }
\begin{align}\label{direct encoding}
    {\ket{\sigma_k}:=\bigotimes_{i=1}^{n}\ket{u_i},\;\mbox{where}\;u_i=\begin{cases}
        1,\:&\mbox{if $v_i\in\sigma_{k}$,}\\
        0,\:&\mbox{otherwise.}
        \end{cases}}
\end{align}

\paragraph{Compact approach.}  Alternatively, Ref.~\cite{McArdle2022AQubits} proposed a representation that uses exponentially fewer qubits when $k=O(\mathrm{polylog}(n))$, 
and is hence referred to as compact encoding. Let $j\in[k+1]{= \{0, \dots, k\}}$ and $q_{j}\in\{1.\cdots,n\}$. A $k$-simplicial state $\lvert\sigma_{k}\rangle$ is a $((k+{1})\lceil\log_{2}(n+1)\rceil)$-qubit register such that for any $k$-simplex $\sigma_{k}=\{v_{q_{j}}\}_{j\in[k+1]}$
\begin{eqnarray}\label{compact encoding}
    \lvert\sigma_k\rangle:=\bigotimes_{j=0}^{k}\:\lvert q_{j}\rangle,
\end{eqnarray}
for $0<q_0<\cdots<q_k\leq n$. We call each $\ket{q_{j}}$ a vertex state. {If the filtering task is such that $d_u > 0$ (i.e. the polynomial of the upper Laplacian is not a constant) then} there will be an additional {register in state $\ket{0}^{\otimes \lceil\log_2(n+1)\rceil}$} to accommodate the additional vertex state that is created by the action of the coboundary operator $\bkdag$. 

\section{Boundary operator construction}\label{sec:boundaryoperatorconstruction}

In this section we provide a description of the projected unitary encoding (PUE) of the boundary operator $\bk$ that we use in the compact encoding scheme, originally given by McArdle, Gily\'en, and Berta~\cite{McArdle2022AQubits} to study QTDA. This is the main ingredient we use for constructing our quantum simplicial filters. We briefly overview their construction and refer the interested reader to their paper for the full details (\cite{McArdle2022AQubits}, Appendix C2). We slightly modify their construction by changing the implementation of the ``membership oracles'' that determine whether an arbitrary basis state represents a simplex in the simplicial complex. We give the specific implementation in Appendix~\ref{sec:proj}. Our membership oracle is an adaptation of the oracle given in Refs.~\cite{Metwalli2021FindingComputer,berry2023analyzing}, which was originally designed for the direct encoding scheme. For the unfamiliar, we also briefly introduce the concept of PUEs defined by Gily\'en, Su, Low, and Wiebe~\cite{Gilyen2019QuantumArithmetics} in Appendix~\ref{PUE of arbitrary matrices}, along with a summary of the quantum singular value transformation framework. 

Recall that $\alpha_k = \sqrt{(n+1)(k+1)}$ and define $a_k=\lceil\log_2(k+1)\rceil$. Then an $(\alpha_k, a_k,0)$-PUE of $\bk$ is given by Ref.~\cite{McArdle2022AQubits} such that
\begin{equation}\label{compact boundary operator}
\begin{aligned}
    \Bigl(\langle 0\rvert^{\otimes\log_2(k+1)}\otimes\brm{I}^{s}\Bigr)\Bigl(\Pi'_{k-1} \brm{U}_{\bm{\mathrm{B}}_k}\Pi_{k}\Bigr)\Bigl(\lvert 0\rangle^{\otimes\log_2(k+1)}\otimes\brm{I}^{s}\Bigr) &=\frac{\bm{\mathrm{B}}_k}{\alpha_{k}}=:\bkn,
\end{aligned}
\end{equation}
where
\begin{eqnarray*}
    \Pi_{k}&:=& {\sum_{\sigma_k \in \mathcal{K}} \ket{\sigma_k}\bra{\sigma_k}}\\
    \Pi'_{k}&:=& {\sum_{\sigma_k \in \mathcal{K}} \ket{\sigma_k}\bra{\sigma_k} \otimes \ket{0}\bra{0}^{\otimes \log_2(n+1)},}
\end{eqnarray*}
{where the summation range $\sigma_k \in \mathcal{K}$ denotes a summation over all $k$-simplices in the complex. For neatness, w}e omit the ceiling functions in the exponents where it is unambiguous. The unitary $\brm{U}_{\bm{\mathrm{B}}_k}$ acts on $(k+1)\lceil \log_2(n+1) \rceil$ qubits, specifically the qubits corresponding to the registers $\ket{q_0}\ket{q_1}\dots\ket{q_k}$ in Eq.~(\ref{compact encoding}), as well as an ancilla register of size $\lceil\log_2(k+1)\rceil$. Equivalently, $\brm{U}_{\bm{\mathrm{B}}_{k+1}}$ acts on $(k+2)\lceil \log_2(n+1) \rceil$ and $\lceil\log_2(k+2)\rceil$ qubit registers, {where the extra $\lceil \log_2(n+1) \rceil$-qubit register begins in state $\ket{0}$ and is used when applying the upper Laplacian}. 

\paragraph{Direct approach.} {Here we give a brief summary of the PUE of the boundary operator for basis states represented in the direct encoding, as well as its complexity.} 
This approach is {used} in some previous QTDA literature~\cite{Akhalwaya2022TowardsComputers,berry2023analyzing,Kerenidis2022QuantumStates}. {At a high level, the action of the boundary operator in} this approach {can be described by} a superposition of {$n$ tensor products of $n$ Pauli} operators{,} in which each $i$-th operator flips {the} qubit in the $i$-th location in $\lvert\sigma_{k}\rangle$ and change{s the sign based on the number of} $1$s before the $i$-th {qubit}. This operator is called {the} \textit{fermionic Dirac operator} $\brm{D}${,} presented in~{Ref.}~\cite{Akhalwaya2022RepresentationOperator} and given by
\begin{eqnarray}
     \bm{\mathrm{D}}&:=&\sum_{i=1}^{n}\brm{Z}^{\otimes(i-1)}\otimes\brm{X}\otimes\brm{I}^{\otimes(n-i)}.
\end{eqnarray}
The action of this operator {on a $k$-simplicial state (basis state corresponding to a $k$-simplex in the direct encoding, given in (\ref{direct encoding}})) gives {a superposition of} $(k-1)$- and $(k+1)$-simplicial state{s}. However, {the output of the boundary operator should only consist of $(k-1)$-simplices, and} not all {the resulting} states {even} correspond to simplices in $\mathcal{K}$. This is because $\brm{D}$ outputs all possible $(k+1)$-simplicial states, which include states that do not necessarily represent co-faces of $\sigma_{k}$. {This operator is transformed into a PUE of $\bk$ by applying the appropriate left and right projectors. Recall the definition of the} \textit{$k$-simplex identifying oracle} $\bm{\mathrm{P}}_{k}$ {with the action $\brm{P}_k\ket{\sigma_k}\ket{0} = \ket{\sigma_k}\ket{\mathds{1}\{\sigma_k \in \mathcal{K}{\}}}$ which checks} if a given $\lvert \sigma_k\rangle$ is in {the simplicial complex} $\mathcal{K}$ {(by an abuse of notation, we use the same notation {for} $\brm{P}_k$ as {in} the compact encoding). {An implementation of $\brm{P}_k$ for the direct encoding for the case where $\mathcal{K} = \mathcal{K}(G)$ is a clique complex is given in} {Refs.}~\cite{Metwalli2021FindingComputer,berry2023analyzing}. Using a Clifford loader circuit given in {Ref.}~\cite{Kerenidis2022QuantumStates}, we can implement {a} $(\alpha_k,0,0)$-PUE of {$\bk$ in the direct encoding} as
\begin{eqnarray}\label{direct boundary operator}
    \Pi_{k-1}\brm{D}\Pi_{k}&=&\frac{\bm{\mathrm{B}}_k}{\alpha_k},
\end{eqnarray}
where $\Pi_{k-1}={\sum_{\sigma_{k-1} \in \mathcal{K}}\ket{\sigma_{k-1}}\bra{\sigma_{k-1}}}$, $\Pi_{k}={\sum_{\sigma_{k} \in \mathcal{K}}\ket{\sigma_{k}}\bra{\sigma_{k}}}$ {(where the sums are over all $(k-1)$- and $k$-simplices respectively)} act as {the} projection operators defined in Definition~\ref{projected unitary encoding}, and  $\alpha_k=\sqrt{n}$ for all $k$. The implementation of $\brm{D}$ requires a non-Clifford gate depth of $O(\log(n))$ and zero ancilla qubits, while implementing {the corresponding} $\brm{P}_{k}$ in the direct encoding uses $O(n\log(n))$ non-Clifford gate depth and $O(n)$ ancilla qubits~\cite{McArdle2022AQubits}.

\paragraph{Compact approach.} We give a summary of the 
quantum circuit for $\brm{U}_{\brm{B}_k}$ given in Ref.~\cite{McArdle2022AQubits}. The general idea is to perform a quantum operation called the \textsc{Select} operation (we refer the readers to the Appendix~\ref{PUE of arbitrary matrices}) using a $\left(1/\sqrt{k+1}\right)\sum_{j=0}^{k}\ket{j}$ ancilla register that
\begin{itemize}
\item[(1)] swaps the desired vertex state, e.g., $\ket{{q}_{j}}$, to the last register of the system qubit registers using controlled-$\brm{SWAP}$ and integer comparator circuits, 
\item[(2)] checks the value of each vertex state register against the values of the adjacent vertex pair registers, 
\item[(3)] uncomputes the $\left(1/\sqrt{k+1}\right)\sum_{j=0}^{k}\ket{j}$ ancilla qubit, and 
\item[(4)] implements Hadamard gates to the last register. 
\end{itemize}
The result of this unitary applied to some state $\ket{q_0}\ket{q_1}\dots\ket{q_k}$ is to create a superposition of states $\ket{q_0}\ket{q_1}\dots\widehat{\ket{q_j}}\dots\ket{q_{k-1}}\ket{x}$ for each $j \in [k+1]$ and all $x \in [n+1]$, where $\widehat{\ket{q_j}}$ denotes that this register is not present. Then, the projector $\Pi'_{k-1}$ zeroes out any basis state where $x \neq 0$. {The} operator $\brm{U}_{\bk}$ has a non-Clifford gate depth of $O(k\log(\log(n)))$ and requires $O(\log(k))$ ancilla qubits.

{To complete the description of this} PUE, we need to give an implementation of {the $\mathrm{C}_{\Pi_k}\mathrm{NOT}$ gate, which has the action}
\begin{align}\label{eq:cpinot}
    \mathrm{C}_{\Pi_k}\mathrm{NOT}&:= \Pi_k\otimes\brm{X}+\left(\brm{I}-\Pi_k\right)\otimes\brm{I},
\end{align}
{as well as an implementation of the $\mathrm{C}_{\Pi'_k}\mathrm{NOT}$ gate.}
The analogous function given in Ref.~\cite{McArdle2022AQubits} does not quite suit our purpose, as it was designed with persistent homology in mind. {Recall {from Eq.~(\ref{eq:projector})} that $\brm{P}_k$ is the so-called membership oracle defined by the action
\begin{eqnarray*}
    \brm{P}_{k}\lvert\sigma_{k}\rangle\lvert a\rangle\ket{0}^{\otimes a_p}&=&\lvert\sigma_{k}\rangle\lvert a \oplus \mathds{1}\{\sigma_{k}\in\mathcal{K}\}\rangle\ket{0}^{\otimes a_p},
\end{eqnarray*}
where $a\in\{0,1\}$, $a_p$ is the number of ancilla qubits, and $\mathds{1}\{\sigma_{k}\in\mathcal{K}\}$ is the indicator function for the event that $\sigma_k \in \mathcal{K}$ for some $\sigma_k \subseteq \mathcal{V}$. Constructions for $\mathrm{C}_{\Pi_k}\mathrm{NOT}$ and $\mathrm{C}_{\Pi'_k}\mathrm{NOT}$ follow immediately from a construction for $\brm{P}_k$.} In Appendix~\ref{sec:proj}, we give an explicit construction for $\brm{P}_k$ when the simplicial states are described by the compact encoding scheme. {This circuit has a non-Clifford gate depth of $O(n^2k\log(n))$ and uses} $O(k)$ ancillas. {Additionally, note} that the procedure and cost of implementing the PUE{s} of the adjoint operators $\bkdag$ and $\bkkdag$ follow from the fact that $\brm{U}_{\brm{A}^\dagger}=\brm{U}_{\brm{A}}^\dagger$.

\section{Membership oracles for 
 compact approach}\label{sec:proj}

In this section, we give an explicit construction for the membership oracles $\mathbf{P}_k$ in the case where $\mathcal{K}$ is a clique complex and $\ket{\sigma_k}$ is represented by the compact encoding. This is an adaptation of the analogous clique-finding oracle given by Berry et al.~\cite{berry2023analyzing}, which applies for basis states in the direct encoding scheme (their clique-finding oracle is itself based on a similar algorithm by Metwalli, Le Gall, and van Meter~\cite{Metwalli2021FindingComputer}).

In general, for an arbitrary basis state ${\ket{q_0}\ket{q_1}\dots\ket{q_{k}}}$ of the system register containing $\ket{\brm{s}^k}$, we want to construct an operator that flags whether the following properties hold:
\begin{enumerate}[label=(\alph*)]
    \item ${0< q_0} < q_1 < \dots < q_k$,
    \item $\{v_{q_0}, \dots, v_{q_k}\}$ is a simplex in $\mathcal{K}$.     
\end{enumerate}
To check property (a), inequality check the register $\ket{q_{0}}$ against a $\lceil\log(n+1)\rceil$-qubit ancilla register in state {$\ket{\bar{0}}:=\ket{00\dots0}$}. Then perform $k$ inequality checks between neighbouring registers $i$ and $i+1$ for $i \in \{0, \dots, k-1\}$ (that is, checking that {$q_0<q_1<...<q_{k}$}). Property (a) is satisfied if and only if $q_0 > 0${,} and all other inequality checks are satisfied. {This is sufficient for $\mathrm{C}_{\Pi_k}\mathrm{NOT}$. For $\mathrm{C}_{\Pi'_k}\mathrm{NOT}$, we also need to check that $\ket{q_{k+1}} = \ket{\bar{0}}$. This can be done by performing another equality check with the ancilla register mentioned above.} Each of these checks requires a circuit of depth and Toffoli count $O(\log k)$ as well as one ancilla per check, along with a final flag qubit (using the inequality check described in Section 4.3 of Ref.~\cite{Cuccaro2004ACircuit}). The checks can be parallelised such that there are exactly two rounds of checks, and thus the depth is $O(\log k)$ with $k+1$ ancillas and a flag qubit for passing the check.

Checking whether $\{v_{q_0}, \dots, v_{q_k}\}$ corresponds to a $k$-simplex is more difficult. For an arbitrary simplicial complex $\mathcal{K}$, it is not known if there is an efficient quantum implementation of the membership oracle
\begin{align*}
    \ket{x}\ket{0} \mapsto \ket{x} \ket{\mathds{1}\{x \in \mathcal{K}\}},
\end{align*}
where $\ket{x}$ is some basis state representing a subset of the vertex set $\mathcal{V}$ (in compact, direct, or other encodings). If $\mathcal{K}$ is a clique complex and $\ket{x}$ is a basis state in the direct encoding, then {Berry et al.}~\cite{berry2023analyzing} give an implementation, based on work by {Refs.~\cite{Hayakawa2022QuantumAnalysis,Metwalli2021FindingComputer}}, of an oracle that implements 
\begin{align*}
    \ket{x}\ket{0}\ket{0} \mapsto \ket{x} \ket{\mathds{1}\{x \in \mathcal{K}\}}\ket{\mbox{garb}(x)}, 
\end{align*}
where $\mbox{garb}(x)$ represents ancilla registers that are entangled with $\ket{x}$. This implementation requires $O(E)$ Toffolis, where $E \leq \binom{n}{2}$ is the number of edges in the underlying graph $G = (\mathcal{V}, \mathcal{E})$. 

Here we adapt this circuit to the compact encoding scheme. Our circuit has depth and Toffoli count $O( E k^2 \log n)$. The main cause for the increase in complexity is that, in the compact encoding, the location of the register containing any particular integer $j \leq n$ is not fixed. As a toy example, consider the representation of the simplices $\{v_1,v_3,v_4\}$ and $\{v_1,v_4,v_5\}$ in each encoding. In the direct encoding, these simplices are represented by the states $\ket{10110}$ and $\ket{10011}$ (assuming $n=5$). Notably, the same qubit is used to denote the presence or absence of a particular vertex $v_i$ in each state. On the other hand, in the compact encoding, these two simplices would be represented by the states $\ket{001}\ket{011}\ket{100}$ and $\ket{001}\ket{100}\ket{101}$. Both simplices contain vertex $v_4$, but the register containing this information is different for each basis state. This means that checking for the existence of edges between vertices in a simplex is requires checking between each pair of vertices in all pairs of registers. This increases the circuit complexity considerably.

The broad idea is as follows. We first define an edge-checking sub-circuit $C_{ij}$ that acts on two registers each of size $\lceil\log (n+1)\rceil$ (e.g. $\ket{q_i}$ and $\ket{q_j}$) and an ancilla qubit initially in state $\ket{0}$. This circuit consists of $E$ different $(2\lceil\log (n+1)\rceil)$-controlled NOT gates, which check whether $\ket{q_i}\ket{q_j} = \ket{a}\ket{b}$ for each $v_av_b \in \mathcal{E}$. {The circuit} $C$ is then implemented across each pair of registers $(i,j)$ where $0 \leq i < j \leq k$. If $v_{q_i} v_{q_j} \in \mathcal{E}$, then the edge checking block $C_{ij}$ flips the ancilla. This in turn controls an increment block on another register of $2\lceil\log(k+1)\rceil$ ancillas. We perform the edge check $C_{ij}$ again to reset the ancilla, and then do the same thing with the next edge-checking block. After all $C_{ij}$ blocks have been run, the ancilla register is in its original state, and the incrementing register contains a binary representation of the number of edges between $\{v_{q_0}, \dots, v_{q_{k}}\}$ in $G$. Finally, we perform an equality check between this result of this register and an ancilla register containing the value $\binom{k+1}{2}${; the result of this is stored in} a flag qubit initially set to $\ket{0}$. This qubit is in state $\ket{1}$ if and only if $\{v_{q_0}, \dots, v_{q_k}\}$ form a clique in $G$. The rest of the circuit can be then uncomputed by performing the operations in reverse. A high-level circuit diagram of this is given in Figure \ref{fig:proj}. 

Each $2\lceil\log (n+1)\rceil$-controlled NOT gate in $C$ can be implemented in $O(\log n)$ Toffoli gates, and thus the depth and Toffoli of each block $C$ is $O(E\log n)$. There are $\binom{k+1}{2}$ pairs of registers, and thus $C$ must be implemented $2\binom{k+1}{2}$ times to account for uncomputation of the ancilla. An increment block on $2\lceil\log (k+1)\rceil$ qubits can be implemented in $O(\log k)$ Toffoli-or-smaller gates with a single ancilla~\cite{Metwalli2021FindingComputer}{.} This also needs to be implemented $\binom{k+1}{2}$ times. After this, {the value of this register} is checked against a register containing the value $\binom{k+1}{2}$. This equality checking can be done as two different inequality checks as outlined in Section 4.3 of Ref.~\cite{Cuccaro2004ACircuit}, which each use one ancilla and $O(\log k)$ gates. The output of these two inequality checks (stored in the ancillas) is then added onto the flag qubit using a $00$-controlled Toffoli gate. The values in the ancillas are all uncomputed by reversing the circuit.
Thus, the total depth of the clique-checking circuit is $O(Ek^2\log n)$, as is the total number of Toffoli gates required, and there are {$O(\log k)$} ancillas used. 

Finally, we apply a single Toffoli gate to the flag qubits for checking properties (a) and (b). The overall action of this is 
\begin{align*}
    {\ket{x} \ket{0}^{\otimes a} \ket{0} \mapsto \ket{x} \ket{0}^{\otimes a} \ket{\mathds{1}\{x \in \mathcal{K}\}}}
\end{align*}
as required, for some $a \in O(k)$. Thus, the depth and Toffoli count of $\brm{P}_k$ in the compact encoding is $O(Ek^2\log n)$, since the cost of checking property (a) is comparably negligible. 

We note that the depth of this circuit can be reduced by increasing the number of ancillas used in edge-checking by $k/2$. This allows us to parallelise many of the $C_{ij}$ applications and check for the existence of all pairs of edges in either $k$ or $k+1$ rounds (depending on the parity of $k$), rather than $\binom{k+1}{2}$. In this case, each increment block is controlled by every ancilla, and the final value of this register is compared against $k$ (or $k+1$ depending on parity), instead of $\binom{k+1}{2}$. In this case the ancilla register containing the increment blocks does not count the number of edges between $\{v_{q_0}, \dots, v_{q_k}\}$ in $G$, instead grouping edges together and counting whether all the blocks have the correct number of edges in them. In this case the number of ancillas increases by approximately $k/2 - 2\log(k+1)$, the number of Toffoli gates remains similar, and the depth changes to $O(Ek\log n)$ (the depth for checking property (a) is still negligible).

\begin{figure*}[t]
    \centering
    \includegraphics[scale=0.55]{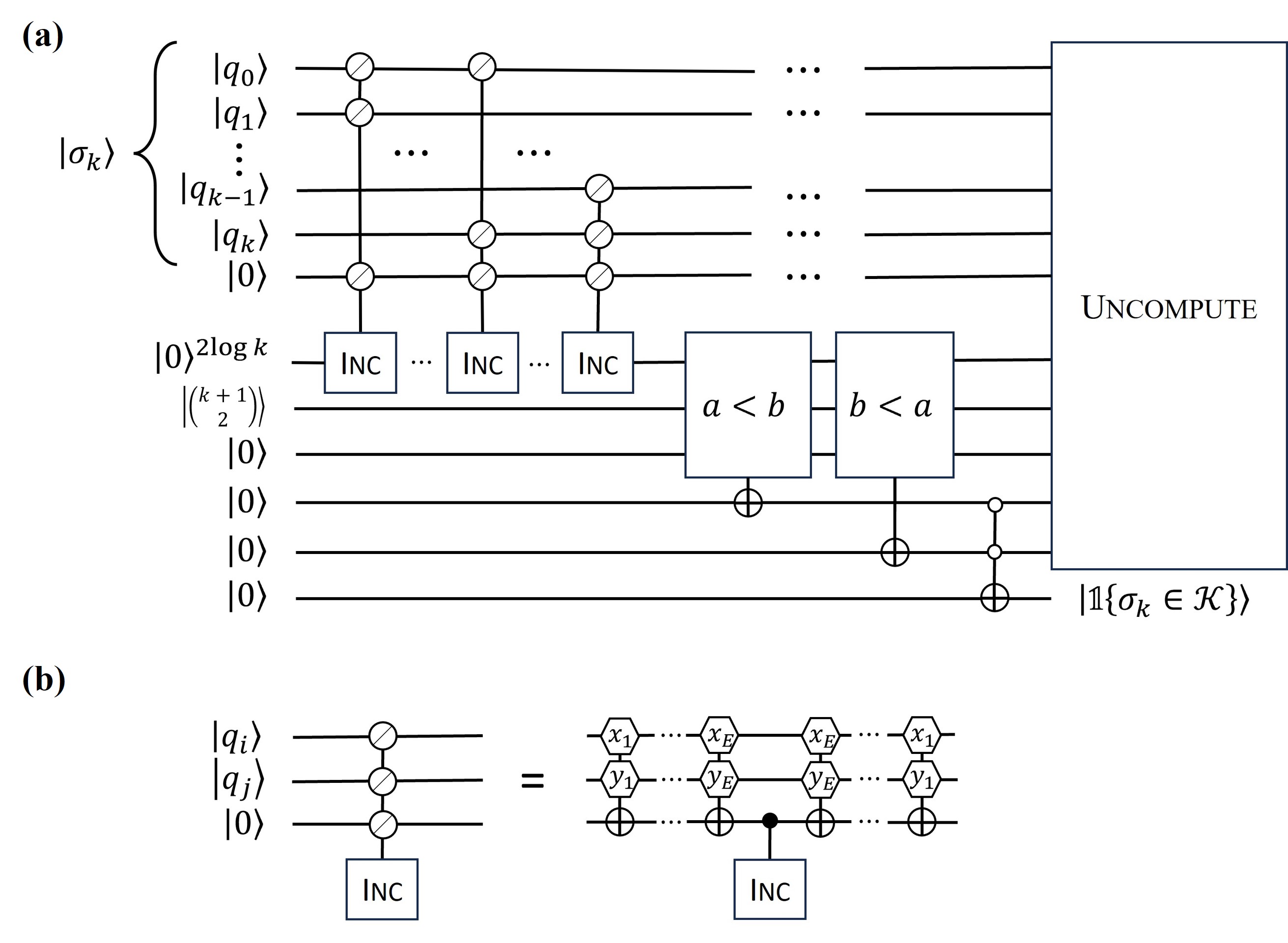}
    \caption{(a) The circuit that implements $\mathbf{P}_k$ which maps $\ket{\sigma_k}\ket{0} \mapsto \ket{\sigma_k}\ket{\mathds{1}\{\sigma_k \in \mathcal{K}\}}$, where $\ket{\sigma_k}$ is a basis state satisfying property (a) and $q_i$ is the binary representation of vertex $v_{q_i}$. The $\oslash$ notation is described in the following part. (b) The edge-checking circuit $C$ checking for the existence of an edge between vertices $q_i$ and $q_j$. The $\oslash$ is shorthand for a sequence of multi-controlled CNOT operations, described in more detail on the right-hand side. The pairs $x_i < y_i$ form a list of all edges in $G$. A number $i$ inscribed in a hexagon indicates that this operation is controlled on that register containing $i$ in binary.}
    \label{fig:proj}
\end{figure*}

\section{Proof of Lemma~\ref{LCU of many polynomials}}\label{Proof of Lemma LCU of many polynomials}
\label{app:lemma1proof}
\begin{proof}
Let $\brm{U}_{\brm{A}_{i}}$,$\Pi'_{i}$,$\Pi_{i}$ be as in Definition~\ref{projected unitary encoding} for all $i\in[m]$. Theorem~\ref{QSVT} and Corollary~\ref{QSVT real polynomial} guarantee that there exist $2m$ sequences of phase factors {$\left\{\bm{\Phi}_{g_{i}},\bm{\Phi}_{h_{i}}\right\}_{\{i \in [m]\}}$} and the corresponding complex-valued polynomials {$\left\{g_{i},h_{i}\right\}_{\{i \in [m]\}}$ such that}
\begin{eqnarray}
    {\Re(g_i)(\brm{A}_i):=}g_{\Re,i}(\brm{A}_{i})&=&\biggl(\langle +\rvert \otimes\Bigl(\langle 0\rvert^{\otimes a_i}\otimes \brm{I}^{s}\Bigr)\biggr)\brm{W}_{g_{\Re,i}}\biggl(\lvert + \rangle\otimes \Bigl(\lvert 0\rangle^{\otimes a_i}\otimes \brm{I}^{s}\Bigr)\biggr),\\
    {\imath\Re(h_i)(\brm{A}_i):=}\imath h_{\Re,i}(\brm{A}_{i})&=&\biggl(\langle +\rvert\otimes\Bigl(\langle 0\rvert^{\otimes a_i}\otimes \brm{I}^{s}\Bigr)\biggr)\brm{W}_{\imath h_{\Im,i}}\biggl(\lvert + \rangle\otimes \Bigl(\lvert 0\rangle^{\otimes a_i}\otimes \brm{I}^{s}\Bigr)\biggr),  
\end{eqnarray}
where the \textsc{Select} operation{s} $\brm{W}_{g_{\Re,i}}$ and $\brm{W}_{\imath h_{\Im,i}}$ are given {by}
\begin{eqnarray}
    \brm{W}_{g_{\Re,i}}&:=&\lvert 0\rangle\langle 0\rvert\otimes\Pi'_{i}\brm{U}_{\brm{A}_i}^{\bm{\Phi}_{g_i}}\Pi_{i}+\lvert 1\rangle\langle 1\rvert\otimes\Pi'_{i}\brm{U}_{\brm{A}_i}^{-\bm{\Phi}_{g_i}}\Pi_{i},\label{Select g}\\
    \brm{W}_{\imath h_{\Im,i}}&:=&\Bigl(\brm{Z}\otimes\brm{I}^{a_i+s+1}\Bigr)\Bigl(\lvert 0\rangle\langle 0\rvert\otimes\Pi'_{i}\brm{U}_{\brm{A}_i}^{\bm{\Phi}_{h_i}}\Pi_{i}+\lvert 1\rangle\langle 1\rvert\otimes\Pi'_{i}\brm{U}_{\brm{A}_i}^{-\bm{\Phi}_{h_i}}\Pi_{i}\Bigr).\label{Select h}
\end{eqnarray}
Then, we define a new \textsc{Select} operation $\brm{W}_{f}$ {as }
\begin{eqnarray}\label{Select f}
    \brm{W}_{f}:=\sum_{i=0}^{m-1}\,\left[\lvert i\rangle\langle i\rvert\otimes \Bigl(\lvert 0\rangle\langle 0\rvert\otimes \brm{W}_{g_{\Re,i}}+\lvert 1\rangle\langle 1\rvert\otimes \brm{W}_{\imath h_{\Im,i}}\Bigr)+\Bigl(\brm{I}^{b}-\lvert i\rangle\langle i\rvert\Bigr)\otimes\brm{I}\otimes\brm{I}^{a+s+2}\right]{,}
\end{eqnarray} 
where $\ketbra{i}{i}$ and $\brm{I}^b - \ketbra{i}{i}$ are acting on a $b=\lceil\log(2m)\rceil$-qubit ancilla register. 
The overall \textsc{Select} operation $\brm{W}_{f}$ requires $(a+b+3)$ ancilla qubits. Let $\brm{G}$ be a \textsc{Prepare} operation as defined in Eq.~(\ref{prepare operation}){,} acting on the $b$-qubit register, and recall that $\beta=\lVert c\rVert_{1}$. Then
\begin{eqnarray*}
   \beta\,\Bigl(\langle \brm{G}\rvert\otimes\langle 0\rvert^{\otimes (a+3)}\otimes \brm{I}^{s}\Bigr)\brm{W}_{f}\Bigl(\lvert \brm{G}\rangle\otimes\lvert 0\rangle^{\otimes (a+3)}\otimes \brm{I}^{s}\Bigr)&=&\sum_{i=0}^{m-1}\,c_{i}\Bigl(g_{\Re,i}(\brm{A}_{i})+\imath h_{\Im,i}(\brm{A}_{i})\Bigr)\\
    &=& \sum_{i=0}^{m-1} c_i\, f(\brm{A}_{i}),
\end{eqnarray*}
proving that $\Bigl(\langle \brm{G}\rvert\otimes\langle 0\rvert^{\otimes (a+3)}\otimes \brm{I}^{s}\Bigr)\brm{W}_{f}\Bigl(\lvert \brm{G}\rangle\otimes\lvert 0\rangle^{\otimes (a+3)}\otimes \brm{I}^{s}\Bigr)$ is $(\beta,a+\lceil\log(2m)\rceil+3,0)$-PUE of $\sum_{i=0}^{m-1} c_i\, f_{i}(\brm{A}_i)$.

Now we prove the second claim in the theorem. {Given a $\delta > 0$ and}  $O({m}\,\mathrm{poly}\left(d,1/\delta\right))$ classical computation {time}, {one} can find $2m$ sequences of phase factors $\left\{\bm{\Phi}'_{g_{i}},\bm{\Phi}'_{h_{i}}\right\}$ and the corresponding complex-valued polynomials $\left\{\mathrm{Poly}_{g_{i}},\mathrm{Poly}_{h_{i}}\right\}$ for all $i\in[m]$, such that $\Re\left(\mathrm{Poly}_{g_{i}}\right)=\mathrm{Poly}_{g_{\Re,i}}$ and $\imath\Im\left(\mathrm{Poly}_{h_{i}}\right)=\mathrm{Poly}_{\imath h_{\Im,i}}$ satisfy  
\begin{eqnarray*}
    \Bigl\lVert g_{\Re,i}(\brm{A}_{i})-\mathrm{Poly}_{g_{\Re,i}}(\brm{A}_{i})\Bigr\rVert\leq {\frac{\delta}{2}},\\
    \Bigl\lVert \imath h_{\Re,i}(\brm{A}_{i})-\mathrm{Poly}_{\imath h_{\Im,i}}(\brm{A}_{i})\Bigr\rVert\leq {\frac{\delta}{2}},
\end{eqnarray*}
{and thus}
\begin{eqnarray*}
    \biggl\lVert f_{i}(\brm{A}_{i})-\Bigl(\mathrm{Poly}_{g_{\Re,i}}(\brm{A}_{i})+\mathrm{Poly}_{\imath h_{\Im,i}}(\brm{A}_{i})\Bigr)\biggr\rVert\leq \frac{\delta}{2} + \frac{\delta}{2}=\delta.
\end{eqnarray*}
Here, we change $\brm{U}_{\brm{A}_{i}}^{\bm{\phi}_{g_{i}}}$ and $\brm{U}_{\brm{A}_{i}}^{\bm{\phi}_{h_{i}}}$ in Eqs.~(\ref{Select g}) and~(\ref{Select h}) {to} $\brm{U}_{\brm{A}_{i}}^{\bm{\phi}'_{g_{i}}}$ and $\brm{U}_{\brm{A}_{i}}^{\bm{\phi}'_{h_{i}}}${, and in turn define the corresponding \textsc{Select} operation $\brm{W}_{f'}$ analogously to $\brm{W}_{f}$ in Eq.~(\ref{Select f}) to build} $\mathrm{Poly}_{g_{\Re,i}}(\brm{A}_{i})$ and $\mathrm{Poly}_{\imath h_{\Im,i}}(\brm{A}_{i})$ for all $i\in[m]$. {Thus,}
\begin{eqnarray*}
    \biggl\lVert \sum_{i=0}^{m-1} c_i\, f(\brm{A}_{i})&-&\beta\,\Bigl(\langle \brm{G}\rvert\otimes\langle 0\rvert^{\otimes (a+3)}\otimes \brm{I}^{s}\Bigr)\brm{W}_{f}\Bigl(\lvert \brm{G}\rangle\otimes\lvert 0\rangle^{\otimes (a+3)}\otimes \brm{I}^{s}\Bigr)\biggr\rVert\\
    &=&\biggl\lVert \sum_{i=0}^{m-1} c_i\, f(\brm{A}_{i})-c_{i}\Bigl(\mathrm{Poly}_{g_{\Re,i}}(\brm{A}_{i})+\mathrm{Poly}_{\imath h_{\Im,i}}(\brm{A}_{i})\Bigr)\biggr\rVert\\
    &\leq&{\sum_{i=0}^{m-1} c_i\,\biggl\lVert f(\brm{A}_{i})-\Bigl(\mathrm{Poly}_{g_{\Re,i}}(\brm{A}_{i})+\mathrm{Poly}_{\imath h_{\Im,i}}(\brm{A}_{i})\Bigr)\biggr\rVert}\\
    &\leq&\sum_{i=0}^{m-1} c_i\delta=\beta\delta{,}
\end{eqnarray*}
proving that $\Bigl(\langle \brm{G}\rvert\otimes\langle 0\rvert^{\otimes (a+3)}\otimes \brm{I}^{s}\Bigr)\brm{W}_{f}\Bigl(\lvert \brm{G}\rangle\otimes\lvert 0\rangle^{\otimes (a+3)}\otimes \brm{I}^{s}\Bigr)$ is $(\beta,a+\lceil\log(2m)\rceil+3,\beta\delta)$-PUE {of $\sum_{i=0}^{m-1} c_i\, f_{i}(\brm{A}_i)$.} 
\end{proof}

\section{Proof of Lemma~\ref{QTSP Chebyshev filter design}}
\begin{proof}
As described in Section~\ref{sec:boundaryoperatorconstruction}, {there exists} a $\left(\alpha_{k},a_{k},0\right)$-PUE of $\bk$ and a $\left(\alpha_{k+1},a_{k+1},0\right)$-PUE of $\bkk$. {We} make use of Lemma~\ref{LCU of many polynomials} {to create a linear combination of} two even polynomials of $\bkn:=\bk/\alpha_{k}$ and $\bkkdagn:=\bkk/\alpha_{k+1}$ and an identity map. Let $f(x) = x^2$; clearly $|f(x)| \in [0,1]$ for all $x \in [-1,1]$. Given 
{$g^{\mathrm{G}},g^{\mathrm{C}}$ as defined by their polynomial coefficients
$\left\{\mathfrak{h}_{0},\mathfrak{h}_{i_{\ell}}^{\ell},\mathfrak{h}_{i_{u}}^{u}\right\}$,
we} define two even polynomials $h^{\mathrm{G}}:=g^{\mathrm{G}}\circ f$ and $h^{\mathrm{C}}:=g^{\mathrm{C}}\circ f$. Since $g^{\mathrm{G}},g^{\mathrm{C}}:[0,1]\to[-1,1]$, it follows that $h^{\mathrm{G}},h^{\mathrm{C}}:[-1,1]\to[-1,1]$. Thus, Lemma~\ref{LCU of many polynomials} implies that there exist two {sequences of phase factors} $\bm{\phi}_{\ell} \in\mathbb{R}^{2d_{\ell}},\bm{\phi}_{u}\in\mathbb{R}^{2d_{u}}$ and the corresponding {alternating phased sequences} $\brm{U}_{\bk}^{\bm{\phi}_{\ell}}$, $\brm{U}_{\bkkdag}^{\bm{\phi}_{u}}$ (these terms are defined explicitly in Append{ices}~\ref{PUE of arbitrary matrices} and~\ref{quantum algorithm for constructing polynomials}) which construct $h^{\mathrm{G}}\left(\bkn\right)=g^{\mathrm{G}}\left(\hlln\right)$ and $h^{\mathrm{C}}\left(\bkkdagn\right)=g^{\mathrm{G}}\left(\hlun\right)$. Applying {the construction in} Lemma~\ref{LCU of many polynomials} with $c_{0}=\mathfrak{h}_{0}$ {and} $c_{1}=c_{2}=1$ {gives} a $\Bigl(\beta,a_{k}+a_{k+1}+a_p+6,0\Bigr)$-PUE of $H_d\left(\hlln,\hlun\right)${, where $\beta = \|c\|_1 = \mathfrak{h}_0+2$}.

To calculate the {precise} cost, Theorem~\ref{QSVT} states that implementing $\brm{U}_{\bk}^{\bm{\Phi}_{\ell}}$, $\brm{U}_{\bkkdag}^{\bm{\Phi}_{u}}$, $\brm{U}_{\bk}^{-\bm{\phi}_{\ell}}$, and $\brm{U}_{\bkkdag}^{-\bm{\phi}_{u}}$ require{s} $4d_{\ell}$ calls to $\brm{U}_{\bk},\brm{U}_{\bk}^{\dagger}$ and $4d_{u}$ calls to $\brm{U}_{\bkk},\brm{U}_{\bkk}^{\dagger}$ respectively, with $8d_{\ell}$ calls of $\brm{C}_{\Pi'_{k-1}}\brm{NOT},\brm{C}_{\Pi_{k}}\brm{NOT}$ and $8d_{u}$ calls to $\brm{C}_{\Pi'_{k}}\brm{NOT},\brm{C}_{\Pi_{k+1}}\brm{NOT}$. Furthermore, {Lemma~\ref{LCU of many polynomials} implies that we require} $a=a_{k}+a_{k+1}+a_{p}$ ancilla qubits for constructing $\brm{U}_{\bk}^{\bm{\phi}_{\ell}}$ and $\brm{U}_{\bkkdag}^{\bm{\phi}_{u}}$ (recall that $a_{p}$ is the ancilla qubits used in the membership oracle {$\brm{P}_k$}, which are uncomputed during each application and can thus be reused), $b=2$ ancilla qubits for the \textsc{Prepare} operation, and two ancilla qubits for constructing the real-valued polynomials. There are two additional qubits introduced because we differentiate the ancilla used in the controlled phase rotations in QSVT to construct $\brm{U}_{\bk}^{\bm{\Phi}_\ell}$ and $\brm{U}_{\bkkdag}^{\bm{\Phi}_u}$.
\end{proof}

\begin{remark}\label{rem:approx-lemma2}
As described in Lemma~\ref{LCU of many polynomials}, {given $\delta > 0$ and} using $O(\mathrm{poly}(d,\log(1/\delta))$ classical computation time one can obtain two sequences of phase factors $\bm{\Phi}'_{\ell} \in\mathbb{R}^{2d_{\ell}},\bm{\Phi}'_{u}\in\mathbb{R}^{2d_{u}}$, and the corresponding alternating phased sequences $\brm{U}_{\bk}^{\bm{\Phi}_{\ell}}$, $\brm{U}_{\bkkdag}^{\bm{\Phi}_{u}}$, which construct polynomials $\p{h^{\mathrm{G}}}{\Bigl(\bkn\Bigr)}$, $\p{h^{\mathrm{C}}}{\Bigl(\bkkdagn\Bigr)}$  of degree $2d_{\ell}$ and $2d_{u}$ respectively such that
\begin{eqnarray*}\label{QTSP lower filter approximation}
    \norm{h^{\mathrm{G}}\left(\bkn\right)}{\p{h^{\mathrm{G}}}{\Bigl(\bkn\Bigr)}}\leq
    \delta/2
    \quad\mbox{and}\quad
    \norm{h^{\mathrm{C}}(\bkkdagn)}{\p{h^{\mathrm{C}}}{\Bigl(\bkkdagn\Bigr)}}\leq
    \delta/2.
\end{eqnarray*}
Implementing the LCU method as in Lemma~\ref{LCU of many polynomials}, we obtain
\begin{eqnarray*}
    \norm{H_d\left(\hlln,\hlun\right)}{\beta\,\left(\p{h^{\mathrm{G}}}{\Bigl(\bkn\Bigr)}+\p{h^{\mathrm{C}}}{\Bigl(\bkkdagn\Bigr)}-\mathfrak{h}_{0}\brm{I}\right)}\leq\beta\delta,
\end{eqnarray*}
which gives a $\beta\delta$-approximation to $H_d/\beta$.
\end{remark}

\section{Proof of Lemma~\ref{Hodge decomposition filter}}\label{Proof of Lemma Hodge decomposition filter}
\begin{proof}
We focus our calculation on approximating the gradient projector, as the calculation for the curl projector is analogous. Let $\varepsilon'_\ell = \frac{\alpha_{k}^2{\varepsilon_{\ell}}}{n}$ and $\varepsilon'_u = \frac{\alpha_{k+1}^2{\varepsilon_u}}{n}$, where $\varepsilon_{\ell},\varepsilon_u\in(0,1/2)$. Recall that $\kappa_\ell$ is such that $ \kappa_\ell^{-1} \in(0, \alpha_{k}/\xi_{\mathrm{min}}^\ell)$ and $\kappa_u$ is such that $ \kappa_u^{-1} \in(0, \alpha_{k+1}/\xi_{\mathrm{min}}^u)$. Thus, all eigenvalues of $\hl^\ell$ and $\hl^u$ lie within $[1/\kappa_\ell^2,1]$ and $[1/\kappa_u^2,1]$, respectively. Theorem 41 of Ref.~\cite{Gilyen2019QuantumArithmetics} states that there exists a real {degree-$d$} polynomial ${g_d}:[-1,1]\to[-1,1]$ that $\frac{\varepsilon'}{2\kappa}$-approximates the function $\frac{1}{2\kappa x}$ for all $x \in [-1,1] \backslash \left( -\frac{1}{\kappa}, \frac{1}{\kappa}\right)$ such that $|{g_d}(x)| \leq 1$ for all $x \in [-1,1]$, ${g_d}(0) = 0$, {and $d = O(\kappa \log(\kappa/\varepsilon'))$}. For approximating the gradient projector, we set $\kappa := \kappa_\ell^2$ {and} $\varepsilon' := \varepsilon_\ell'$, (naturally setting $\kappa := \kappa_u^2$ {and} $\varepsilon' := \varepsilon_u'$, and adjusting $d$ accordingly when approximating the curl projector). {Note that the nonzero spectra of $\hl^\ell$ and $\brm{L}_{k-1}^u$ are the same~\cite{Steenbergen2012AComplexes}, and so their minimum singular values are equal.}

Now consider the function ${H_{d+1}}(x) = x {g_{d}}(x)$, defined on $[-1,1]$. The aforementioned properties of $g_{d}(x)$ imply that 
\begin{align*}
   {\sup_{|x|\leq 1} |{H_{d+1}}(x^2)| = \sup_{|x|\leq 1} |x^2 {g_d}(x^2)| \leq \sup_{|x|\leq 1} |{g_d}(x^2)| \leq 1.}
\end{align*}
If ${g_d}(x)$ is a degree-$d$ polynomial of $x^2$ then ${H_{d+1}}(x^2)$ is a degree-$(d+1)$ polynomial of $x^2$ (equivalently a degree-($2d+2$) polynomial of $x$). {Suppose that $g_d(x) = \sum_{i=0}^d c_i x^i$ for some real constants $c_i$. We also suppose that the singular value decomposition of $\bk$ is given by $\bk = \sum_{j} \xi_j\ketbra{\tilde{\psi}_j}{\psi_j}$ where $\xi_j \geq 0$. It follows that $\brm{L}_{k-1}^u = \bk\bk^\dagger = \sum_{j} \lambda_j \ketbra{\tilde{\psi}_j}{\tilde{\psi}_j}$ where $\lambda_j = \xi_j^2$. This implies that }
\begin{align*}
    H_{d+1}\left( \frac{\hll}{\alpha_{k}^2} \right) &= \frac{1}{\alpha_{k}^2} \bk^\dagger \bk g_d\left( \frac{\bk^\dagger\bk}{\alpha_{k}^2} \right)
    = \frac{1}{\alpha_{k}^2} \bk^\dagger \bk \left(\sum_{i=0}^d c_i \left( \frac{\bk^\dagger\bk}{\alpha_{k}^2} \right)^i\right)
    = \frac{1}{\alpha_{k}^2} \bk^\dagger \left(\sum_{i=0}^d c_i \left( \frac{\bk\bk^\dagger}{\alpha_{k}^2} \right)^i\right) \bk\\
    &= \frac{1}{\alpha_{k}^2} \bk^\dagger \left(\sum_{i=0}^d c_i \left( \frac{\brm{L}_{k-1}^u}{\alpha_{k}^2} \right)^i\right) \bk
    = \frac{1}{\alpha_{k}^2} \bk^\dagger \frac{\alpha_{k}^2}{2\kappa_{k}^2}\left( \brm{L}_{k-1}^u \right)^+ \bk + \frac{1}{\alpha_{k}^2}\bk^\dagger\sum_{j} \delta_j'\ketbra{\tilde{\psi}_j}{\tilde{\psi}_j} \bk
\end{align*}
{where $\delta_j' \in \left[-\frac{\varepsilon'_\ell}{2\kappa_\ell^2}, \frac{\varepsilon'_\ell}{2\kappa_\ell^2} \right]$. Thus,}
\begin{align*}
    H_{d+1}\left( \frac{\hll}{\alpha_{k}^2} \right) &= \frac{1}{2\kappa_{\ell}^2} \bk^\dagger (\brm{L}_{k-1}^u)^+ \bk + \bk^\dagger\left(\sum_{j} \delta_j \ketbra{\tilde{\psi}_j}{\tilde{\psi}_j}\right) \bk = \frac{1}{2\kappa_{\ell}^2} \bk^\dagger (\brm{L}_{k-1}^u)^+ \bk +\sum_{j} \lambda_j \delta_j \ketbra{\psi_j}{\psi_j},
\end{align*}
{for $\delta \in \left[ -\frac{{\varepsilon_\ell}}{2n\kappa_{\ell}^2}, \frac{{\varepsilon_\ell}}{2n\kappa_\ell^2} \right]$. It is shown in Ref.~\cite{Horak2013SpectraComplexes} that $\lambda_i \leq n$ for all $i$. Recall that $\bm{\mathcal{B}}_k = \bk/\alpha_{k}$. Thus,}
\begin{align*}
    \left\|  \Pi^\mathrm{G} - 2\kappa_\ell^2\, H_{d+1}\left( \frac{\hll}{\alpha_{k}^2} \right)\right\| &= \left\|2\kappa_\ell^2\sum_{j} \lambda_j \delta_j \ketbra{\psi_j}{\psi_j}\right\| \leq \left\|2\kappa_\ell^2 \sum_{j} \frac{{\varepsilon_\ell}}{2\kappa_\ell^2} \ketbra{\psi_j}{\psi_j}\right\| \leq {\varepsilon_\ell}.
\end{align*}
{Note that ${H_{d+1}}$ is a polynomial of degree $2d+2$ where $d = O({\kappa_\ell^2}\log((n\kappa_\ell)^2/(\alpha^2_{k}\varepsilon_\ell)))$. This completes the proof of the approximation of the gradient projector{;} as mentioned above, the approximation of the curl projector is analogous.}
\end{proof}

\begin{remark}\label{rem:alt-projector}
    We can also define $H'_{d+1}(x) = xg_d(x^2)$. This approximates a modified projector $\Pi^{\mathrm{G}'}$ with the action
    \begin{align*}
        \Pi^{\mathrm{G}'} \brm{s}^k = \brm{s}^{k-1}
    \end{align*}
    where $\brm{s}^{k-1}$ is the (minimum Euclidean norm) vector such that $\brm{B}^T_k \brm{s}^{k-1} = \brm{s}^k$.
    The function $H'_{d+1}(x)$ is a degree-$(2d+1)$ polynomial that also satisfies $|H'_{d+1}(x)| \leq 1$ for all $x \in [-1,1]$. Performing a similar calculation to above, it follows that there exists some $d \in O(\kappa_\ell^2\log(n\kappa_\ell^2/(\alpha_{k}\varepsilon_\ell)))$ such that $\big\|\Pi^{\mathrm{G}'} - 2\kappa_\ell^2/\alpha_{k}\,H'_{d+1}(\bkn)\big\| \leq \varepsilon$.  However, note that this filter is not defined similarly to the simplicial filter that is given in Eq.~\eqref{TSPfilter}. This is because $\Pi^{\mathrm{G}'}$ projects a simplicial signal to the lower dimension, which is not the case generally considered in TSP filtering tasks. Nevertheless, this projection is beneficial for analyzing pairwise comparison data in higher-order networks.
\end{remark}

\end{document}